\documentclass[a4paper,fleqn,usenatbib]{mnras}

\usepackage{newtxtext,newtxmath}
\usepackage[T1]{fontenc}
\usepackage{ae,aecompl}

\usepackage{graphicx}	% Including figure files
\usepackage{amsmath}	% Advanced maths commands
\usepackage{amssymb}	% Extra maths symbols
\usepackage{pifont}
\usepackage{afterpage}
\usepackage{verbatim}
\usepackage{caption}
\usepackage{subfig, float}

\title[AGN feedback in a multiphase ISM]{Impact of AGN feedback on galaxies and their
multiphase ISM across cosmic time}

\author[M. Valentini et al.]
{Milena Valentini$^{1,2,3,4}$\thanks{E-mail: valentini@usm.lmu.de}, 
Giuseppe Murante$^{3}$\thanks{E-mail: murante@oats.inaf.it},
Stefano Borgani$^{3,5,6,7}$\thanks{E-mail: borgani@oats.inaf.it},
\newauthor
Gian Luigi Granato$^{3}$, 
Pierluigi Monaco$^{3,5,7}$, 
Fabrizio Brighenti$^{8}$, 
Luca Tornatore$^{3,7}$, 
\newauthor
Alessandro Bressan$^{4,9}$, and 
Andrea Lapi$^{4}$
\\ ~ \\
\footnotesize
$^{1}$ Universit{\"a}ts-Sternwarte M{\"u}nchen, Fakult{\"a}t f{\"u}r Physik, LMU Munich, Scheinerstr. 1, 81679 M{\"u}nchen, Germany\\
$^{2}$ Scuola Normale Superiore, Piazza dei Cavalieri 7, I-56126 Pisa, Italy\\ 
$^{3}$ INAF - Osservatorio Astronomico di Trieste, via Tiepolo 11, I-34131 Trieste, Italy\\
$^{4}$ SISSA - International School for Advanced Studies, via Bonomea 265, I-34136 Trieste, Italy\\
$^{5}$ Astronomy Unit, Department of Physics, University of Trieste, via Tiepolo 11, I-34131 Trieste, Italy\\
$^{6}$ INFN - National Institute for Nuclear Physics, Via Valerio 2, I-34127 Trieste, Italy\\
$^{7}$ IFPU - Institute for Fundamental Physics of the Universe, Via Beirut 2, 34014 Trieste, Italy\\
$^{8}$ Dipartimento di Fisica e Astronomia, Universit\`a di Bologna, Via Gobetti 93/2, 40122, Bologna, Italy\\
$^{9}$ INAF - Osservatorio Astronomico di Padova, Vicolo dell'Osservatorio 5, I-35122 Padova, Italy\\
}

\date{Accepted 2019 November 05. Received 2019 November 05; in original form 2019 June 04}

\pubyear{2018}

\begin{document}
\label{firstpage}
\pagerange{\pageref{firstpage}--\pageref{lastpage}}
\maketitle

% Abstract of the paper
\begin{abstract}

\noindent  
We present simulations of galaxy formation, based on the GADGET-3 code, in which a sub-resolution model 
for star formation and stellar feedback is interfaced with a new model for AGN feedback. 
Our sub-resolution model describes a multiphase ISM, 
accounting for hot and cold gas within the same resolution element: we exploit 
this feature to investigate the impact of coupling AGN feedback energy to the different phases 
of the ISM over cosmic time. Our fiducial model considers that AGN feedback energy coupling is driven by the 
covering factors of the hot and cold phases. 
We perform a suite of cosmological hydrodynamical simulations of disc galaxies 
($M_{\rm halo, \, DM} \simeq 2 \cdot 10^{12}$~M$_{\odot}$, at $z=0$), to investigate: 
$(i)$ the effect of different ways of coupling AGN feedback energy to the 
multiphase ISM; 
$(ii)$ the impact of different prescriptions for gas accretion (i.e. only cold gas, both cold and hot gas, with the 
additional possibility of limiting gas accretion from cold gas with high angular momentum); 
$(iii)$ how different models of gas accretion and coupling of AGN feedback energy affect the 
coevolution of supermassive BHs and their host galaxy. 
We find that at least a share of the AGN feedback energy has to couple with 
the diffuse gas, in order to avoid an excessive growth of the BH mass.  
When the BH only accretes cold gas, it experiences a growth that is 
faster than in the case in which both cold and hot gas are accreted. 
If the accretion of cold gas with high angular momentum is reduced, the 
BH mass growth is delayed, the BH mass at $z=0$ is reduced by up to an order of magnitude, and the 
BH is prevented from accreting below $z \lesssim 2$, when the galaxy disc forms. 
\end{abstract}

% Select between one and six entries from the list of approved keywords.
% Don't make up new ones.

%%%%%%%%%%%%%%%%%%%%%%%%%%%%%%%%%%%%%%%%%%%%%%%%%%

%%%%%%%%%%%%%%%%% BODY OF PAPER %%%%%%%%%%%%%%%%%%

\section{Introduction} 
\label{sec:introduction}

AGN (Active Galactic Nucleus) activity is observed across cosmic time, and the 
role of AGN feedback is fundamental in regulating the formation and evolution of galaxies. 
The existence of tight correlations between properties of SMBHs (supermassive black holes) 
and their host galaxies \citep[or better, their host galaxy bulges - e.g.][]{Magorrian1998, Ferrarese2000, 
Gebhardt2000, Merritt2001, Tremaine2002, Marconi2003, Haring2004, Gaspari2019} 
is commonly interpreted as the evidence of a coevolution of BHs and host galaxies. 
According to this scenario, the host galaxy evolution and the physical properties 
of its interstellar medium (ISM) regulate BH feeding and growth; 
conversely, feedback from BHs determines and shapes general properties of the host galaxy. 
However, there is no general consensus on the scenario of BH-galaxy coevolution and SMBH self-regulation: 
rather, observed scaling relations could be explained as the result of common mechanisms 
(e.g. mergers and/or gas accretion) which drive the formation of both SMBHs and their host galaxies 
\citep[e.g.][]{Croton2006, Alexander2012, Dekel2019}. 
Hydrodynamical simulations that model structure formation and evolution in a cosmological context 
have to take into account the effect of AGN. Indeed, nuclear galactic activity is deemed fundamental 
to simulate structures whose properties are in agreement with observations at different redshifts. 
 
In particular, the role of AGN feedback is key in controlling the star formation and the gas
cooling processes in galaxies. AGN can have both a positive 
\citep[e.g.][]{Silk2013, Bieri2015, Cresci2015b, Wagner2016, Cresci2018Nat}
and a negative \citep[e.g.][]{Croton2006, mcnamara2007, fabian2012, Wylezalek2016}
impact on the star formation of their hosts. They can stimulate some degree of cooling, 
enhancing the star formation (the so-called positive feedback), or they can produce an overall heating 
and/or mechanical ejection of the gas from the central regions of the galaxy, 
ultimately quenching the star formation 
\citep[][]{Lapi2006, Lapi2014, Lapi2018, Peterson2006, Cresci2015, Carniani2016, McNamara2016}. The relative importance of these processes is still under debate.

In recent years, a wealth of multiwavelength observations has revealed the presence 
of gas spanning a wide range of densities, temperatures, and ionisation states in and around galaxies. 
This multiphase gas is ubiquitous not only in spiral galaxies, commonly recognised as systems rich in cold gas, 
but also in ellipticals and in the innermost regions of galaxy groups and clusters, environments commonly known 
to be dominated by X-ray emitting hot gas \citep[e.g.][]{werner2014, david2014}. 
This multiphase component has been observed to be present also in galactic-scale outflows, which represent one 
of the most characteristic imprints of the AGN presence in a system \citep[e.g.][]{Chartas2003, Rupke2011, 
cicone2014, Feruglio2015, Tombesi2015, Morganti2016, Russell2019}.

Multiphase outflows powered by AGN are a direct consequence of the fact that the energy generated by the 
accreting SMBH is coupled to the surrounding ISM in what is commonly referred to as AGN feedback. 
It is still debated how cold gas gets involved into galactic scale outflows, if by outward acceleration of 
cold gas already present in the innermost regions of the host system, or by condensation of outflowing hot gas, 
resulting in a cold outflow. These possibilities have been considered both by observational 
\citep[e.g.][]{Alatalo2011, Combes2013, Morganti2013, Russell2014} 
and numerical \citep[e.g.][]{Gaspari2012, li2014, Costa2015, Valentini2015} studies. 
Whatever the origin of the cold outflowing gas, observed cold and molecular outflows are thought to be mainly 
accelerated directly by the AGN, as it is unlikely that cold gas has been induced to outflow 
by entrainment by the hot gas phase outflow. 
As a consequence, the AGN feedback energy has to be transferred to both the diffuse and cold phases. 
This complex process is still far from being fully understood, 
and thus an accurate modelling in cosmological hydrodynamical simulations is still missing.

SMBHs accrete surrounding gas and the released gravitational energy provides feedback energy. 
AGN feedback develops through the interaction between the mechanical, thermal and radiative energy supplied 
by accretion and the gas in the host galaxy. BH feedback operates through 
two main distinct modes (although this distinction is purely phenomenological and conventional): 
quasar (or radiative) mode, and radio (or kinetic) mode \citep[e.g.][]{fabian2012}. 
During the quasar mode the AGN is highly luminous, its luminosity approaching the Eddington limit, 
i.e. $L_{\rm Edd} \simeq 1.3 \cdot 10^{38}$~(M$_{\rm BH}$/M$_{\odot}$)~erg~s$^{-1}$ \citep{Frank2002}.
Quasar radiation likely originates from an accretion disc; 
at large scales, gas-rich mergers and cold flows are supposed to be 
the main mechanisms by which the BH is fed during this phase, as they can sustain high BH accretion rates. 
Feedback energy is released through winds and by radiation when AGN are in quasar-mode. 
On the other hand, the accreting BH acts through the mechanical energy of its radio-emitting
jets during the radio mode. These collimated jets can inflate cavities and bubbles 
in the hot atmosphere of dark matter (DM) haloes, and entrain ambient gas 
resulting in massive outflows, that are sub-relativistic on kpc scales. The latter mode is dominant among 
low-power AGN at redshift $z \lesssim 2$ 
(unless we consider Seyfert galaxies), 
where BHs are characterised by lower accretion rates and mainly sustained by the secular evolution of the host system 
\citep[e.g. reviews by][and references therein]{Ferrarese2005, mcnamara2007, fabian2012, 
KormendyHo2013, Morganti2017}. 
Radiation pressure can also power outflows \citep{Proga2007}.
AGN feedback energy also affects the accretion and growth of the BH itself, thus controlling its duty-cycle and 
making the system reach the self-regulation.

A key point which is under debate is the best way to capture an effective description of AGN feeding and feedback 
in cosmological simulations. Sub-resolution prescriptions adopted to simulate both the mechanism 
through which the gas is accreted onto the SMBH and the way of releasing energy are burning issues. 
As for AGN feedback, the commonly pursued approaches consist in providing the feedback energy to 
the surrounding medium in the form of thermal or kinetic energy 
\citep[e.g.][]{Springel2005e361, Sijacki2007, Dubois2010, Barai2016}, 
or with a combination of the two \citep{Dave2019}. 
The recently pursued direction of investigation aims at simulating the effect of the radiative power of the AGN 
via the injection of thermal energy, while modelling the outcome of the mechanical power of the AGN 
by means of outflows in the form of kinetic feedback 
\citep[e.g.][and references therein]{Steinborn2015, Weinberger2017}. 

As for AGN feeding, the most common way to model gas accretion onto SMBHs is to assume the Bondi accretion 
(see Section~\ref{BHgrowth} for details). 
However, due to the inability of resolving the Bondi radius in cosmological hydrodynamical simulations, the estimate 
of the Bondi accretion needs to be done by sampling gas properties over quite large volumes in the proximity of the BH. 
Indeed, in order to properly represent the Bondi accretion, one has to resolve the Bondi radius 
$r_{\rm B} = G\, M_{\rm BH} / c_{\rm s}^2 
\sim 0.04 \, (M_{\rm BH}/10^6 \, \text{M}_{\odot}) \, (c_{\rm s}/ 10 \, \text{km s}^{-1})^{-2}$~kpc, 
where $G$, $M_{\rm BH}$, and $c_{\rm s}$ are the gravitational constant, the BH mass and 
the sound speed of the ambient gas, respectively \citep{Edgar2004, BoothSchaye2009}; 
on the other hand, cosmological simulations generally have spatial resolutions spanning from few kpc down to few hundreds of pc 
(for instance, simulations in this paper have a force resolution which is from two to three orders of magnitude 
larger than the typical Bondi radius of BHs in our galaxies). 
This lack of resolution causes gas density to be underestimated, while gas temperature is overestimated. 
This leads to an underestimation 
of the accretion rate, that is commonly boosted in order to have an effective AGN feedback and to match the 
observations \citep{DiMatteo2005, BoothSchaye2009, Negri2017}. 
To overcome this limitation, challenging mass-refinement techniques \citep{CurtisSijacki2015, Beckmann2019} 
have been recently developed to increase resolution in the BH surroundings, but till now they have been employed 
in simulations of isolated galaxies only. 
Moreover, cold gas that accretes onto SMBHs is expected to deviate considerably from 
the idealised Bondi assumptions \citep[e.g.][and Section~\ref{BHmango}]{gaspari2013}.

%In addition to this, further uncertainties, such as those coming from the prescriptions for seeding 
%BHs at high redshift within simulations (see Section~\ref{BHseeding}), 
%enter in the physical description and the numerical modelling of AGN feedback. 
%This framework opens to new challenging tasks. 
The properties of the ISM surrounding SMBHs in the centre of galaxies, 
galaxy groups and clusters are thought to regulate the BH feeding. Also, the presence of a multiphase medium in the 
innermost regions of cosmic structures poses a challenging question: how different gas phases 
experience AGN feedback? 

The key questions that we want to address in this Paper are the following: 
how do accreting BHs transfer feedback energy to the surrounding multiphase ISM? 
How do they determine the properties of their host galaxy? 
How do different models and regimes of gas accretion affect the BH-galaxy coevolution? 
Does AGN feedback affect significantly the circulation of heavy elements within the galaxy? 

The sub-resolution model MUPPI \citep[MUlti Phase Particle Integrator,][]{muppi2010, muppi2014} that we 
adopt for our cosmological simulations of galaxy formation is crucial to carry out this investigation. 
Indeed, it describes a multiphase ISM (Section~\ref{MUPPI}) and solves the set of equations 
accounting for mass and energy flows among the different phases within the SPH time-step itself: 
these features are key to explicitely and effectively model the effect of AGN feedback energy 
within the resolution element (i.e. the multiphase gas particle). 
%For instance, with respect to our model, in the effective model for star formation and stellar feedback 
%by \citet{SpringelHernquist2003}, the AGN feedback energy provided to multiphase particles is 
%almost entirely lost \citep{Springel2005e361}. The reason for this stems from the fast convergence of the energy of 
%the multiphase particle to the equilibrium energy set within the regime of quiescent and 
%self-regulated star formation \citep{SpringelHernquist2003}. 
%In that model \citep{Springel2005e361}, deviations from the equilibrium energy due to the 
%AGN feedback decay over a timescale which is shorter than the typical SPH timestep of multiphase particles. 

This Paper is organised as follows. Section~\ref{MUPPI} describes the main features of 
the original sub-resolution model MUPPI. Section~\ref{AGNmodelling} is devoted 
to introduce the AGN feedback model that we implemented within the code and the sub-resolution model 
adopted for cosmological simulations. In Section~\ref{simus}, we introduce the suite of simulations that we 
carried out, and in Section~\ref{Results} we present and discuss results. 
The main conclusions are drawn in Section~\ref{sec:conclusions}. 

AGN operate in systems with different mass residing in different environments, from isolate spiral galaxies 
to massive ellipticals located at the centre of bright groups and clusters of galaxies.
This work is focused on late-type galaxies: we introduce our AGN feedback model and explore 
how it works within the scenario of disc galaxy formation and evolution. 
The investigation of the effect of AGN feedback in elliptical galaxies is 
postponed to a forthcoming work.

\section{The sub-resolution model: star formation and stellar feedback in a multiphase ISM} 
\label{MUPPI}

In this Section, we outline the most relevant features of the model, while a more comprehensive description 
and further details can be found in the introductory papers by \citet{muppi2010, muppi2014}. 

\begin{figure}
\begin{center}
\includegraphics[trim=8.8cm 6.8cm 8.4cm 2.5cm, clip, width=.48\textwidth]{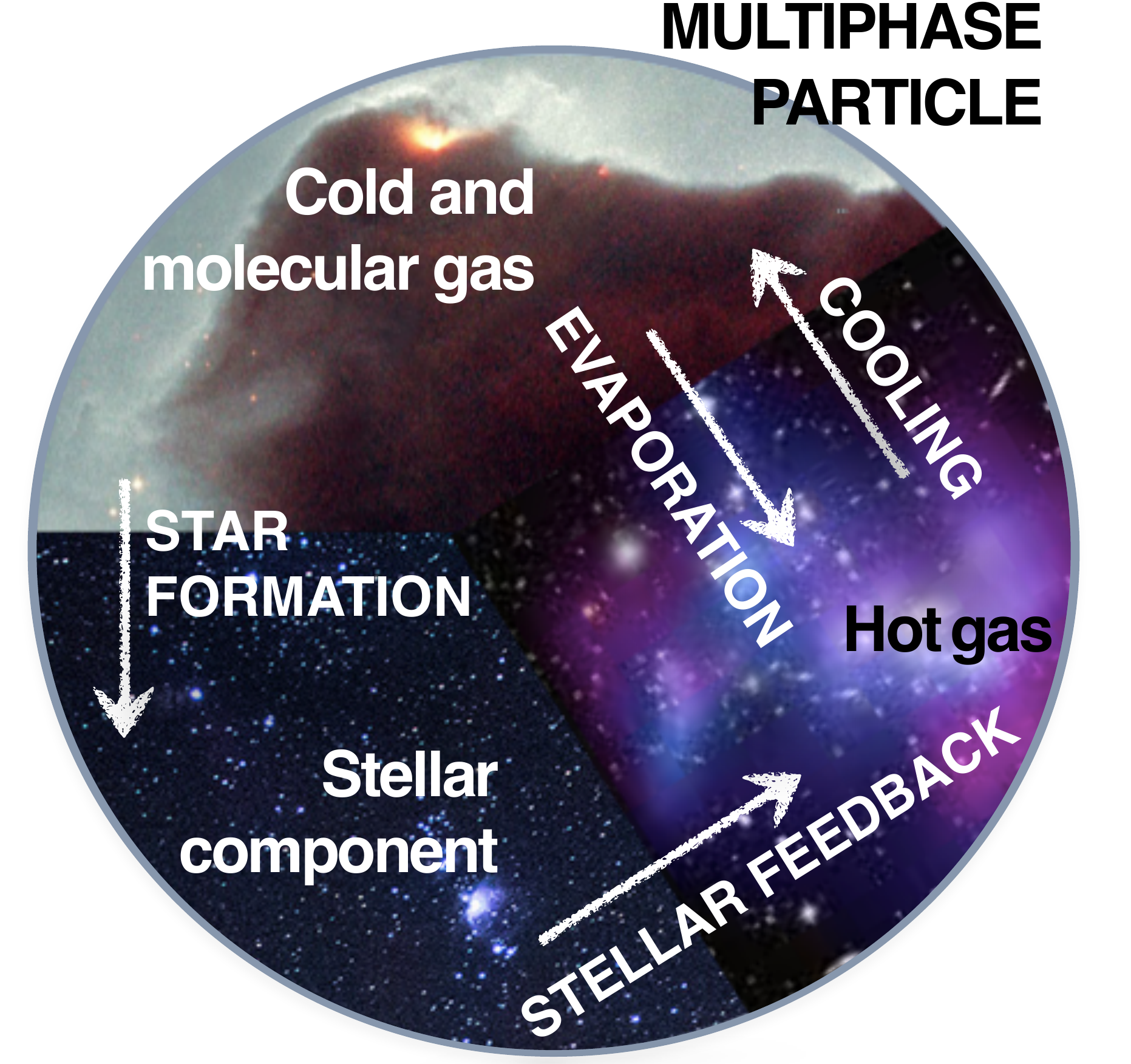}
\end{center}
\caption[Multiphase particle: cartoon]{Cartoon showing the composition of a multiphase gas particle within 
the MUPPI model. Mass and energy flows among different components are highlighted with arrows.
}
\label{fig:particellaMP}
\end{figure}

The sub-resolution model MUPPI represents a multiphase ISM and accounts for star formation and stellar feedback, 
both in thermal and kinetic forms. The constitutive element of the model is the multiphase particle: it is made up 
of a hot and a cold gas component in pressure equilibrium, and a possible stellar component 
(see Figure~\ref{fig:particellaMP}). Considering a multiphase particle whose total mass is $M_{\rm P}$, 
the mass of its hot, cold, and stellar components are $M_{\rm h}$, $M_{\rm c}$, $M_{\ast}$, respectively. 
The pressure equilibrium between the hot and cold phases implies that 
\begin{equation}
\centering
n_{\rm h} \, T_{\rm h} =  n_{\rm c} \, T_{\rm c} \,\,\,,
\label{ch5:pressureEq}
\end{equation}
where $n_{\rm h}$, $T_{\rm h}$, $n_{\rm c}$, and $T_{\rm c}$ are the number density and temperature of 
the hot and cold phases, respectively.

A gas particle is eligible to become multiphase whenever its density rises above a density threshold and 
its temperature falls below a temperature threshold ({$T_{\rm thresh}=5 \cdot 10^4$~K}). 
We choose $n_{\rm thres}=0.01$ cm$^{-3}$ as the particle number density threshold \citep[see][]{muppi2010}. 
The aforementioned number density threshold corresponds to 
a number density of hydrogen atoms of $n_{\rm H} \sim 0.0045$~cm$^{-3}$, 
the assumed fraction of neutral hydrogen being $0.76$ (adopting a mean molecular weight $\mu \sim 0.6$).  
Note that within MUPPI, this is not the density threshold for the star formation, 
but for enabling the gas particle to sample the multiphase ISM (see below). 
When a gas particle becomes multiphase, it is considered to be made of hot gas only (so that 
$M_{\rm h}=M_{\rm P}$, and $T_{\rm h}$ is set to the temperature of the gas particle).  
This hot component then cools down according to its density and metallicity (see Section~\ref{ch5:cooling}),
thus generating the cold component of the multiphase particle, whose temperature is fixed to 
$T_{\rm c}=300$ K.
The fraction of gas mass in the hot phase within the multiphase particle, labelled $F_{\rm h}$, is related to 
the filling factor $f_{\rm h}$ of the hot gas through:
\begin{equation}
\centering
f_{\rm h}  =  \frac{1}{1+ \frac{F_{\rm c}}{F_{\rm h}}\,\frac{\mu_{\rm h}}{\mu_{\rm c}}\,\frac{T_{\rm c}}{T_{\rm h}}} \,\,\,,
\label{ch5:fillingFactor}
\end{equation}
where $F_{\rm c} = 1 - F_{\rm h}$ is the mass fraction of cold gas, 
$\mu_{\rm h} \simeq 0.6$ and $\mu_{\rm c} \simeq 1.2$ are the molecular weights of the hot and cold phase,  respectively. The filling factor of the cold phase is $f_{\rm c}=1-f_{\rm h}$. 
The hot gas number density is therefore computed as:
\begin{equation}
\centering
n_{\rm h}  =  \frac{\rho  \, F_{\rm h}}{f_{\rm h} \, \mu_{\rm h} \, m_{\rm p}} \,\,\,,
\label{ch5:numberDensity}
\end{equation}
where $\rho$ is the SPH density of the gas particle, and $m_{\rm p}$ the mass of the proton. 
A similar relation holds for the cold gas number density $n_{\rm c}$.

The following set of ordinary differential equations describes mass and energy flows between the different components: 
\begin{align}
\dot{M}_{\rm h}  &=  - \dot{M}_{\rm cool} + \dot{M}_{\rm ev} \,\,\,, \label{ch5:muppi1} \\
\dot{M}_{\rm c}  &=   \dot{M}_{\rm cool}  - \dot{M}_{\rm sf} - \dot{M}_{\rm ev} \,\,\,,  \label{ch5:muppi2}  \\
\dot{M}_{\ast}  &=  \dot{M}_{\rm sf}  \,\,\,,  \label{ch5:muppi3} \\
\dot{E}_{\rm h}  &=  \dot{E}_{\rm fb, local} - \dot{E}_{\rm cool} + \dot{E}_{\rm hydro} \,\,\,. \label{ch5:muppi4}
\end{align}
Equations (\ref{ch5:muppi1}),~(\ref{ch5:muppi2}),~and~(\ref{ch5:muppi3}) describe the evolution of the hot 
and cold gas masses, and of the mass of the stellar component, respectively. Equation (\ref{ch5:muppi4}) 
accounts for the evolution of the thermal energy of the hot phase $E_{\rm h}$. 
These equations model the following processes.

Hot gas condenses into a cold phase due to radiative cooling (see Section~\ref{ch5:cooling}), so that: 
\begin{equation}
\centering
\dot{M}_{\rm cool}  =   \frac{{M}_{\rm h}}{t_{\rm cool}}  \,\,\,, 
\label{ch5:muppi5}
\end{equation}
where $t_{\rm cool}$ is the cooling time of the hot phase. 
In turn, a tiny part $f_{\rm ev}$ of the cold gas evaporates because of the destruction of molecular clouds:
\begin{equation}
\centering
\dot{M}_{\rm ev}  =  f_{\rm ev} \,  \dot{M}_{\rm sf}  \,\,\,,
\label{ch5:muppi6}
\end{equation}
$\dot{M}_{\rm sf}$ being the star formation rate (SFR, see below).
Table~\ref{ch5:TableParameters} lists the values for the main model's parameters adopted in the simulations 
presented in this paper.

\begin{table*}
\centering
\begin{minipage}{176mm}
\caption[Relevant parameters of the sub-resolution model]{Relevant parameters of the sub-resolution model.\\ 
{\sl {Column~1:}} number density threshold for multiphase particles. 
{\sl {Column~2:}} temperature of the cold phase. 
{\sl {Column~3:}} pressure at which the molecular fraction is $f_{\rm mol}=0.5$. 
{\sl {Column~4:}} gas particle's probability of becoming a wind particle. 
{\sl {Column~5:}} maximum lifetime of a wind particle.
{\sl {Columns~6:}} half-opening angle of the cone for thermal feedback, in degrees. 
{\sl {Columns~7~and~8:}} thermal and kinetic SN feedback energy efficiencies, respectively.
{\sl {Column~9:}} fraction of SN energy directly injected into the hot phase of the ISM.
{\sl {Column~10:}} evaporation fraction. 
{\sl {Column~11:}} star formation efficiency, as a fraction of the molecular gas.
{\sl {Column~12:}} number of stellar generations, i.e. number of star particles generated by each gas particle.
{\sl {Column~13:}} average stellar masses of stars formed per each SN~II.} 
\renewcommand\tabcolsep{3.3mm}
\begin{tabular}{@{}lcccccccccccc@{}}
\hline
$n_{\rm thresh}$  & $T_{\rm c}$  & $P_{\rm 0}$    &   $P_{\rm kin}$ 
& $t_{\rm wind}$ &  $\theta$   & 
$f_{\rm fb, therm}$ & $f_{\rm fb, kin}$ 
& $f_{\rm fb, local}$ & $f_{\rm ev}$ & $f_{\ast}$ & $N_{\ast}$ & $M_{\ast, \rm SN}$ \\ 
(cm$^{-3}$)          & (K)                & (k$_{\rm B}$ K cm$^{-3}$)       &  
&  (Myr)                  & ($^{\circ}$)  & 
 & &  &  & & & (M$_{\odot}$)  \\ 
\hline
\hline
0.01 &  300 &  2 $\cdot 10^4$ &  0.05 & 15 - $t_{\rm dyn, c \,\, [Myr]}$ & 30 &
0.2 & 0.26 & 0.02 & 0.1 & 0.02 & 4 & 120\\  
\hline
\hline
\end{tabular}
\label{ch5:TableParameters}
\end{minipage}
\end{table*}

As for the SFR $\dot{M}_{\rm sf}$, a fraction $f_{\rm mol}$ of the cold gas mass $M_{\rm c}$ is expected 
to be in the molecular phase: it is converted into stars with an efficiency $f_{\ast}$. 
Therefore, the SFR associated to a multiphase particle is:
\begin{equation}
\centering
\dot{M}_{\rm sf} = f_{\ast} \, \frac{f_{\rm mol} \, M_{\rm c}}{t_{\rm dyn, c}} \,.
\label{eq:sfr}
\end{equation}
\noindent
Here, $t_{\rm dyn, c} = [ 3 \, \pi / (32 \, G \, \rho_{\rm c})]^{1/2}$ is the dynamical time of the cold phase. 
The SFR is directly proportional to the molecular fraction $f_{\rm mol}$, that is computed according to the 
phenomenological prescription by \citet[][]{blitz2006}:
\begin{equation}
\centering
f_{\rm mol} \:=\: \frac{1}{1+P_0/P} \,\,, 
\label{eq:f_mol}
\end{equation}
\noindent
where $P$ is the hydrodynamic pressure of the gas particle and the parameter $P_0$ (i.e. the pressure 
of the ISM at which $f_{\rm mol}=0.5$) is derived from observations. 
The galaxy sample of \citet{blitz2006}, for instance, suggests that values of $P_0 / {\text {k}}_{\rm B}$, 
${\text {k}}_{\rm B}$ being the Boltzmann constant, range 
between {$0.4$$\cdot$$10^4$~and~$7.1$$\cdot$$10^4$~K~cm$^{-3}$}: 
we adopt a constant value $P_0 / {\text {k}}_{\rm B}= 2 \cdot 10^4$~K~cm$^{-3}$ 
(see Table~\ref{ch5:TableParameters}), that is in keeping with observations. 
According to equation~(\ref{eq:f_mol}), the hydrodynamic pressure of a gas particle is
used to estimate the ISM pressure entering in the phenomenological relation by \citet{blitz2006}. 
Equation~(\ref{eq:f_mol}) can be used to estimate an effective density threshold for the star formation, 
$n_{\rm thresh, sf}$, as follows. 
Assuming the latter threshold as the number density of the cold gas phase for which $f_{\rm mol}=0.5$, 
and considering {$P_{\rm 0}/ {\text {k}}_{\rm B}= 2 \cdot 10^4$~K~cm$^{-3}$} and $T_{\rm c}=300$~K 
(see Table~\ref{ch5:TableParameters}), then equation~(\ref{eq:f_mol}) implies that 
{$\, n_{\rm thresh, sf} \, T_{\rm c} = 2 \cdot 10^4$~K~cm$^{-3}$}, so that $n_{\rm thresh, sf} \simeq 66.7$~cm$^{-3}$. 
This number density is by far higher than $n_{\rm thres}$, that rather represents the number density threshold 
for a particle to become multiphase, as discussed above. As a consequence, multiphase particles with low pressure 
are characterised by very low SFR. 

Star formation is implemented according to the stochastic model introduced by \citet{SpringelHernquist2003}. 
A multiphase gas particle with mass $M_{\rm P}$ generates a star particle of 
mass $M_{\ast, \rm init}$ if the probability: 
\begin{equation}
\centering
p = \frac{M_{\rm P}}{M_{\ast, \rm init}} \Biggl[ 1 - {\text {exp} } \Biggl( - \frac{\Delta M_{\ast}}{M_{\rm P}}  \Biggr) \Biggr] \,\,, 
\label{eq:SF}
\end{equation}
\noindent
exceeds a randomly generated number in the interval $[0,1]$. 
In equation (\ref{eq:SF}), $\Delta M_{\ast}$ is the mass of the multiphase particle that has been
converted into stars in a time-step according to equation (\ref{eq:sfr}). 
Each star particle is spawned with mass $M_{\ast, \rm init} = M_{\rm P}/N_{\ast}$, 
$N_{\ast}$ being the number of stellar generations, i.e. the number of star particles generated by each gas 
particle. Note that $M_{\rm P}$ is smaller than the initial mass of the gas particle 
if the gas particle has already spawned stars. 
	Also, $M_{\ast, \rm init}$ is the mass of the new {\sl {star particle}} that is generated and should not be 
	confused with $M_{\ast}$, that is the mass of the {\sl {stellar component within}} the multiphase particle itself. 
	The mass of the star particle that is generated is subtracted from the mass of 
	the stellar component $M_{\ast}$ of the spawning multiphase particle; should $M_{\ast}$ be smaller than 
	$M_{\ast, \rm init}$, additional mass is taken from the cold phase $M_{\rm c}$ 
	\citep[see][for details]{muppi2010}.
The number $N_{\ast}$ is a numerical parameter: we choose $N_{\ast}=4$ in order to have an accurate 
representation of the star formation process, but no significant variations are observed 
for small deviations from this number \citep{tornatore2007}.

Equation~(\ref{ch5:muppi4}) describes the evolution of the thermal energy of the hot gas, that is 
related to the hot gas temperature by:
\begin{equation}
\centering
T_{\rm h} = \frac{E_{\rm h}}{M_{\rm h}} \, \frac{(\gamma -1) \, \mu_{\rm h} \, m_{\rm p}}{{\text {k}}_{\rm B}}  \,\,\,, 
\label{ch5:muppi7}
\end{equation}
\noindent
where $\gamma=5/3$ is the adiabatic index and ${\text {k}}_{\rm B}$ is the Boltzmann constant. 
The right-hand side of equation~(\ref{ch5:muppi4}) shows that two sources of energy counterbalance the 
cooling process ($\dot{E}_{\rm cool} =  E_{\rm h}  / t_{\rm cool} $). 
The first is the energy rate directly injected into the hot phase by SN explosions within the multiphase particle 
itself (therefore, {\sl {locally}}): 
\begin{equation}
\centering
\dot{E}_{\rm fb, local} = f_{\rm fb, local} \, E_{\rm SN} \, \frac{\dot{M}_{\rm sf}}{M_{\ast, \rm SN}} \,\,\,,
\end{equation}
where $E_{\rm SN} = 10^{51}$~erg is the energy provided by each SN, 
$f_{\rm fb, local}$ a feedback efficiency, and $M_{\ast, \rm SN}$ is the mass in stars that is required, 
on average, to have a single SN~II event (it depends on the assumed IMF; see Table~\ref{ch5:TableParameters} 
for the adopted value). 

The second source term is $\dot{E}_{\rm hydro}$: it accounts for the energy contributed by 
neighbour particles because of thermal feedback from dying massive stars (see below), 
and also considers shocks and heating or cooling due to gravitational compression or expansion of gas. 

Stellar feedback is taken into account both in thermal \citep{muppi2010} and 
kinetic \citep{muppi2014, Valentini2017} forms. 
As for thermal feedback, each star-forming particle delivers to neighbours the 
following amount of thermal energy in a given time-step: 
\begin{equation}
\centering
\Delta E_{\rm fb, therm}= f_{\rm fb, therm} \, E_{\rm SN} \, \frac{\Delta M_{\ast}}{M_{\ast, \rm SN}}\,\,\,,
\label{eq:thFB}
\end{equation}
where $\Delta M_{\ast}$ is the mass of the multiphase star-forming particle that has been converted into stars 
within the time-step. The star-forming particle shares its thermal feedback energy among neighbours within a
cone whose half-opening angle is $\theta$. The origin of the cone lies on the particle itself and its axis is aligned according to minus the particle's density gradient. 
Each energy donor weights its contribution to eligible particles 
using the Wendland $C^4$ SPH kernel 
$C^4(\widetilde{q}) = h_{\rm i}^3 \, {W_{\rm i j} (a_{\rm i j}, h_{\rm i})}$ \citep{dehnen2012}, 
$h_{\rm i}$ being the smoothing length. 
Here, $\widetilde{q}= a_{\rm i j} / h_{\rm i}$, where the distance $a_{\rm i j}$ of the neighbour $j$ from the 
axis of the cone (aligned as $- (\nabla \rho)_{\rm i}$, see Figure~1 of \citet{Valentini2017} for the geometry) 
is considered instead of the radial distance $x_{\rm i j}= | \mathbf{x}_{\rm i} - \mathbf{x}_{\rm j} |$ 
between particle pairs. 
If there are no particles in the cone, the total amount of thermal energy is given to the particle nearest to the 
axis \citep{muppi2010, muppi2014}. 

%In the original version of the model, kinetic feedback is implemented so that each star-forming particle can
%provide $f_{\rm fb, kin} \, E_{\rm SN}$ as feedback energy. This
%amount of energy is distributed to wind particles lying inside both the cone and
%the smoothing length of the star-forming particle (see Figure~\ref{fig:completeFBorig}), 
%and the delivering mechanism is the same as the thermal scheme. 
%Thus, outflowing energy is modelled so as to leave the star-forming particle through the least-resistance 
%direction \citep{Monaco2004}. Note that only gas particles that were selected to become wind particles are allowed 
%to receive kinetic energy.
%If there are no particles in the cone, the total amount of thermal energy is given to the particle nearest to the 
%axis \citep{muppi2010,muppi2014}. This does not happen with the kinetic energy; in this case, if no eligible wind 
%particle can receive it, the energy is not assigned. 
As for kinetic stellar feedback, 
the fiducial version of our sub-resolution model adopts the galactic outflow model introduced in \citet{Valentini2017}.
%, at variance with the original version of the sub-resolution model MUPPI \citep{muppi2014}. 
According to this model, the ISM is isotropically provided with kinetic stellar feedback energy. 
By analogy with equation~(\ref{eq:thFB}), each star-forming particle supplies the energy 
\begin{equation}
\centering
\Delta E_{\rm fb, kin}= f_{\rm fb, kin} \, E_{\rm SN} \, \frac{\Delta M_{\ast}}{M_{\ast, \rm SN}}\,\,\,
\label{eq:kinFB}
\end{equation}
isotropically, to all the wind particles (see below) within the smoothing 
length, with kernel-weighted contributions. Here, $f_{\rm fb, kin}$ describes the kinetic stellar feedback efficiency 
(see Table~\ref{ch5:TableParameters}). 
The contributions $\Delta E_{\rm fb, therm}$ and $\Delta E_{\rm fb, kin}$ enter in the source term $E_{\rm hydro}$, 
along with the further contribution of the thermal energy which is isotropically provided 
by star particles (see Section~\ref{ch5:cooling}). 
Wind particles receiving energy use it to increase their velocity along their least resistance
path, since they are kicked against their own density gradient \citep[see Figure~3 of][]{Valentini2017}. 
The directionality to the outflow is in this way ensured; this is at variance with the thermal stellar feedback scheme, 
which has been designed as well to produce outflows that are perpendicular to the galaxy disc, but where 
the energy is provided within a cone as there is no way to exploit the direction of the velocity kick. 
We adopt this model for triggering galactic outflows as it promotes the formation of disc galaxies with 
morphological, kinematic and chemical properties in agreement with observations 
\citep{Valentini2018, Valentini2019}. 

A gas particle exits its multiphase stage after a maximum allowed time 
given by the dynamical time of the cold gas ($t_{\rm dyn, c}$).   
When a gas particle exits a multiphase stage, it has a probability $P_{\rm kin}$ of being 
{\sl {kicked}} and to become a wind particle for a time interval $t_{\rm wind}$. 
Both $P_{\rm kin}$ and $t_{\rm wind}$ are parameters of the model (Table~\ref{ch5:TableParameters}). 
This scheme relies on the physical idea that galactic winds are powered by SN~II explosions,
once the molecular cloud out of which stars formed has been
destroyed. Wind particles are decoupled from the surrounding medium
for the aforementioned interval $t_{\rm wind}$. During this
time, they receive kinetic energy from neighbouring star-forming gas
particles. The wind stage can be concluded before $t_{\rm wind}$ whenever the particle 
density drops below a chosen density threshold, $0.3 \, \rho_{\rm thresh}$, meaning that a wind particle has 
finally gone away from star-forming regions. We note that a multiphase particle is forced to exit the 
multiphase stage if its density drops below $0.2 \, \rho_{\rm thresh}$, as star formation is not expected to occur 
anymore in the ISM that it samples. 

The system of equations~(\ref{ch5:muppi1}),~(\ref{ch5:muppi2}),~(\ref{ch5:muppi3}),~and~(\ref{ch5:muppi4}) is 
integrated with a Runge-Kutta algorithm within each SPH time-step \citep[see][for details]{muppi2010, muppi2014}. 

The original release of the sub-resolution model MUPPI does not include the effect of AGN feedback. 
In Section \ref{AGNmodelling}, we introduce the implementation of AGN feedback within our sub-resolution model.

%%%%%%%%%%%%%%%%%%%%%%%%%%%%%%%%%%%%%%%%%%%%%%%%%%%%%%%%%%
%%%%%%%%%%%%%%%%%%%%%%%%%%%%%%%%%%%%%%%%%%%%%%%%%%%%%%%%%%

\subsection{Additional physics: cooling and chemical enrichment}
\label{ch5:cooling} 

Chemical enrichment and radiative cooling are self-consistently included in our simulations. 
Metal-dependent radiative cooling is implemented according to the model by \citet{wiersma2009}. 
Cooling rates are estimated on an element-by-element basis, by adopting pre-computed tables where 
rates are functions of density, temperature, and redshift. Tables have been compiled using the 
spectral synthesis code CLOUDY \citep{Ferland1998}.
The gas is considered to be optically thin and exposed to a spatially uniform, redshift-dependent ionising 
background radiation from star-forming galaxies and quasars \citep{HaardtMadau2001}. 
When computing cooling rates, photoionization equilibrium is thus assumed 
\citep[see][for details]{wiersma2009, Wiersma2009b}. 

Besides providing the ISM with energy, stellar feedback resulting from 
star formation and evolution also supplies heavy elements (chemical feedback), and 
galactic outflows foster metal spread and circulation throughout the galaxy. Our model self-consistently accounts 
for the chemical evolution and enrichment processes, following \citet{tornatore2007}, where 
a thorough description can be found. Here, we only highlight the most crucial features of the model. 

Each star particle initially shares the chemical composition of the gas particle from which it has been originated. 
Star particles are considered to be simple stellar populations (SSPs), i.e. ensembles of coeval stars that share the 
same initial metallicity. By assuming an IMF (initial mass function, see below)
and adopting predictions for stellar lifetimes and stellar yields (see below), 
our model evaluates the number of stars aging and eventually exploding as SNe (according 
to a mass-dependent time-delay function), 
as well as the amount of metals polluting the surrounding ISM. 
In all the simulations presented in this work, we adopt the \citet{kroupa93} IMF. 
This IMF is characterised by three slopes, as $\alpha$ in the equation $\phi (m) =  \beta m^{- \alpha}$ 
which defines the IMF has the following values according to the mass interval:
\begin{equation}
\begin{array}{l}
\alpha = 1.3    $\,\,\,\,\,\,\,\,\,\,$ {\text {for}} $\,\,\,\,\,\,$ 0.1 {\text { M}}_{\odot} \le m \le 0.5 {\text { M}}_{\odot}, \\
\alpha = 2.2    $\,\,\,\,\,\,\,\,\,\,$ {\text {for}} $\,\,\,\,\,\,$ 0.5 {\text { M}}_{\odot} < m \le 1.0 {\text { M}}_{\odot},  \\
\alpha = 2.7    $\,\,\,\,\,\,\,\,\,\,$ {\text {for}} $\,\,\,\,\,\,$ 1.0 {\text { M}}_{\odot} < m \le 100 {\text { M}}_{\odot}.  \\
\label{IMFslopes}
\end{array}
\end{equation}
It is defined in the mass range $[0.1, 100]$ M$_{\odot}$. 
The effect of the choice of the IMF in our simulations is thouroughly investigated in \citet{Valentini2019}. 

The model accounts for different timescales of evolving stars with different masses by adopting the mass-dependent 
lifetimes by \citet{PadovaniMatteucci1993}. 
The minimum mass giving rise to stellar BHs is considered to be $8$~M$_{\odot}$. Stars that are more massive 
than $40$~M$_{\odot}$ directly implode into BHs, thus not contributing to further chemical enrichment nor to 
stellar feedback.

A fraction of stars relative to the entire mass range in which the IMF is defined (see Section~\ref{simus})
is assumed to be located in binary systems suitable for being progenitors of SNe~Ia. It is set to $0.03$: 
the effect of the value of this fraction is extensively explored in \citet{Valentini2019}. 
Energy contributed by SNe~Ia which is provided to multiphase particles enters in the 
source term $\dot{E}_{\rm hydro}$ in equation~(\ref{ch5:muppi4}). 

The production of different metals by aging and exploding stars is followed by assuming sets of stellar yields. 
We adopt the stellar yields provided by \citet{Thielemann2003} for SNe~Ia and 
the mass- and metallicity-dependent yields by \citet{Karakas2010} for 
intermediate and low mass stars that undergo the AGB (asymptotic giant branch) phase.
As for SNe~II, I use the mass- and metallicity-dependent yields by \citet{WoosleyW1995}, 
combined with those provided by \citet{Romano2010}. Also, the effect of adopted stellar yields is 
addressed in detail in \citet{Valentini2019}.

Different heavy elements produced and released by star particles are distributed to neighbouring gas particles 
with kernel-weighted contributions, so that subsequently generated star particles are richer in metals. 
The chemical evolution process is therefore responsible for the gradual reduction of the initial mass 
of stellar particles, too. 
We follow in detail the chemical evolution of $15$ elements (H, He, C, N, O, Ne, Na, Mg, Al, Si, S, Ar, Ca, Fe and Ni) 
produced by different sources, namely AGB stars, SNe~Ia and SNe~II.
Each atomic species independently contributes to the cooling rate, as discussed above. 

Note that the mass of gas particles is not constant throughout the simulation: the initial mass can indeed 
decrease due to star formation (i.e. spawning of star particles), and it can increase because of gas return 
by neighbour star particles.

%%%%%%%%%%%%%%%%%%%%%%%%%%%%%%%%%%%%%%%%%%%%%%%%%%%%%%%%%%
%%%%%%%%%%%%%%%%%%%%%%%%%%%%%%%%%%%%%%%%%%%%%%%%%%%%%%%%%%

\section{AGN feedback modelling} 
\label{AGNmodelling}

In this Section, we describe the AGN feedback model adopted to carry out the simulations presented in this paper. 
BH accretion and ensuing feedback are modelled resorting to sub-resolution prescriptions, as for star formation 
and stellar feedback (see Section~\ref{MUPPI}). 
The prescriptions adopted for BH seeding and accretion are predominantly based, despite a number of 
differences that are detailed in the following, on the original model by \citet{Springel2005e361} and 
largely inherited from simulations of galaxy clusters 
\citep[e.g.][]{Fabjan2010, RagoneFigueroa2013, Steinborn2015, Rasia2015}. 
As for the modelling of the release of AGN feedback energy, 
since the sub-resolution model MUPPI describes a multiphase ISM, we exploit this feature 
in order to study the coupling of AGN feedback energy to different phases of the ISM 
(by modelling the energy distribution within multiphase particles).

%%%%%%%%%%%%%%%%%%%%%%%%%%%%%%%%%%%%%%%%%%%%%%%%%%%%%%%%%%
%%%%%%%%%%%%%%%%%%%%%%%%%%%%%%%%%%%%%%%%%%%%%%%%%%%%%%%%%%

\subsection{Including BHs: seeding and pinning}
\label{BHseeding}

%%%%     BH seeding
BHs in cosmological hydrodynamical simulations are represented by means of collisionless sink 
particles of mass $M_{\rm BH}$. 
BH particles are introduced in massive haloes at relatively high-redshift in cosmological simulations, 
and they are then allowed to grow and increase their initial or {\sl {seed}} mass. 
As we are still lacking a solid understanding of the formation of first SMBHs 
\citep[see e.g.][for possible scenarios]{Bromm2003, Begelman2006, Mayer2010, Volonteri2012, Maio2018}, 
BHs are first inserted according to seeding prescriptions.

\begin{table*}
\centering
\begin{minipage}{176mm}
\caption[Relevant parameters of the AGN feedback model]{Relevant parameters of the AGN model.\\ 
{\sl {Column~1:}} BH seed mass in the reference simulations. 
{\sl {Column~2:}} halo mass for BH seeding. 
{\sl {Column~3:}} minimum stellar mass fraction for BH seeding. 
{\sl {Column~4:}} threshold temperature to distinguish between hot and cold accreting gas. 
{\sl {Column~5:}} radiative efficiency. 
{\sl {Column~6:}} feedback efficiency.
{\sl {Column~7:}} reference value for the parameter regulating the angular momentum dependent accretion of cold gas.} 
\renewcommand\tabcolsep{8.6mm}
\begin{tabular}{@{}lcccccc@{}}
\hline
$M_{\rm BH, \, seed}$  & $M_{\rm DM, \, thresh}$  &   $f_{\rm \star, \, seed}$   & $T_{\rm split}$    &    $\epsilon_{\rm r}$      
&  $\epsilon_{\rm f}$  &   $C_{\rm visc}$                  \\ 
(M$_{\odot}$)               & (M$_{\odot}$)             &  &   (K)             &                                    &                                &                      \\ 
\hline
\hline
$1.1 \cdot 10^5$ &  $1.7 \cdot 10^{10}$  &   $0.02 $  &  $5 \cdot 10^5 $ &  $0.1$ &  $0.01$  & $2 \, \pi \cdot (1, \, 10^2, \, {\text{or }} 10^3)$ \\  
\hline
\hline
\end{tabular}
\label{ch10:AGNmodelParamm}
\end{minipage}
\end{table*}

In the simulations presented in this section, a BH of theoretical mass $M_{\rm BH, \, seed}$ is seeded within 
DM haloes whose mass exceeds the threshold mass $M_{\rm DM, thresh}$, and that do not already have a BH. 
The commonly quoted mass $M_{\rm BH}$ is the {\sl {theoretical mass}} of the BH, 
modelled at the sub-resolution level; its {\sl {dynamical}} mass is the actual gravitational mass 
of the BH particle in the simulation (the two masses are in general not equal, see Section~\ref{BHgrowth}). 
We adopt $M_{\rm BH, \, seed}=1.1 \cdot 10^5$~M$_{\odot}$ %%%%%%%   seed
and $M_{\rm DM, thresh}=1.7 \cdot 10^{10}$~M$_{\odot}$ for %%%%%%%   halo mass for seed
the fiducial simulations (see Table~\ref{ch10:AGNmodelParamm} and Section~\ref{simus}). 
We exploited the $M_{\rm bulge}$-$M_{\rm BH}$ relation to choose the reference $M_{\rm BH, \, seed}$ 
(see Appendix~\ref{ch10:CalibMago}). 
Also, the eligible DM halo is required to have a minimum stellar mass fraction ($f_{\rm \star, \, seed} = 0.02$) 
for BH seeding. The latter requirement ensures that BHs are seeded only within haloes that host adequately 
resolved galaxies \citep{Hirschmann2014}.

DM haloes are identified by means of the Friends-of-Friends (FoF) algorithm. 
The FoF is performed on-the-fly, on DM particles alone, 
and a linking length of $0.16$ times the mean inter-particle spacing is adopted. 
To achieve an accurate centering of the BH particle in the pinpointed halo, the BH is seeded in 
the position of the minimum potential of the halo, by identifying the star particle which has the 
highest binding energy. The selected star particle is thus converted into a BH sink particle of 
theoretical mass $M_{\rm BH, \, seed}$. The dynamical mass of the BH at the seeding is the mass of the star particle 
which has been converted into it. The initial mass of gas particles can be thus considered 
as a typical dynamical mass of a seed BH. 

%%%%     BH reposition
The location of the BH is crucial to determine physical properties of the gas that undergoes accretion and to 
compute quantities involved in the ensuing feedback. In simulations, BHs can generally move away from the innermost 
regions of the forming galaxy and wander becuse of numerical artefacts \citep[][]{Wurster2013}: indeed, they 
can be dragged by surrounding particles, especially in highly-dynamical, high-redshift environments. Also, 
the dynamical friction that is expected to promote the settling of massive BHs in the centre of halos usually 
is not adequately captured at the resolution achieved in cosmological simulations \citep{Weinberger2017}.
In order to avoid that BHs move from the centre of the halo in which they reside because of numerical spurious 
effects, we re-position the BH on the minimum of the gravitational potential. To this end, at each time-step 
the BH is shifted towards the position of the particle (DM, stellar or gas particle) with the absolute 
minimum value of the local gravitational potential within the gravitational softening of the BH 
\citep[as done, among others, by][]{RagoneFigueroa2013, Schaye2015, 
Weinberger2017, Pillepich2018}.

%%%%%%%%%%%%%%%%%%%%%%%%%%%%%%%%%%%%%%%%%%%%%%%%%%%%%%%%%%
%%%%%%%%%%%%%%%%%%%%%%%%%%%%%%%%%%%%%%%%%%%%%%%%%%%%%%%%%%

\subsection{BH accretion}
\label{BHgrowth}

BHs grow because of gas accretion and mergers with other BHs. 
%%%%     BH gas accretion 
Gas accretion onto the central BH is commonly modelled through the Bondi-Hoyle-Lyttleton accretion 
solution \citep{Hoyle1939, Bondi1944, Bondi1952}.
%The Bondi accretion rate reads:
%\begin{equation}    
%\dot{M}_{\rm B} = \frac{4 \, \pi \, \lambda \, G^2 \, M_{\rm BH}^2 \, \rho_{\infty}}{c_{\rm s, \, \infty}^3}  \,\,\,,
%\label{BondiMdot}
%\end{equation}
%where $G$ is the gravitational constant, and $\lambda$ is a dimensionless parameter of order unity, 
%that is a function of the adiabatic index $\gamma$. 
%In equation~(\ref{BondiMdot}), $\rho_{\infty}$ and $c_{\rm s, \, \infty}$ are the density and the sound speed of the gas, 
%respectively, evaluated at very large distance (formally infinity) from the accreting BH, where the gas is also supposed 
%to be at rest.
Following \citet{Springel2005e361}, the Bondi-like BH accretion rate is numerically estimated as:
\begin{equation}    
\dot{M}_{\rm B} = \frac{4 \, \pi \, G^2 \, M_{\rm BH}^2 \, \langle \rho \rangle}{(\langle c_{\rm s}\rangle^2 + \langle v\rangle^2)^{3/2}}  \,\,\,,
\label{BondiMdotAve}
\end{equation}
where $G$ is the gravitational constant. In equation~(\ref{BondiMdotAve}), 
the density of the gas $\rho$, its sound speed $c_{\rm s}$, and the velocity $v$ of the BH relative to the gas is 
computed by averaging over SPH quantities of the gas particles within the smoothing length of the BH, with 
kernel-weighted contributions. 
Although the smoothing length usually pertains only to gas particles, a smoothing length for the 
BH particle is generally defined to estimate hydrodynamical quantities in the proximity of the BH. 
By analogy with gas particles, the mass within the sphere whose radius is the BH smoothing length $h_{\rm i}$ 
is required to be constant (and equal to that enclosed within the gas particles' smoothing sphere). 

We do not assume any boost factor in equation~(\ref{BondiMdotAve}), neither for the hot nor for the cold gas accretion. 
While being still far from resolving the Bondi radius (see Section~\ref{sec:introduction}), our simulations 
have indeed a resolution which allows us to explore this possibility, at variance with previous, 
lower-resolution cosmological simulations. Indeed, we resolve quite low temperatures and gas densities high 
enough to avoid the underestimate of the BH accretion rate that several previous simulations suffered from 
(see Section~\ref{sec:introduction}). 
A number of improvements in recent simulations 
\citep[e.g.][]{Pelupessy2007, Khandai2015, Schaye2015, Weinberger2017, Pillepich2018}, 
such as the increase of resolution and sub-resolution description of the accretion process, 
remove the need to compensate for low accretion rates by means of the boost factor, 
that had been introduced in previous simulations 
\citep[e.g.][]{Springel2005e361, DiMatteo2005,  Sijacki2007, Khalatyan2008, BoothSchaye2009, Dubois2013}. 
Extensive tests have shown that the BHs in the simulation presented in Section~\ref{Results} 
grow to masses that are in agreement with observations without the need for boosting the accretion. 
Rather, we will explore the possibility to accrete cold gas only, thus neglecting hot gas accretion, as 
discussed in Section~\ref{simus}.

As suggested by \citet{gaspari2013} and \citet{Steinborn2015}, 
gas accretion is estimated by considering separately hot 
and cold gas. The temperature $T_{\rm split} = 5 \cdot 10^5 $~K is assumed to distinguish between 
hot and cold accreting gas. 
The accretion rates for the hot and cold phases are thus computed according to 
equation~(\ref{BondiMdotAve}), and $\dot{M}_{\rm B, \, h}$ and $\dot{M}_{\rm B, \, c}$ are estimated. 
As for the way in which multiphase particles contribute to gas accretion onto the BH, 
we consider the mass-weighted temperature of multiphase particles to decide whether each of them 
(entirely) contributes to hot or cold accretion. In the majority of cases, multiphase particles contribute 
to cold accretion because the cold gas mass (whose $T = 300$~K) usually represents by up to 
$90$\% of the total mass of a typical multiphase particle. 

In our simulations we assume that the accretion rate cannot exceed the Eddington accretion rate:
\begin{equation}    
\dot{M}_{\rm Edd} = \frac{4 \, \pi \, G \, M_{\rm BH} \, m_{\rm p}}{\epsilon_{\rm r} \, \sigma_{\rm T} \, c}  \,\,\,,
\label{EddMdot}
\end{equation}
where $m_{\rm p}$, $\sigma_{\rm T}$, and $c$ are the proton mass, the Thompson cross-section, 
and the speed of light, respectively, and $\epsilon_{\rm r}$ is the radiative efficiency. 
As for the radiative efficiency, we adopt a constant value $\epsilon_{\rm r}=0.1$ (see also 
Table~\ref{ch10:AGNmodelParamm}); 
this value is slightly larger than the maximum value for a non-rotating (or Schwarzschild) BH \citep{Shakura1973}, 
but well below the maximum value attainable by a rotating BH. 
%, which corresponds to the mean value for the radiatively 
%efficient accretion onto a Schwarzschild BH \citep{Shakura1973}. 

Therefore, the accretion rate of the BH is given by the sum of both hot and cold gas accretion, and is 
capped to the Eddington accretion rate, i.e.:
\begin{equation}    
\dot{M}_{\rm BH} = {\text{min}}  (\dot{M}_{\rm B, \, h} + \dot{M}_{\rm B, \, c}, \dot{M}_{\rm Edd}) \,\,\,.
\label{Mdot_limited}
\end{equation}

Computing a separated accretion rate for hot and cold gas drives a faster BH growth 
during the high-accretion rate mode of AGN, when they are expected to be surrounded mainly 
by cold gas. Indeed, averaging velocities and sound speeds of (almost completely) cold gas and hot gas 
separately leads to a higher estimate for $\dot{M}_{\rm B}$ in 
equation~(\ref{BondiMdotAve}) \citep{Steinborn2015}. 

Gas accretion is modelled according to the stochastic scheme originally proposed by \citet{Springel2005e361}. 
Should the theoretical mass of the BH exceed the dynamical one, the BH can absorb gas particles. 
Each BH neighbouring gas particle has a probability to be swallowed, which is proportional to the 
difference between the theoretical and dynamical mass of the BH over 
the kernel-smoothed mass (i.e. the ratio between its density and the kernel function) of the gas particle itself. 
The original model was marginally modified in order to achieve 
a more continuous sampling of the accretion process: each selected gas particle contributes to BH feeding with 
a fraction of its mass \citep{Fabjan2010, Hirschmann2014}. In this way, a larger number of gas particles are 
involved in sampling the accretion and selected gas particles are not always entirely swallowed, their mass being rather 
decreased. We assume a value of $1/4$ for the slice of the gas particle mass to be accreted. 
When stochastically accreting particles, the total accretion rate is computed according 
to equation~(\ref{Mdot_limited}) and all the gas particles within the BH smoothing length 
are eligible for the stochastic accretion process. 

The stochastic accretion scheme determines the increase of the dynamical mass of the BH. On the other hand, 
the sub-resolution continuous increase of the theoretical mass of the BH is smooth and computed according to 
equation~(\ref{BondiMdotAve}). The accurate numerical description of the accretion process ensures that 
the increase of the dynamical mass faithfully reproduces that of the theoretical mass, with marginal 
fluctuations around it. 

%%%%     BH merging
As for BH merging which contributes to BH growth, two BHs are merged whenever their distance 
is smaller than twice their gravitational softening length, and if their relative velocity is smaller than 
a fraction (assumed to be $0.5$) of the sound speed of the surrounding gas (i.e. the average of 
the sound speed of gas particles within the smoothing length of the BH, with kernel-weighted 
contributions). 
The resulting BH is located at the position of the most massive one between the two BHs that 
undergo merging\footnote{This scheme for repositioning the merged BH leads to a violation 
of the momentum conservation law \citep[see e.g.][]{Hirschmann2014}.}.

%%%%%%%%%%%%%%%%%%%%%%%%%%%%%%%%%%%%%%%%%%%%%%%%%%%%%%%%%%
%%%%%%%%%%%%%%%%%%%%%%%%%%%%%%%%%%%%%%%%%%%%%%%%%%%%%%%%%%

\subsection{Limiting BH accretion}
\label{BHmango}

The Bondi model for gas accretion onto BHs relies, among others, on the assumptions 
of spherical simmetry and of zero angular momentum for the inflowing gas. 
However, accreting gas does have some angular momentum: therefore, it settles onto a circular 
orbit whose radius is determined by its angular momentum, and the accretion proceeds 
through an accretion disc \citep{King2010, Hobbs2011}. 
This is especially true for cold gas, that is expected to depart significantly from the Bondi assumptions 
because of cooling and turbulence
\citep[][and Section~\ref{sec:introduction}]{BoothSchaye2009, gasp12, gaspari2013, Gaspari2015}.
The angular momentum represents a natural barrier to accretion: as a consequence, only gas 
with the lowest angular momentum is effectively accreted and feeds the BHs \citep{Power2011}. 

For this reason, we consider the possibility of reducing the gas accretion onto the central BH for gas which 
has a high angular momentum. To this end, our implementation of BH feeding allows to adopt an 
angular momentum dependent accretion rate for the cold gas, following the phenomenological correction 
introduced by \citet{RosasGuevara2015} and also adopted by \citet{Schaye2015}. Note that the limiter 
to the gas accretion reduces the overall BH accretion rate in \citet{RosasGuevara2015, Schaye2015}, 
as they do not distinguish between hot and cold gas accretion. On the other hand (see Section~\ref{BHgrowth}), 
we prefer to limit only cold gas accretion, for the reasons outlined above. 
Also, as extensively discussed in Section~\ref{ch10:Mango}, properties of the warm and cold ISM 
are expected to crucially impact on the evolution of the accretion rate of SMBHs in the centre of 
late-type galaxies, where cold gas is rotationally supported. 
When the BH accretion rate is suppressed according to the angular momentum of the cold gas, %that is undergoing accretion, 
the contribution to the BH accretion rate from the cold gas (entering in 
equation~(\ref{Mdot_limited})) reads: 
\begin{equation}    
\dot{M}_{\rm B, \, c} = \dot{M}_{\rm B, \, c} \cdot {\text{min}}  (1, \mathcal{L}_{\rm AM}) \,\,\,, 
\label{Mdot_limited_mango}
\end{equation}
where $\mathcal{L}_{\rm AM}$ is the BH accretion rate limiter, i.e.: 
\begin{equation}    
\mathcal{L}_{\rm AM} = \frac{1}{C_{\rm visc}} \, \biggl( \frac{c_{\rm s, \, c}}{V_{\phi}} \biggr)^3   \,\,\,.
\label{AngMomLimiter}
\end{equation}
In equation~(\ref{AngMomLimiter}), $C_{\rm visc}$ is a constant parameter (see below), 
$c_{\rm s, \, c}$ is the sound speed of the cold 
($T < T_{\rm split} = 5 \cdot 10^5 $~K, see Section~\ref{BHgrowth}) gas, 
and $V_{\phi}$ is the rotational velocity of the cold gas surrounding the BH, 
that can be cast as \citep{RosasGuevara2015}: 
\begin{equation}    
V_{\phi} = \biggl\lvert \, \sum_{\rm i=0}^{N_{\rm ngb}} { \mathbf{x}_{\rm i} \times \mathbf{v}_{\rm i} \, m_{\rm i} \, 
W(x_{\rm i}, h) \, \frac{1}{\rho_{\rm c} \, h}} \, \biggr\rvert  \,\,\,.
\label{VphiMod}
\end{equation}
In equation~(\ref{VphiMod}), $W(x_{\rm i}, h)$ is the smoothing kernel, 
$h$ is the smoothing length of the BH, and the sum spreads over the BH neighbour gas particles whose 
position with respect to the BH is $\mathbf{x}_{\rm i}$ and whose velocity is $\mathbf{v}_{\rm i}$. 
Also, $\rho_{\rm c}$ is the smoothed density of the cold gas surrounding the BH. Note that only gas particles 
whose SPH temperature is lower than $T_{\rm split} = 5 \cdot 10^5 $~K (see Section~\ref{BHgrowth}) enter 
in the computation of equation~(\ref{VphiMod}). 

$C_{\rm visc}$ is a constant parameter that has been introduced to parametrise at the sub-resolution level 
the viscosity of the accretion disc 
\citep[see Section~\ref{ch10:Mango}, and][for further details]{RosasGuevara2015}. This parameter regulates 
how the BH accretion rate is sensitive to the angular momentum of the accreting gas. 

As a consequence, the limiter to the cold gas accretion rate is switched off 
unless $\, C_{\rm visc}^{1/3} \, V_{\phi}  > c_{\rm s, \, c} \,$. 
We adopt $C_{\rm visc}/ 2 \, \pi = 1, \, 10^2, \, 10^3$ 
\citep[as suggested by][see also Table~\ref{ch10:AGNmodelParamm}]{Schaye2015}. 
In Section~\ref{ch10:Mango}, we will discuss how variations of this parameter impact on final results. 
%Note that the suppression of the accretion rate is controlled by the ratio $\,V_{\phi} / c_{\rm s, \, c}\,$ 
%%and it is effective only when $\, C_{\rm visc}^{1/3} \, V_{\phi} > c_{\rm s, \, c}\,$  
%\citep[see also][]{RosasGuevara2015}.

%%%%%%%%%%%%%%%%%%%%%%%%%%%%%%%%%%%%%%%%%%%%%%%%%%%%%%%%%%
%%%%%%%%%%%%%%%%%%%%%%%%%%%%%%%%%%%%%%%%%%%%%%%%%%%%%%%%%%

\subsection{AGN feedback}
\label{AGNmuppi}

Each BH radiates away a small part $\epsilon_{\rm r}$ 
(see Section~\ref{BHgrowth} and Table~\ref{ch10:AGNmodelParamm}) 
of its accreted rest-mass energy. Its bolometric luminosity can be thus cast as: 
\begin{equation}    
L_{\rm r} = \epsilon_{\rm r} \,  \dot{M}_{\rm BH}  \, c^2   \,\,\,.
\label{Luminosity}
\end{equation}
A tiny fraction of the radiated luminosity $L_{\rm r}$ is provided to the ISM in the form of AGN feedback energy, 
so that the feedback energy per unit time is:
\begin{equation}    
\dot{E}^{\rm AGN}_{\rm fb, \rm tot} = \epsilon_{\rm f} \, L_{\rm r} = \epsilon_{\rm f} \, \epsilon_{\rm r} \,  \dot{M}_{\rm BH}  \, c^2 \,\,\,, 
\label{Luminosity2}
\end{equation}
where $\epsilon_{\rm f}$ is the feedback efficiency, quantifying the radiated luminosity that is actually coupled to the 
surrounding gas. 
This AGN feedback energy is coupled thermally and isotropically to the BH neighbouring gas particles, 
as detailed in Section~\ref{AGNmuppi}. We assume $\epsilon_{\rm f}=0.01$ (see also Table~\ref{ch10:AGNmodelParamm}): this value is smaller than commonly adopted feedback efficiencies 
\citep[which usually span the range $0.05 - 0.1$, e.g.][]{Vogelsberger2013, Rasia2015, Pillepich2018}, 
and highlights a quite effective response of the ISM described by our sub-resolution model 
to the injection of AGN feedback energy. 

The rate of total AGN feedback energy $\dot{E}^{\rm AGN}_{\rm fb, \rm tot}$ available is distributed to all gas particles 
within the smoothing kernel of the BH, and kernel-weighted contributions are assigned to both single-phase and 
multiphase particles. The rate of AGN feedback energy pertaining to each considered particle 
is $\dot{E}^{\rm AGN}_{\rm fb}$. 

For single-phase particles, the AGN feedback energy received in the SPH time-step is a source term contributing 
to the heating rate, that enters the hydrodynamic equation for the evolution of the internal energy. 
AGN feedback energy is therefore used to increase their specific internal energy, and hence their entropy. 

Multiphase particles selected to receive feedback energy pose a non-trivial question: how does AGN feedback 
energy couple to the different components of a multiphase ISM?

%%%%%%%%%%%%%%%%%%%%%%%%%%%%%%%%%%%%%%%%%%%%%%%%%%%%%%%%%%
%%%%%%%%%%%%%%%%%%%%%%%%%%%%%%%%%%%%%%%%%%%%%%%%%%%%%%%%%%

\subsection{Including AGN feedback within MUPPI}
\label{AGNmuppi}
  
The sub-resolution model MUPPI accounts for the evolution of multiphase particles that have been provided with 
AGN feedback energy. 
We consider that a fraction $\mathcal{A}_{\rm h}$ of the rate of feedback 
energy $\dot{E}^{\rm AGN}_{\rm fb}$ is coupled with the hot phase of each multiphase particle, while the remaining 
fraction $\mathcal{A}_{\rm c} = 1- \mathcal{A}_{\rm h}$ of the energy budget per unit time is supplied to the cold component. 
The values and modelling of $\mathcal{A}_{\rm h}$ and $\mathcal{A}_{\rm c}$ are extensively discussed in 
Section~\ref{ch10:couplingFactors}. In this way, the feedback energy per unit time available to the hot phase is: 
\begin{equation}    % BHEhot    sfr_muppi : 1455
\dot{E}^{\rm AGN}_{\rm h} = \mathcal{A}_{\rm h} \, \dot{E}^{\rm AGN}_{\rm fb}  \,\,\,,
\label{couplingHot}
\end{equation}
while the rate of feedback energy of the cold phase is: 
\begin{equation}     % BHEcold    sfr_muppi : 1455
\dot{E}^{\rm AGN}_{\rm c} = \mathcal{A}_{\rm c} \, \dot{E}^{\rm AGN}_{\rm fb}  \,\,\,.
\label{couplingCold}
\end{equation}
The energy contributions $E^{\rm AGN}_{\rm h}$ and $E^{\rm AGN}_{\rm c}$ corresponding to 
equations~(\ref{couplingHot})~and~(\ref{couplingCold}) are used as follows: the AGN feedback 
energy $E^{\rm AGN}_{\rm h}$ provided to the hot gas is used to increase its temperature $T_{\rm h}$. 
On the other hand, the AGN feedback energy $E^{\rm AGN}_{\rm c}$ coupled to the cold phase is 
employed to bring cold gas mass to the hot phase. As a consequence, the initial mass of cold gas $M_{\rm c}$ of 
the multiphase particle (whose temperature remains fixed at $T_{\rm c}= 300$~K, see Table~\ref{ch5:TableParameters}) 
is progressively eroded because of the effect of AGN.

%%         The following set if NO AGNmuppi: 
%%  \begin{align}
%%  \dot{M}_{\rm h}  &=  - \dot{M}_{\rm cool} + \dot{M}_{\rm ev} \,\,\,, \label{ch5:muppi1} \\
%%  \dot{M}_{\rm c}  &=   \dot{M}_{\rm cool}  - \dot{M}_{\rm sf} - \dot{M}_{\rm ev} \,\,\,,  \label{ch5:muppi2}  \\
%%  \dot{M}_{\ast}  &=  \dot{M}_{\rm sf}  \,\,\,,  \label{ch5:muppi3} \\
%%  \dot{E}_{\rm h}  &=  \dot{E}_{\rm fb, local} - \dot{E}_{\rm cool} + \dot{E}_{\rm hydro} \,\,\,. \label{ch5:muppi4}
%%  \end{align}

When the effect of AGN feedback is included within the sub-resolution model MUPPI, 
the following set of ordinary differential equations describes mass and energy flows between the different components: 
\begin{align}
\dot{M}_{\rm h}  &=  - \dot{M}_{\rm cool} + \dot{M}_{\rm ev} 
				+  \dot{M}^{\rm AGN}_{\rm c \rightarrow h}  \,\,\,, \label{ch9:AGNmuppi1} \\
\dot{M}_{\rm c}  &=   \dot{M}_{\rm cool}  - \dot{M}_{\rm sf} - \dot{M}_{\rm ev} 
				-  \dot{M}^{\rm AGN}_{\rm c \rightarrow h} \,\,\,,  \label{ch9:AGNmuppi2}  \\
\dot{M}_{\ast}  &=  \dot{M}_{\rm sf}  \,\,\,,  \label{ch9:AGNmuppi3} \\
\dot{E}_{\rm h}  &=  \dot{E}_{\rm fb, local} - \dot{E}_{\rm cool} + \dot{E}_{\rm hydro} 
			    + \dot{E}^{\rm AGN}_{\rm h} + \dot{E}^{\rm AGN}_{\rm c \rightarrow h} \,\,\,, \label{ch9:AGNmuppi4} \\
\dot{E}^{\rm AGN}_{\rm c, \, used}  &=  \dot{E}^{\rm AGN}_{\rm c \rightarrow h} \,\,\,. \label{ch9:AGNmuppi5} 
\end{align}

Equations~(\ref{ch9:AGNmuppi1}),~(\ref{ch9:AGNmuppi2}),~(\ref{ch9:AGNmuppi3}),~(\ref{ch9:AGNmuppi4}),~and~(\ref{ch9:AGNmuppi5}) 
are integrated instead of 
equations~(\ref{ch5:muppi1}),~(\ref{ch5:muppi2}),~(\ref{ch5:muppi3}),~and~(\ref{ch5:muppi4}) 
introduced in Section~\ref{MUPPI}. The new contributions that account for the AGN feedback are 
labelled with the superscript~{\sl{AGN}}. We detail each of the new terms below. 

The term $\dot{M}^{\rm AGN}_{\rm c \rightarrow h}$ in equation~(\ref{ch9:AGNmuppi1}) accounts for the 
mass of cold gas that is brought to the hot phase due to the AGN feedback energy $E^{\rm AGN}_{\rm c}$ coupled 
to the cold component. 

The set of 
equations~(\ref{ch9:AGNmuppi1}),~(\ref{ch9:AGNmuppi2}),~(\ref{ch9:AGNmuppi3}),~and~(\ref{ch9:AGNmuppi4}) is 
integrated with a Runge-Kutta algorithm (whose time-step we refer to as $\Delta t_{\rm MUPPI}$) within 
each SPH time-step $\Delta t_{\rm SPH} > \Delta t_{\rm MUPPI}$, 
as explained in Section~\ref{MUPPI} \citep[see][for details]{muppi2010, muppi2014}. 

Therefore, first, the code evaluates the amount of the cold gas mass that can be brought to the hot phase within 
the SPH time-step $\Delta t_{\rm SPH}$ using the entire energy budget $E^{\rm AGN}_{\rm c}$ available, i.e.:
\begin{equation}     
\dot{M}^{\rm AGN}_{\rm c, \, th } = \dot{E}^{\rm AGN}_{\rm c } 
\, \frac{(\gamma -1) \, \mu_{\rm c} \, m_{\rm p}}{{\text {k}}_{\rm B} \, (T_{\rm h} - T_{\rm c})}  \,\,\,, 
\label{ch9:AGNmuppi6_th}
\end{equation}
where $\gamma=5/3$ is the adiabatic index, and ${\text {k}}_{\rm B}$ and $m_{\rm p}$ are 
the Boltzmann constant and the mass of the proton, respectively. 
In equation~(\ref{ch9:AGNmuppi6_th}), we neglect the work done against the hot phase to let the cold gas 
expand as soon as it is brought into the hot component\footnote{When the cold gas evaporates, we 
	increase for simplicity its internal energy and neglect the $P \, d V$ work contribution, assuming that 
	hydrodynamical forces account for it in the following timestep. 
	Equation~(\ref{ch9:AGNmuppi6_th}) indeed provides a slight overestimation of $\dot{M}^{\rm AGN}_{\rm c, \, th }$. 
	The energy rate $\dot{E}^{\rm AGN}_{\rm c }$ should actually be divided by
	 $\,\, \frac{{\text {k}}_{\rm B} \, (T_{\rm h} - T_{\rm c})} {(\gamma -1) \, \mu_{\rm c} \, m_{\rm p}} + \frac{P_{\rm c}}{\rho_{\rm h}}\,\,$, where $P_{\rm c} / \rho_{\rm h}$ represents the $P \, d V$ work done against the hot gas, 
	 and $P_{\rm c} = P_{\rm h}$ due to the pressure equilibrium between the gas phases. 
	 Should this correction be taken into account, equation~(\ref{ch9:AGNmuppi6_th}) becomes:  
	 $\,\, \dot{M}^{\rm AGN}_{\rm c, \, th } = \dot{E}^{\rm AGN}_{\rm c } 
 	 \, \frac{(\gamma -1) \, \mu_{\rm c} \, m_{\rm p}}{{\text {k}}_{\rm B} \, (\gamma \, T_{\rm h} - T_{\rm c})}  \,\,$.
	 The factor $1.67$ by which the hot gas temperature would be increased introduces a contribution which 
	 can be considered within the uncertainty of the parameters of the model.}. 
Then, should the cold gas mass to be (in {\sl {theory}}) evaporated $\dot{M}^{\rm AGN}_{\rm c, \, th }$ exceed 
the gas of the cold phase $M_{\rm c}$ available in the MUPPI time-step $\Delta t_{\rm MUPPI}$, 
we limit $\dot{M}^{\rm AGN}_{\rm c, \, th }$ to: 
\begin{equation}  
\dot{M}^{\rm AGN}_{\rm c \rightarrow h} = \frac{M_{\rm c} }{ \Delta t_{\rm MUPPI}}   \,\,\,. 
\end{equation}
Therefore:
\begin{equation}
\dot{M}^{\rm AGN}_{\rm c \rightarrow h} = \begin{cases}
               \dot{M}^{\rm AGN}_{\rm c, \, th } {\,\,\,\,\,\,\,\,\,\,\,\,\,\,\,\,\,\,\,\,\,\,\,\,\,\,\,\,\,\,\,\,\,\,\,\,\,\,\,\,\,\,\,\,\,\,\,{\text {for}}\,\,\,\, \dot{M}^{\rm AGN}_{\rm c, \, th } \, \Delta t_{\rm MUPPI} \leq M_{\rm c}\,\,\,,}\\
               M_{\rm c} / \Delta t_{\rm MUPPI} {\,\,\,\,\,\,\,\,\,\,\,\,\,\,\,\,\,\,\,\,\,\,\,\,\,\,\,\,\,\,\,\,{\text {for}}\,\,\,\, \dot{M}^{\rm AGN}_{\rm c, \, th } \, \Delta t_{\rm MUPPI} > M_{\rm c} \,\,\,,}
            \end{cases}
\label{ch10:sph4k}
\end{equation}
and $\dot{M}^{\rm AGN}_{\rm c \rightarrow h}$ is lower than or equal to $\dot{M}^{\rm AGN}_{\rm c, \, th }$. 
$\dot{M}^{\rm AGN}_{\rm c \rightarrow h}$ in equation~(\ref{ch9:AGNmuppi5}) represents a source term 
for the evolution of the hot gas mass in equation~(\ref{ch9:AGNmuppi1}), 
and a sink term for the evolution of the mass of the cold phase in equation~(\ref{ch9:AGNmuppi2}).

Once $\dot{M}^{\rm AGN}_{\rm c \rightarrow h}$ is retrieved, 
it is adopted to compute the amount of AGN feedback energy $E^{\rm AGN}_{\rm c \rightarrow h}$ that is actually 
used to evaporate cold gas. It reads: 
\begin{equation}     
\dot{E}^{\rm AGN}_{\rm c \rightarrow h}    =  \dot{M}^{\rm AGN}_{\rm c \rightarrow h}
\, \frac{k_{\rm B} \, (T_{\rm h} - T_{\rm c})} {(\gamma -1) \, \mu_{\rm c} \, m_{\rm p}} \,\,\,. 
\label{ch9:AGNmuppi6}
\end{equation}
The energy contribution $E^{\rm AGN}_{\rm c \rightarrow h}$ accounts for the energy that is supplied 
to the hot component by the cold mass that is evaporated and enters the hot phase. 
In this way, besides the term $\dot{E}^{\rm AGN}_{\rm h}$ described in equation~(\ref{couplingHot}), 
$\dot{E}^{\rm AGN}_{\rm c \rightarrow h} $ also increases the hot gas energy (see 
equation~(\ref{ch5:muppi7})). 

Equation~(\ref{ch9:AGNmuppi5}) describes the evolution of the AGN feedback energy that the cold gas is 
provided with and that is actually consumed to evaporate cold gas. The energy rate 
$\dot{E}^{\rm AGN}_{\rm c, \, used}$ records the rate of consumed energy $\dot{E}^{\rm AGN}_{\rm c \rightarrow h}$, 
that is lower than or equal to $\dot{E}^{\rm AGN}_{\rm c }$ (see equation~(\ref{couplingCold})) due to the fact that 
$\dot{M}^{\rm AGN}_{\rm c \rightarrow h}$ is lower than or equal to $\dot{M}^{\rm AGN}_{\rm c, \, th }$. 

Then, at the end of the SPH time-step, $\dot{E}^{\rm AGN}_{\rm c, \, used}$ from equation~(\ref{ch9:AGNmuppi5}) 
is contrasted to the originally available $\dot{E}^{\rm AGN}_{\rm c}$. Should:
\begin{equation}     % SphP[i].E_h += InitialBHEcold - y[4];    sfr_muppi : 1988
\dot{E}^{\rm AGN}_{\rm c, \, extra} = \dot{E}^{\rm AGN}_{\rm c} - \dot{E}^{\rm AGN}_{\rm c, \, used} > 0  \,\,\,, 
\label{ch9:extraEnergy}
\end{equation}
the corresponding further energy contribution $E^{\rm AGN}_{\rm extra, c}$ is provided to 
the energy of the hot gas component $E^{\rm AGN}_{\rm h}$. 
This addition amounts to the energy that has not been used during the entire SPH time-step to bring cold gas from 
the cold to the hot phase, because the multiphase particle was already devoid of the cold gas mass. 
Also, this contribution ensures that no feedback energy is lost: indeed, whenever $E^{\rm AGN}_{\rm c}$ 
exceeds the maximum energy that can be used to lead all the available cold gas to the hot phase, the 
remaining energy is coupled to the hot gas, that is the only component left. 
We integrate equation~(\ref{ch9:AGNmuppi5}) and then provide the extra 
energy $E^{\rm AGN}_{\rm c, \, extra}$ to the hot gas at the end of the SPH time-step rather than 
estimating the extra energy budget at the beginning of the SPH time-step for the following reason: 
the temperature $T_{\rm h}$ of the hot gas changes during the integration of the 
equations~(\ref{ch9:AGNmuppi1}),~(\ref{ch9:AGNmuppi2}),~(\ref{ch9:AGNmuppi3}),~and~(\ref{ch9:AGNmuppi4}), 
so that the precise amount of $M^{\rm AGN}_{\rm c \rightarrow h}$ is known only at the end of the integration.  

%   FB.dE += BHEhot*All.HubbleParam/dtime; //BH energy coupled with the hot phase
%   dMBH is the amount of cold mass brought to the temperature of the hot phase (in dtime)
%   dBHEcold_used is the BH energy required to bring to T_h a mass dMDB*dt_rungekutta of cold gas    : 2361
%   InitialBHEcold è l’energia che è inizialmente destinata dall’AGN alla fase fredda;    sfr_muppi : 1986 
%   y[4] è l’energia effettivamente usata per scaldare una porzione o tutta la massa fredda presente nella particella. Se inferiore a InitialBHEcold, allora (InitialBHEcold-y[4]) viene data alla fase calda;

The general description of the AGN feedback model outlined so far accounts for AGN feedback energy that is 
distributed to both the hot and the cold gas. The way in which the feedback energy is shared among the 
different phases of the multiphase ISM is established by the coupling parameters $\mathcal{A}_{\rm h}$ and 
$\mathcal{A}_{\rm c}$ (see equations~(\ref{couplingHot})~and~(\ref{couplingCold})), and different scenarios 
arise when specific values or parametrizations for them are adopted. 
This is discussed in Section~\ref{ch10:couplingFactors}.

%%%%%%%%%%%%%%%%%%%%%%%%%%%%%%%%%%%%%%%%%%%%%%%%%%%%%%%%%%
%%%%%%%%%%%%%%%%%%%%%%%%%%%%%%%%%%%%%%%%%%%%%%%%%%%%%%%%%%

\subsection{Coupling AGN feedback energy to a multiphase ISM}
\label{ch10:couplingFactors}

As explained in Section~\ref{AGNmuppi}, the AGN feedback energy assigned to multiphase 
particles can be shared between their hot and cold components. Therefore, by designing different ways 
of distributing the available feedback energy to the hot and cold gas phases, it is possible to investigate 
how feedback energy couples to a multiphase ISM, and how various possibilities impact on the 
BH-galaxy coevolution. 
The way in which feedback energy is distributed between the hot and cold gas is controlled by 
the coupling parameters $\mathcal{A}_{\rm h}$ and $\mathcal{A}_{\rm c} = 1 - \mathcal{A}_{\rm h}$ 
(see equations~(\ref{couplingHot})~and~(\ref{couplingCold})). 

Different combinations can be explored, and they can be broadly divided into two different categories: 
{\sl {(i)}} constant values of the coupling parameters, that we arbitrarily set to either $0$, $0.5$, or $1$ 
(see Section~\ref{simus} and also Appendix~\ref{ch10:ConstantcouplingFactors}); 
and {\sl {(ii)}} coupling parameters modelled according to the physical properties of the ISM, i.e. of the 
multiphase particle which is provided with feedback energy (Section~\ref{ch10:CoveringFactors}).

%%%%%%%%%%%%%%%%%%%%%%%%%%%%%%%%%%%%%%%%%%%%%%%%%%%%%%%%%%

%\subsubsection{Constant coupling parameters}
%\label{ch10:ConstantcouplingFactors}

%%%%%%%%%%%%%%%%%%%%%%%%%%%%%%%%%%%%%%%%%%%%%%%%%%%%%%%%%%

\subsubsection{Locally varying energy coupling}
\label{ch10:CoveringFactors}

Our approach to determine the coupling factors according to the physical properties of 
the multiphase particles is based on computing the covering 
factors of the hot and cold phases. The physical idea behind this modelling considers that the larger is the 
cross section of the cold clouds embedded in the cold phase 
(and thus the surface that they expose to the AGN incident radiation), 
the larger is the amount of energy that they can intercept and absorb. 

A multiphase particle in our sub-resolution model samples a portion of the ISM, where the diffuse hot phase 
coexists with a cold component. The cold component also accounts for the presence of molecular gas, that 
we assume as a share of a giant molecular cloud. Moreover, we consider that the molecular content of the 
multiphase particle is made up of a given number $N$ of cold cloudlets or clumps (see below). Observations 
\citep[e.g.][and references therein]{Williams1994, Bergin2007, Munoz2007, Gomez2014} 
suggest that the clumps that constitute a giant molecular cloud have a distribution of masses and sizes, 
ranging between a few tenth to few~pc and spanning the mass range~$10 - 10^4$~M$_{\odot}$. 

The filling factor $f_{\rm h}$ of the hot gas (equation~(\ref{ch5:fillingFactor})) is related to the fraction of gas 
mass in the hot phase within the multiphase particle, labelled $F_{\rm h}$, and quantifies its clumpiness. 
Note that the formulation provided by equation~(\ref{ch5:fillingFactor}) is equivalent to 
express the filling factor of the hot phase as the ratio between the volume filled by the hot gas and the 
volumes occupied by both the hot and cold components. 

Being the filling factor of the cold phase $f_{\rm c}=1-f_{\rm h}$, the covering factor of the cold phase 
can be cast as:
\begin{equation}
\mathcal{C}_{\rm c} = f_{\rm c} \, \frac{L_{\rm P}}{\ell_{\rm MC}} \,\,\,.
\label{ch9:GL10}
\end{equation}
In equation~(\ref{ch9:GL10}), $\ell_{\rm MC}$ is the typical size of molecular (cold) cloudlets, 
while $L_{\rm P}$ is the size of the multiphase particle, i.e.: 
\begin{equation}
L_{\rm P} = \biggl( \frac{3}{4 \, \pi} \, V_{\rm P} \biggr)^{1/3} =  \biggl( \frac{3}{4 \, \pi} \, 
\frac{M_{\rm P} - M_{\ast} }{\rho} \biggr)^{1/3}  \,\,\,, 
\label{ch9:GL11}
\end{equation}
where $V_{\rm P}$, $M_{\rm P}$, $M_{\ast}$, and $\rho$ are the volume of the multiphase particle 
occupied by the gas phases, the mass of the multiphase particle, the mass of its stellar component, 
and the SPH density of the multiphase particle, respectively. 
$\ell_{\rm MC}$ is a parameter of the model: 
after carrying out extensive tests (see Appendix~\ref{CalibLmc}), 
we adopt $\ell_{\rm MC} = 1$~pc for our fiducial model, 
this value being in keeping with the aforementioned observations. 

Equation~(\ref{ch9:GL10}) is obtained by considering that the filling factor can be expressed as
$ \, f_{\rm c} = N \, \ell_{\rm MC}^3  / L_{\rm P}^3 \,$, $N$ being the number of cold cloudlets within 
the multiphase particle (see above), while $\, \mathcal{C}_{\rm c} = N \, \ell_{\rm MC}^2  / L_{\rm P}^2 \,$, 
and by dividing the latter equation by the former one. 
The covering factor of the hot phase is: $\mathcal{C}_{\rm h} = 1- \mathcal{C}_{\rm c}$. 

Therefore, within this model we assume: $\mathcal{A}_{\rm h}= \mathcal{C}_{\rm h}$ 
and $\mathcal{A}_{\rm c} = \mathcal{C}_{\rm c}$. 
When computing $\mathcal{C}_{\rm c}$, we first check whether $f_{\rm h} < 1$; 
should the multiphase particle be entirely filled by hot gas (i.e. $f_{\rm h} = 1$), 
then $\mathcal{C}_{\rm c} = 0$. 
Also, should $\mathcal{C}_{\rm c} > 1$ happen if $L_{\rm P} \gg \ell_{\rm MC}$, 
the covering factor is limited to unity, i.e. $\mathcal{C}_{\rm c} = 1$, and thus $\mathcal{C}_{\rm h} = 0$. 
This situation can be associated to the case in which cold clouds overlap with each other, 
clouds at small radii shielding clouds at large radii, thus reducing the fraction of energy they receive.

%%%%%%%%%%%%%%%%%%%%%%%%%%%%%%%%%%%%%%%%%%%%%%%%%%%%%%%%%%
%%%%%%%%%%%%%%%%%%%%%%%%%%%%%%%%%%%%%%%%%%%%%%%%%%%%%%%%%%

\begin{table}
\centering
\begin{minipage}{84mm}
\vskip -0.5em
\caption[Relevant parameters of simulation with AGN feedback]{Relevant parameters of the simulations with AGN feedback.\\ 
{\sl {Column~1:}} simulation label. 
{\sl {Column~2:}} AGN feedback: included or not. 
{\sl {Column~3:}} hot and/or cold gas accretion. 
{\sl {Column~4:}} angular momentum limiter: included or not. 
{\sl {Column~5:}} AGN feedback energy coupling to the multiphase ISM. 
{\sl {Column~6:}} BH seed mass. 
{\sl {Column~7:}} size of cold clumps in molecular clouds.
Other parameters as in Table~\ref{ch10:AGNmodelParamm}.} 
\renewcommand\tabcolsep{1.4mm}
\newcommand{\cmark}{\ding{51}}
\newcommand{\xmark}{\ding{55}}
\vskip 1.5em
\begin{tabular}{@{}lcccccc@{}}
Label    &     AGN     &    $\dot{M}_{\rm BH}$     &    $\mathcal{L}_{\rm AM}$              &  Energy          & $M_{\rm BH, \, seed}$   &  $\ell_{\rm MC}$  \\ 
             &                  &    &   $\frac{C_{\rm visc}}{2 \, \pi}$     & coupling    & (M$_{\odot}$)            &   (pc)     \\ 
\hline
\hline
noAGN--reference &  \xmark  & & & &  & \\  
\hline
fiducial--hcAL--cf &   \cmark   &  hot+ &  \cmark & $\mathcal{A}_{\rm h}= \mathcal{C}_{\rm h}$, &  $1.1$$\cdot$$10^5$ & 1\\ 
			 &          &  cold & $1$  &  $\mathcal{A}_{\rm c} = \mathcal{C}_{\rm c}$  &   &    \\
\hline			 
hcA--hot &  \cmark   &  hot+  &  \xmark & $\mathcal{A}_{\rm h}=1$,  &  $1.1$$\cdot$$10^5$    & \\  
 			&             &  cold  &   &  $\mathcal{A}_{\rm c} = 0$ &    &\\
\hline
hcA--both &  \cmark  &  hot+  &  \xmark & $\mathcal{A}_{\rm h}=0.5$,  &  $1.1$$\cdot$$10^5$     &\\ 
			 &           &  cold  &   &  $\mathcal{A}_{\rm c} = 0.5$ &    &  \\ 
\hline
hcA--cold &  \cmark   &  hot+  &  \xmark & $\mathcal{A}_{\rm h}=0$, &  $1.1$$\cdot$$10^5$    &  \\  
			 &             &  cold  &  &  $\mathcal{A}_{\rm c} = 1$  &     & \\
\hline

hcAL2--both &  \cmark   &  hot+  &  \cmark & $\mathcal{A}_{\rm h}=0.5$, &  $1.1$$\cdot$$10^5$     &\\ 
			 &           &  cold  &  $10^2$ &  $\mathcal{A}_{\rm c} = 0.5$ &   & \\ 
\hline
hcAL3--both &  \cmark   &  hot+  &  \cmark & $\mathcal{A}_{\rm h}=0.5$, &  $1.1$$\cdot$$10^5$  & \\  
			 &           &  cold  &  $10^3$ &  $\mathcal{A}_{\rm c} = 0.5$  &     & \\
\hline
ocA--both  &  \cmark   &  cold  &  \xmark & $\mathcal{A}_{\rm h}=0.5$, &  $1.1$$\cdot$$10^5$    & \\ 
			 &           &     &  &  $\mathcal{A}_{\rm c} = 0.5$  &     & \\ 
\hline
ocAL--both  &  \cmark   &  cold  &  \cmark & $\mathcal{A}_{\rm h}=0.5$, &  $1.1$$\cdot$$10^5$    & \\ 
			 &           &     &  $1$ &  $\mathcal{A}_{\rm c} = 0.5$ &      \\
\hline
hcA--cf &   \cmark   &  hot+ &  \xmark & $\mathcal{A}_{\rm h}= \mathcal{C}_{\rm h}$, &  $1.1$$\cdot$$10^5$ & 1 \\ 
			 &          &  cold &  &  $\mathcal{A}_{\rm c} = \mathcal{C}_{\rm c}$   &  &     \\
\hline			 
hcAL2--cf &   \cmark   &  hot+ &  \cmark & $\mathcal{A}_{\rm h}= \mathcal{C}_{\rm h}$, &  $1.1$$\cdot$$10^5$ & 1 \\ 
			 &           &  cold & $10^2$  &  $\mathcal{A}_{\rm c} = \mathcal{C}_{\rm c}$ &  &     \\
\hline		
	 
hcA--cf--20pc &   \cmark   &  hot+ &  \xmark & $\mathcal{A}_{\rm h}= \mathcal{C}_{\rm h}$, &  $1.1$$\cdot$$10^5$ & 20 \\ 
			 &           &  cold & &  $\mathcal{A}_{\rm c} = \mathcal{C}_{\rm c}$   &   &   \\
\hline			 
hcAL--cf--20pc &   \cmark   &  hot+ &  \cmark & $\mathcal{A}_{\rm h}= \mathcal{C}_{\rm h}$, &  $1.1$$\cdot$$10^5$  &  20  \\ 
			 &          &  cold & $1$  &  $\mathcal{A}_{\rm c} = \mathcal{C}_{\rm c}$  &   &    \\
\hline			 

hcA--both--S0.5x  &  \cmark   &  hot+  &  \xmark & $\mathcal{A}_{\rm h}=0.5$, &  $5.5$$\cdot$$10^4$  &  \\ 
			 &          &  cold  &   &  $\mathcal{A}_{\rm c} = 0.5$  &   &  \\
\hline
hcA--both--S2x &  \cmark   &  hot+  &  \xmark & $\mathcal{A}_{\rm h}=0.5$,  &  $2.7$$\cdot$$10^5$  &  \\ 
			 &           &  cold  & &  $\mathcal{A}_{\rm c} = 0.5$   &     & \\ 
\hline
ocA--both--S0.5x  &  \cmark   &  cold  &  \xmark & $\mathcal{A}_{\rm h}=0.5$, &  $5.5$$\cdot$$10^4$   &  \\ 
			 &            &     &   &  $\mathcal{A}_{\rm c} = 0.5$ &       &  \\
\hline
ocA--both--S2x &  \cmark   &  cold  &  \xmark & $\mathcal{A}_{\rm h}=0.5$, &  $2.7$$\cdot$$10^5$    &   \\ 
			 &           &     &  &  $\mathcal{A}_{\rm c} = 0.5$  &      &   \\  
\hline
\hline
\end{tabular}
\label{simList}
\end{minipage}
\end{table}

%%%%%%%%%%%%%%%%%%%%%%%%%%%%%%%%%%%%%%%%%%%%%%%%%%%%%%%%%%

    %%%          %      %          %        %         %      %%%      
  %%               %     % %     % %      %         %    %%               
   %%%           %     %   % %   %      %         %     %%%     
      %%%        %     %     %     %      %         %        %%%  
            %%     %     %             %      %         %              %% 
           %%      %     %             %      %         %            %% 
     %%%         %     %             %       %%%%        %%%
     
%%%%%%%%%%%%%%%%%%%%%%%%%%%%%%%%%%%%%%%%%%%%%%%%%%%%%%%%%%

\section{The suite of simulations}
\label{simus}

In this Section, we introduce the set of simulations carried out to investigate the 
impact of AGN feedback on the evolution of late-type galaxies. 
Simulations have been performed with the GADGET3 code, 
a non-public evolution of the GADGET2 code \citep{springel2005}.  
We use the improved formulation of SPH presented in \citet{beck2015} and introduced 
in cosmological simulations adopting the sub-resolution model MUPPI by \citet{Valentini2017}. 
The initial conditions (ICs) are the $AqC5$~ICs introduced by \citet{Springel2008}. 
They are zoomed-in ICs and describe an isolated DM halo of 
mass $M_{\rm halo, \, DM} \simeq 1.8 \cdot 10^{12}$~M$_{\odot}$ at redshift $z=0$. 
The Plummer-equivalent softening length for the computation of the gravitational force is 
$\varepsilon_{\rm Pl} = 325 \, h^{-1}$~pc, DM particles have a mass of $1.6 \cdot 10^6 \, h^{-1}$~M$_{\odot}$, 
and the initial mass of gas particles is $3.0 \cdot 10^5 \, h^{-1}$~M$_{\odot}$. 

Besides the reference simulation without BHs and the ensuing feedback used as control simulation, 
the simulations carried out for the present analysis include the implementation of AGN that we described 
in Section~\ref{AGNmodelling}. 
We designed a number of simulations (see Table~\ref{simList}) 
aimed at investigating the effect of the following aspects of the numerical 
implementation (we highlight within brackets the reference acronym encoded in the simulation label):

\begin{itemize}
\item  {\sl {Gas accretion:}} We consider the possibility for the BH to accrete: 
\begin{enumerate}%[wide, labelwidth=!, labelindent=0pt]
\item [--] {\sl {(hcA)}} --  both hot and cold gas, according to equation~(\ref{Mdot_limited});
\item [--] {\sl {(ocA)}} --  cold gas only, according to $\, \dot{M}_{\rm BH} = {\text{min}}  (\dot{M}_{\rm B, \, c}, \dot{M}_{\rm Edd}) \,$. 
\end{enumerate}
In addition, we explore the effect of the {\sl {angular momentum limiter}} to cold gas accretion 
introduced in Section~\ref{BHmango}, taking into account: 
\begin{enumerate}
\item [--] {\sl {(hcAL)}} --  both cold (limited) plus hot accretion; 
\item [--] {\sl {(ocAL)}} --  only cold gas accretion, limited by $\mathcal{L}_{\rm AM}$ 
(equations~(\ref{Mdot_limited_mango})~and~(\ref{AngMomLimiter})).
\end{enumerate}

\item  {\sl {Energy coupling:}} We explore the different possibilities described in Section~\ref{ch10:couplingFactors}, considering:
\begin{enumerate}
\item [-- {\sl {(cf)}} --]  $\mathcal{A}_{\rm h}= \mathcal{C}_{\rm h}$ and $\mathcal{A}_{\rm c} = \mathcal{C}_{\rm c}$; 
\item [-- {\sl {(both)}} --]  $\mathcal{A}_{\rm h}=0.5$ and $\mathcal{A}_{\rm c} = 0.5$;
\item [-- {\sl {(hot)}} --]  $\mathcal{A}_{\rm h}=1$ and $\mathcal{A}_{\rm c} = 0$;
\item [-- {\sl {(cold)}} --]  $\mathcal{A}_{\rm h}=0$ and $\mathcal{A}_{\rm c} = 1$.
\end{enumerate} 
  
\item  {\sl {BH seed mass:}} To investigate the effect of the initial BH mass on the BH-galaxy 
coevolution and select the fiducial value, we choose the following BH seed masses:
\begin{enumerate}
\item [-- {\sl {(Sref)}} --]  $M_{\rm BH, \, seed} = 1.1 \cdot 10^5$~M$_{\odot}$, reference value;
\item [-- {\sl {(S0.5x)}} --]  $M_{\rm BH, \, seed} = 5.5 \cdot 10^4$~M$_{\odot}$; 
%, BH seed mass roughly halved with respect to the fiducial value;
\item [-- {\sl {(S2x)}} --]  $M_{\rm BH, \, seed} = 2.7 \cdot 10^5$~M$_{\odot}$. 
%, BH seed mass roughly twice as massive as the reference value.
\end{enumerate} 

\item  {\sl {Cold clump size:}} Also, we examine two possible values for the characteristic size of cold clumps 
in molecular clouds $\ell_{\rm MC}$, i.e.:
\begin{enumerate}
\item [-- {\sl {(1pc)}} --]  $\ell_{\rm MC}=1$~pc, our fiducial value;  
\item [-- {\sl {(20pc)}} --]  $\ell_{\rm MC}=20$~pc, to quantify the impact of the variation of this parameter 
on final results (see Appendix~\ref{CalibLmc} for details). 
\end{enumerate}
\end{itemize}        
       
Table~\ref{simList} lists all the simulations that we present in this work (and contains also those which will be 
discussed in the Appendices). 
The name of the simulations encodes the implementation options described above 
and uses the following template: simulation names are in the form 
{\sl {(noAGN-)aaA(L)-cccc-(Sssx-nnpc)}}. 

%%%
All the simulations described in this section include AGN feedback, 
while the label {\sl {noAGN}} is only present for the reference one where the evolution of SMBHs is not included. 
%%%
The label {\sl {aaA(L)}} stands for the modelling of the accretion: {\sl {hcA}} if both hot and cold gas are accreted, 
{\sl {ocA}} if only cold gas is accreted. In addition, should the simulations include the angular momentum {\sl {limiter}} 
for the cold gas accretion, their label reads {\sl {hcAL}} or {\sl {ocAL}}, 
with the possibility of showing a number that encodes the value of the parameter $C_{\rm visc}$ 
(see Section~\ref{BHmango}). In this way, {\sl {hcAL2}} refers to $C_{\rm visc}/2 \, \pi = 10^2$, 
{\sl {hcAL3}} refers to $C_{\rm visc}/2 \, \pi = 10^3$, while {\sl {hcAL}} and {\sl {ocAL}} 
assume $C_{\rm visc}/2 \, \pi = 1$. 
%%%
The label {\sl {cccc}} refers to the coupling, and it can take the values {\sl {cf}}, {\sl {both}}, {\sl {cold}}, or {\sl {hot}}. 

The accretion and coupling labels are missing only for the control simulation with no AGN feedback. 
The seed label is in the form {\sl {Sssx}}, where {\sl {ss}} can take the values {\sl {ss}} $=2$ or {\sl {ss}} $=0.5$. 
The seed label is absent if the reference seed is used ($M_{\rm BH, \, seed} = 1.1 \cdot 10^5$~M$_{\odot}$). 
Simulations where the coupling of the AGN feedback energy is set according to 
the physical properties of the multiphase particles have a label which indicates the value of $\ell_{\rm MC}$, 
and can be either $1$ or $20$~pc. When not present, $1$~pc is assumed. 
Our fiducial model is fiducial--hcAL--cf. 
We can support or discard one or some among our models by comparing their predictions 
to observations (e.g. Figure~\ref{ch10:mago_ref}) and possible scenarios for BH-galaxy evolution.

%%%%%%%%%%%%%%%%%%%%%%%%%%%%%%%%%%%%%%%%%%%%%%%%%%%%%%%%%%
%%%%%%%%%%%%%%%%%%%%%%%%%%%%%%%%%%%%%%%%%%%%%%%%%%%%%%%%%%

\section{Results}
\label{Results}

In this section we present the results. 
We show how the coupling of AGN feedback energy to the multiphase ISM determines 
the main features of simulated galaxies (Section~\ref{ch10:GenRes}), the evolution of BHs 
(Section~\ref{ch10:BHevo}), and the BH-galaxy coevolution (Section~\ref{ch10:Galaxyevo}). 
In Section~\ref{ch10:galacticOutflows}, we explore the effect of the stellar and BH feedback on galactic outflows, 
while Section~\ref{ch10:Metals} is devoted to investigate the effect of AGN feedback on metallicity profiles. 
In Section~\ref{ch10:Mango}, we discuss the effect of the modelling of BH gas accretion on final results. 
Throughout the paper we often use the term coevolution to refer to SMBHs that evolve within and along with 
their host galaxy. In Section~\ref{ch10:Mango}, we discuss in detail the timing of BH growth with respect to 
that of the different components of the galaxy, and the way in which galaxy scaling relations are set.

%%%%%%%%%%%%%%%%%%%%%%%%%%%%%%%%%%%%%%%%%%%%%%%%%%%%%%%%%%
%%%%%%%%%%%%%%%%%%%%%%%%%%%%%%%%%%%%%%%%%%%%%%%%%%%%%%%%%%

\subsection{Disc galaxies with AGN feedback} 
\label{ch10:GenRes}

%%%%%%%%%%%%%%%%%%%%%%%%%%%%% MAPS
\begin{figure*}
\newcommand{\captionfonts}{\small}
\begin{minipage}{\linewidth}%{1.0\linewidth}
\centering
\vspace{-2.1ex}
\includegraphics[trim=0.cm 0.cm 0.5cm 0.cm, clip, angle=270, width=1.\textwidth]{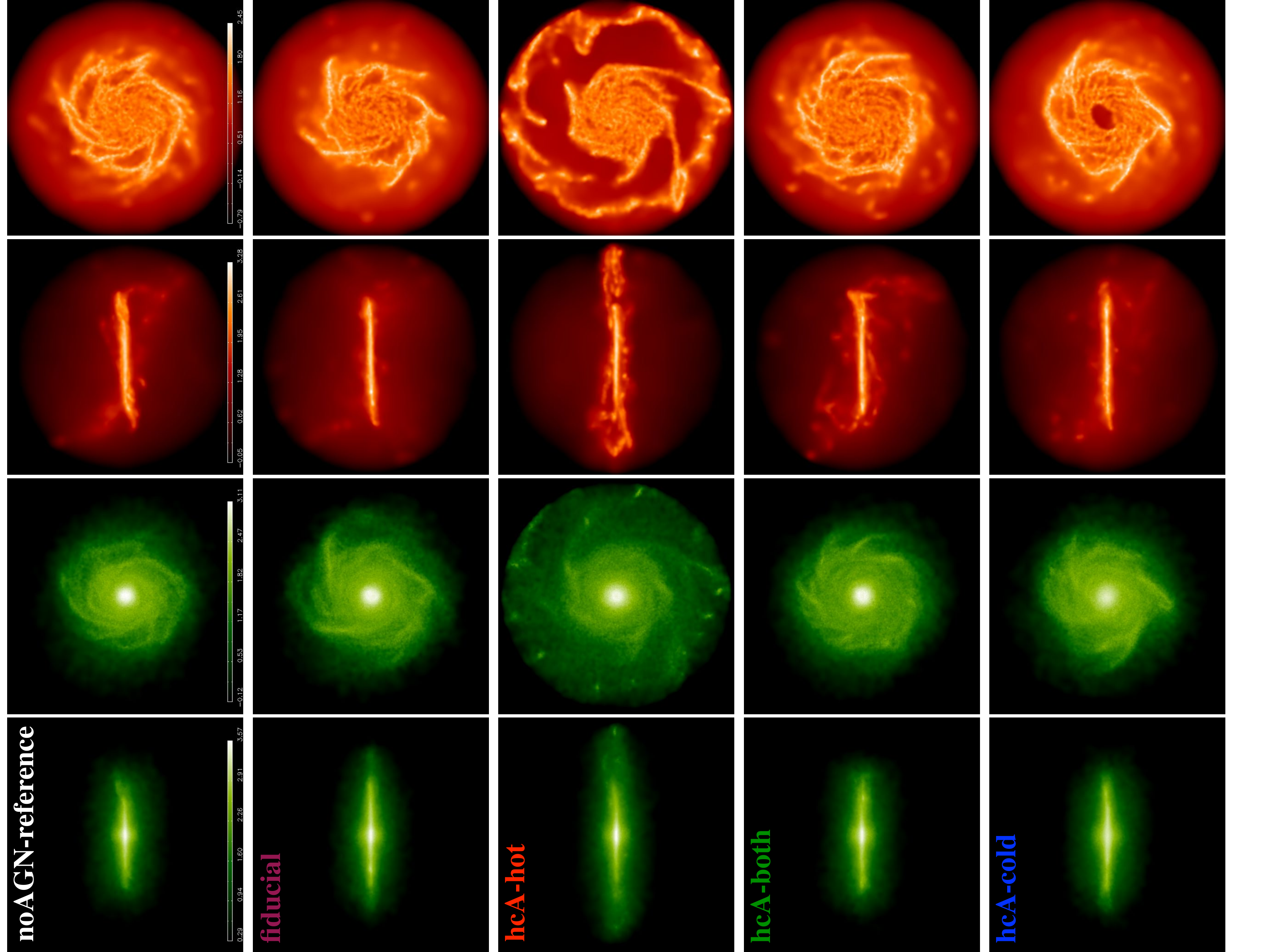} 
\end{minipage} 
\vspace{-2.1ex}
\caption{Projected stellar (first and second columns) and gas (third and forth columns) density maps for four of 
	the simulated galaxies listed in Table \ref{simList}, at redshift $z = 0$. 
	Each row shows a galaxy, whose name is indicated in the first column panel. First and third columns 
	show edge-on galaxies, second and forth columns depict face-on maps. The size of each box is $50$~kpc a side. 
	Colour bars encode the logarithm of projected densities (M$_{\odot}$/pc$^2$). Panels in each column share the same colour bar.}
\label{StellarDensityMaps} 
\end{figure*}
%%%%%%%%%%%%%%%%%%%%%%%%%%%%% MAPS

We start to investigate the BH-galaxy coevolution by focussing on the following five galaxies: 
\begin{itemize}
\item {\sl {noAGN--reference:}} this is the fiducial simulation without BHs and their ensuing feedback (see 
	Table~\ref{simList}). It is identified by the black colour and used to quantify 
	the effect of AGN feedback, that is included in all the other simulations involved in the comparison. 
\item {\sl {fiducial--hcAL--cf:}} identified by the purple colour, this galaxy is our fiducial model. 
	AGN feedback energy provided to the multiphase particles is coupled to the hot and cold gas according 
	to their physical properties, and we limit the accretion of rotationally supported cold gas onto the BH; 	
\item {\sl {hcA--hot:}} identified by the red colour, in this model the AGN feedback energy is coupled 
	entirely to the hot gas; 
\item {\sl {hcA--both:}} for this galaxy model (green) the AGN feedback energy is evenly provided to 
	the hot and cold phases; 
\item {\sl {hcA--cold:}} pinpointed by the blue colour, the AGN supplies all the feedback energy to the 
	cold gas. 
\end{itemize} 

Figure \ref{StellarDensityMaps} introduces projected stellar (first and second columns) and 
gas (third and forth ones) density maps of each galaxy, at redshift $z = 0$. We show 
edge-on (first and third columns) and face-on (second and forth columns) views. 
Galaxies have been rotated in order to align the z-axis of their reference system with the angular momentum
of star and (cold and multiphase) gas particles located within $8$~kpc from the minimum 
of the gravitational potential. 
The origin of the reference system is set on the centre of the galaxy, which is assumed to be the centre of mass 
of the aformentioned particles.
Throughout this paper, we focus our analysis on star and gas particles that are located within the 
galactic radius\footnote{We define here the galactic 
	radius as one tenth of the virial radius, i.e. $R_{\rm gal}= 0.1 R_{\rm vir}$. 
  	The radius $R_{\rm gal}$ is chosen to select the region of the 
  	computational domain where the central galaxy resides. 
  	We consider virial quantities as those computed in a sphere that encloses 
	an overdensity of 200 times the {\sl critical} density at present time and 
	that is centred on the minimum of the gravitational potential of the halo.}, unless otherwise specified. The 
galactic radius of these galaxies ranges between $R_{\rm gal} = 24.03$~and~$24.18$~kpc. 
When analysing radial profiles (see Figure~\ref{ch10:SurfaceDensity}), 
we will consider gas and star particles out to a distance of $r = 30$~kpc. 
	
Density maps in Figure \ref{StellarDensityMaps} show that all the galaxies have a dominant, 
extended disc and a limited bulge component. Gaseous discs are more extended than stellar ones. 
A well-defined spiral pattern is evident in the majority of the discs. The morphology and the extent of the disc vary: 
with respect to the noAGN--reference simulation, galaxies simulated accounting for AGN feedback have 
more extended gaseous and stellar discs (see also Figure~\ref{ch10:SurfaceDensity}). 
However, the morphology of these galaxies usually appear more disturbed, especially in the outermost regions. 
This is the result of a highly dynamical environment. The most characteristic case is represented by 
the galaxy hcA--hot, that exhibits a regular, inner disc and an outer ring-like structure, 
which appears as the natural extension of the internal disc. The outermost gas 
is the result of recently accreted (and re-accreted, after previous ejection by galactic outflows) gas, 
that still has to settle down on the disc and that is characterised by a high angular momentum. 
The recent accretion phase experienced by this galaxy can be seen also by analysing the mass accretion of gas 
below $z\lesssim 0.5$ (see Figure~\ref{ch10:MAH}). 

The galaxy hcA--both, with a more regular morphology, also has an irregular distribution of gas above 
and around the galactic plane, suggesting ongoing gas accretion (see also Figure~\ref{ch10:MAH}). As for the galaxy 
hcA--cold, it certainly has the most evident signature of the presence of a SMBH, that accreted all the 
available gas in its surrounding, leaving a hole in the gas density map. 
The radius of the central region deficient in gas is $r \sim 2.5$~kpc (see also Figure~\ref{ch10:SurfaceDensity}). 
The numerical explanation of the hole which surrounds the BH in the simulation hcA-cold is that the size 
of the hole matches that of the BH accretion length: there are gas particles within the sphere centred on the BH 
and whose radius is the BH accretion length, but their density is not high enough for the accretion to be effective. 
The BH smoothing length increases so as to contain a fixed (kernel weighted) number of gas particles in our code. 
As BH accretion proceeds and particles are removed, other particles enter the BH smoothing sphere. 

By analyzing the distribution of the coupling factors for the hot and cold phase 
(i.e. $\mathcal{C}_{\rm h}$ and $\mathcal{C}_{\rm c}$) of all the multiphase particles that have been selected 
to receive AGN feedback energy down to $z=0$, we found the following mean values for 
$\mathcal{C}_{\rm h}$ and $\mathcal{C}_{\rm c}$: $0.41$ and $0.59$, respectively. 
Mean values for the covering factors predict an evolution for the fiducial--hcAL--cf galaxy 
and for its SMBH that is close to that experienced by hcA--hot.  
Before presenting an extensive analysis of the main features of the simulated galaxies (Section~\ref{ch10:Galaxyevo}), 
we investigate the properties of the central BHs, that drive their host galaxy evolution.

\subsection{BH evolution} 
\label{ch10:BHevo}

In this section, we study the evolution and the properties of the SMBHs of the galaxies: 
fiducial--hcAL--cf, hcA--hot, hcA--both, and hcA--cold. We focus on the evolution 
of their mass and accretion rate, and consider whether they fulfil observed scaling relations. 

\begin{figure}
\newcommand{\captionfonts}{\small}
\vspace{-1.5ex}
\centering
\includegraphics[trim=0.1cm 0.1cm 0.35cm 0.6cm, clip, width=.49\textwidth]{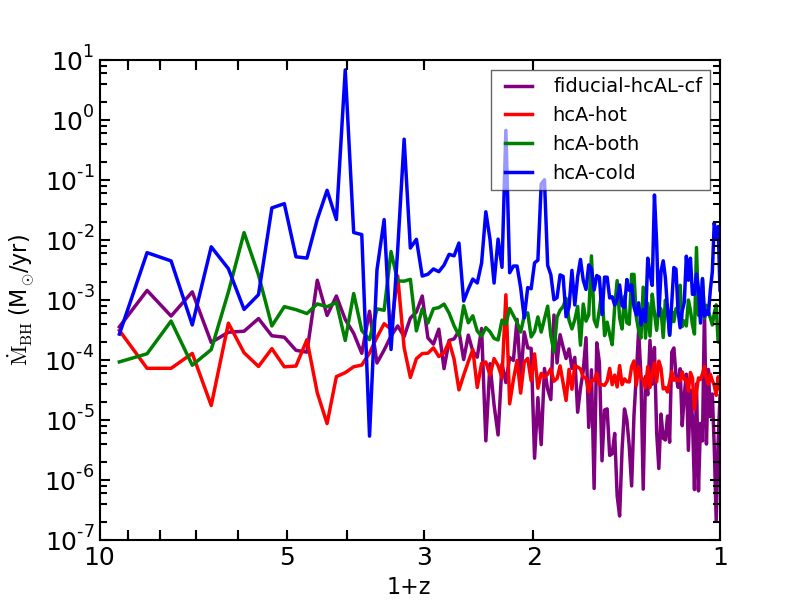} 
\includegraphics[trim=0.1cm 0.1cm 0.35cm 0.9cm, clip, width=.49\textwidth]{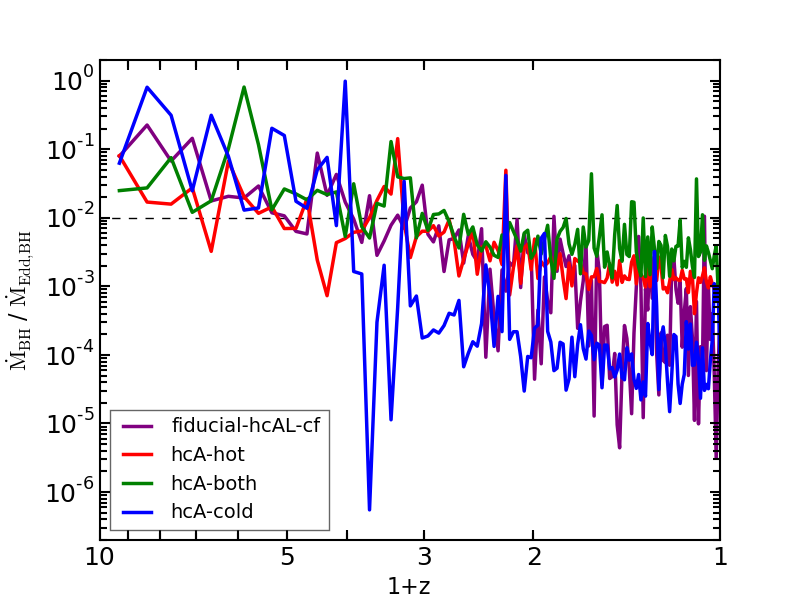} 
\caption[]{Evolution of the accretion rate of the most massive BH in fiducial--hcAL--cf (purple), 
	hcA--hot (red), hcA--both (green), and hcA--cold (blue). The same evolution is 
	shown both in units of M$_{\odot}$~yr$^{-1}$ 
	{\sl {(top panel)}} and in units of the Eddington accretion rate $\dot{M}_{\rm Edd}$
	{\sl {(bottom panel)}}. 
	The dashed black line where $\dot{M}_{\rm BH}/\dot{M}_{\rm Edd} = 0.01$ marks the transition 
	from high- to low-accretion mode. }
\label{ch10:BHMD} 
\end{figure}

Figure~\ref{ch10:BHMD} shows the evolution of the BH accretion rates. 
The top panel describes the redshift evolution of the accretion rate (in units of~M$_{\odot}$~yr$^{-1}$) 
of the most massive BH within each galaxy (i.e. located within $100$~kpc from the galaxy centre). 
The most massive BH within each galaxy in these simulations is always located at the galaxy 
centre, as a consequence of the procedure adopted for the BH pinning (see Section~\ref{BHseeding}) 
and of the relatively quiet dynamical environment within which the galaxy forms. 
The bottom panel shows the same evolution in units of the Eddington accretion rate~$\dot{M}_{\rm Edd}$. 
The central BH is seeded at $z\simeq 8.5$ in all the galaxies. By focussing on the bottom panel 
of Figure~\ref{ch10:BHMD}, it is possible to see 
that BHs experience a high-redshift, high-accretion rate phase 
and then they enter a lower-accretion rate stage 
at a redshift spanning the range $3 \gtrsim z \gtrsim 2$ (see also Section~\ref{sec:introduction}). 
The commonly adopted threshold to discriminate between high-accretion rate and low-accretion rate mode 
feedback is $\dot{M}_{\rm BH}/ \dot{M}_{\rm Edd} = 0.01$ \citep[e.g.][]{Churazov2005, Sijacki2007}. 
During the high-accretion rate phase, the BHs in hcA--both and and hcA--cold experience a few episodes 
of enhanced accretion, with $\dot{M}_{\rm BH}/ \dot{M}_{\rm Edd} \sim 1$. In particular, the SMBH in 
hcA--cold is characterised by several episodes of accretion where 
$0.1 \lesssim \dot{M}_{\rm BH}/ \dot{M}_{\rm Edd} \lesssim 1$, while the accretion is remarkably suppressed later. 
Throughout the BH evolution, the accretion of cold gas dominates 
the total BH accretion rate~$\dot{M}_{\rm BH}$ (see equation~(\ref{Mdot_limited})) over the accretion of the hot gas.

%%%%%%%%%%%%%%%%%%%%%%%%%%%%% BHMD
%%%%%%%%%%%%%%%%%%%%%%%%%%%%% BHMD

%%%%%%%%%%%%%%%%%%%%%%%%%%%%% BHMA
\begin{figure}
\newcommand{\captionfonts}{\small}
%\vspace{-1.75ex}
\centering
\includegraphics[trim=0.1cm 0.1cm 0.35cm 1.cm, clip, width=.49\textwidth]{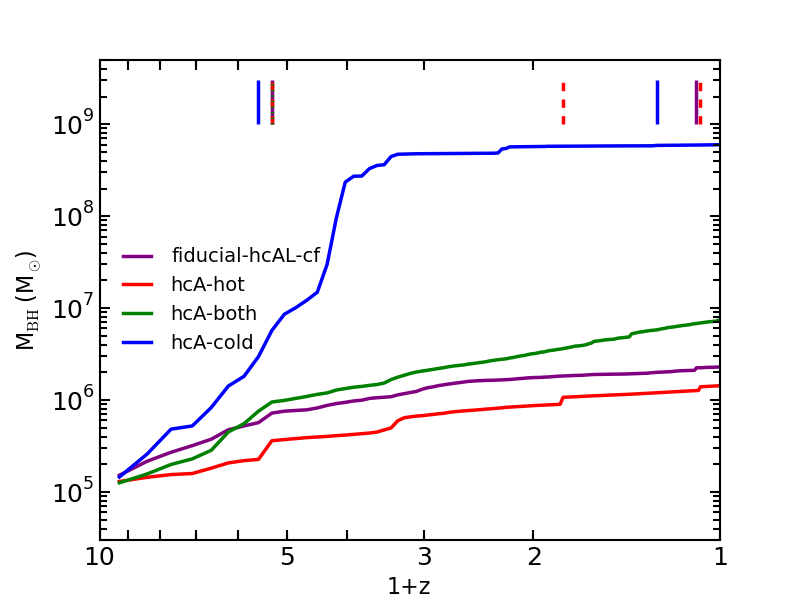} 
\caption[]{Evolution of the BH mass growth for the most massive BH in each of the 
	simulated galaxies. Segments at the top of the figure highlight the redshift at which the considered BH 
	experienced a BH merger. Note that at $z \sim 4.2$ the BHs in fiducial--hcAL--cf, 
	hcA--hot, and hcA--both have a merger.} 
\label{ch10:BHMA} 
\end{figure}
%%%%%%%%%%%%%%%%%%%%%%%%%%%%% BHMA

We computed the duty cycle of the models introduced in Figure~\ref{ch10:BHMD}. 
The duty cycle is the ratio between the time during which the SMBH is active and can be deemed as an AGN 
over the total time of the simulation. Following \citet[][their equation $12$]{Weinberger2018}, we estimated 
the AGN luminosity $L_{\rm AGN}$. We consider that a SMBH is an AGN whenever its $L_{\rm AGN}$ 
exceeds a fraction of the Eddington luminosity $L_{\rm Edd}$ (see Section~\ref{sec:introduction}). 
We adopted $L_{\rm AGN} / L_{\rm Edd} > 0.01$ as a conventional threshold to distinguish between 
active and inactive stages of the SMBH. 
The duty cycle of the models fiducial--hcAL--cf, hcA--hot, hcA--both, and hcA--cold are as follows: 
$0.061$, $0.040$, $0.084$, and $0.097$, respectively. 

The evolution of the BH accretion rates shown in Figure~\ref{ch10:BHMD} produces the growth of BH masses, 
as illustrated in Figure~\ref{ch10:BHMA}.
Figure~\ref{ch10:BHMA} describes the evolution of the mass of the most massive BH within the simulated galaxies. 
Vertical segments highlight the redshift at which BH mergers involving the central SMBH occur. 
Mergers usually appear as jumps in the track of the BH mass evolution, unless the merger is between a low-mass 
BH which has just been seeded, and an already massive one, thus contributing a negligible increase to the 
BH growth (see for instance the slight jump at $z\lesssim 0.2$ for the BH of hcA--cold). 
The mass of the BHs at $z=0$ are 
$2.29 \cdot 10^6$~M$_{\odot}$ (fiducial--hcAL--cf), 
$1.43 \cdot 10^6$~M$_{\odot}$ (hcA--hot), 
$7.37 \cdot 10^6$~M$_{\odot}$ (hcA--both), and 
$5.99 \cdot 10^8$~M$_{\odot}$ (hcA--cold). 
Even if the ICs are the same, the timing of BH mergers can be different from simulation to simulation, 
due to the perturbations that the BHs themselves introduce within the system. 

The evolution of the central BH in fiducial--hcAL--cf is rather moderate: it spans roughly an order of 
magnitude in mass from $z \sim 8.5$ (redshift at which the BH is seeded) to $z=0$: 
it proceeds mainly via gas accretion until $z\sim2$, when it reaches a mass which is comparable to the final one. 
The low-redshift ($z\lesssim2$) difference in the mass evolution of the BH between 
fiducial--hcAL--cf and models hcA--hot and hcA--both is due to the different model of gas accretion onto the BH. 
The reason for the difference between BHs which continue accreting gas and increase their mass 
(hcA--hot and hcA--both) and the BH which does not (fiducial--hcAL--cf) stems from the suppressed accretion 
of cold gas with high angular momentum, rather than from the adopted feedback model. 
In Section~\ref{ch10:Mango}, we discuss how the BH evolution 
and final results are sensitive to the details of gas accretion. 

The different evolution of the three BHs in the models hcA--hot, hcA--both, and 
hcA--cold (the BHs of all these galaxies accrete both hot and cold gas 
according to equation~(\ref{Mdot_limited})) is due to the effect of the AGN when different models 
for coupling feedback energy to the multiphase ISM are adopted. In order to understand how feedback energy coupling 
affects the BH accretion and growth, we focus on the extreme cases represented by 
hcA--hot and hcA--cold. The reason for the intermediate behaviour of hcA--both follows directly. 

When the AGN feedback energy is entirely coupled to the hot phase of the ISM, it increases its temperature 
(see Section~\ref{AGNmuppi}). This causes an increase of pressure, that pushes the heated particle through 
nearly adiabatic expansion, thus triggering an outflow. 
%The corresponding thermal energy increase is converted into momentum 
%for the multiphase particles, and thus can originate outflows. 
Multiphase particles are hence displaced from the innermost regions of the galaxy: as a consequence, 
the density of the central regions feeding the BH decreases and the BH accretion rate is moderate. 
On the other hand, when all the feedback energy supplied to the multiphase ISM is provided to the cold 
component of multiphase particles, it is used to evaporate cold gas and to move its mass to the hot phase. 
However, this AGN-induced mass transfer does not produce a significant increase of the SPH temperature of 
the multiphase particle (see Appendix \ref{ch10:ConstantcouplingFactors}). As a consequence, the multiphase gas 
remains close to the BH and enhances its accretion. Therefore, the BH experiences a rapid phase of mass growth, 
that will be halted when all the gas available within its surroundings is consumed. The BH mass growth is stopped 
(see Figure~\ref{ch10:BHMA}, below $z \lesssim 2$) and the central region of the galaxy appears devoid of gas 
(see the gas density map of hcA--cold in Figure~\ref{StellarDensityMaps}, 
where the central density depression has a radius which is comparable to the smoothing length of the BH at $z=0$, 
see below). 
In this way, it is possible to explain why the evolution of the BH masses of hcA--hot and hcA--cold 
differ significantly from each other. In particular, from redshift $z \gtrsim 4$ on, the way in which 
the BH impacts on the overall evolution of the galaxy is 
remarkably different, especially because of the feedback energy budget involved.

\begin{table}
\centering
\begin{minipage}{82mm}
\caption[]{Mass of gas outflowing from the innermost regions of 
hcA--hot, hcA--cold, and noAGN--reference. 
{\sl {Column~1:}} simulation label. 
{\sl {Column~2:}} redshift. 
{\sl {Column~3:}} total mass of gas outflowing from $r \le 5$~kpc. 
{\sl {Column~4:}} mass of multiphase gas outflowing from $r \le 5$~kpc. 
{\sl {Column~5:}} mass of single-phase gas outflowing from $r \le 5$~kpc.} 
\renewcommand\tabcolsep{3.1mm}
\begin{tabular}{@{}lcccc@{}}
\hline
Simulation  & $z$  &   $M_{\rm outf, \, tot}$   & $M_{\rm outf, \, mp}$    &    $M_{\rm outf, \, sp}$         \\ 
               &              &  (M$_{\odot}$)    &   (M$_{\odot}$)            &       (M$_{\odot}$)                                                   \\ 
\hline
\hline
hcA--hot &  $z=4$  &   $1.4 \cdot 10^9$  &  $1.2 \cdot 10^9$ &  $2.1 \cdot 10^8$   \\  
\hline
                        &  $z=3$  &   $1.5 \cdot 10^9$  &  $1.3 \cdot 10^9$ &  $1.6 \cdot 10^8$   \\  
\hline
hcA--cold &  $z=4$  &   $7.4 \cdot 10^8$  &  $6.2 \cdot 10^8$ &  $1.2 \cdot 10^8$   \\  
\hline
                          &  $z=3$  &   $6.2 \cdot 10^8$  &  $4.9 \cdot 10^8$ &  $1.3 \cdot 10^8$   \\  
\hline
noAGN--reference &  $z=4$  &   $8.2 \cdot 10^8$  &  $7.0 \cdot 10^8$ &  $1.2 \cdot 10^8$   \\  
\hline
                          &  $z=3$  &   $8.5 \cdot 10^8$  &  $7.3 \cdot 10^8$ &  $1.2 \cdot 10^8$   \\  
\hline
\hline
\end{tabular}
\label{ch10:AGNmasseOutflow}
\end{minipage}
\end{table}

We provide a more quantitative explanation by computing the mass of the gas that is located within $5$~kpc 
from the galaxy centre and that is outflowing (i.e. with $v_{\rm r} > 0$, $v_{\rm r}$ being the radial component 
of the particle velocity). 
The size of the region (a sphere with $r=5$~kpc) that we choose to study the gas dynamics is roughly twice 
as large as the smoothing length of the BHs in hcA--hot and hcA--cold in the redshift range 
that we consider, i.e. $3 \lesssim z \lesssim 4$. 
This redshift range identifies the time interval within which the difference between the evolution of the 
BH mass of hcA--hot and hcA--cold is magnified (even if the two evolutions are already different since $z\sim8$). 
Note that the BH smoothing length decreases below this redshift, reaching a size of $\sim2.3$~kpc at $z=0$ 
for hcA--cold, while it is as small as $0.8$~kpc at $z=0$ for hcA--hot. 
The total mass of gas outflowing from the innermost regions at $z=4$ and $z=3$ 
for hcA--hot is detailed in Table~\ref{ch10:AGNmasseOutflow}, together with the 
corresponding shares of multiphase and single-phase outflowing gas. 

Besides considering the simulation noAGN--reference, 
Table~\ref{ch10:AGNmasseOutflow} also shows the same quantities for hcA--cold. 
These latter values are lower than those of hcA--hot by roughly a factor $\sim 2$. 
This supports the interpretation that we provided. 

\begin{figure}
\newcommand{\captionfonts}{\small}
%\vspace{-1.ex}
\centering
\includegraphics[trim=0.4cm 0.4cm 0.35cm 0.2cm, clip, width=.47\textwidth]{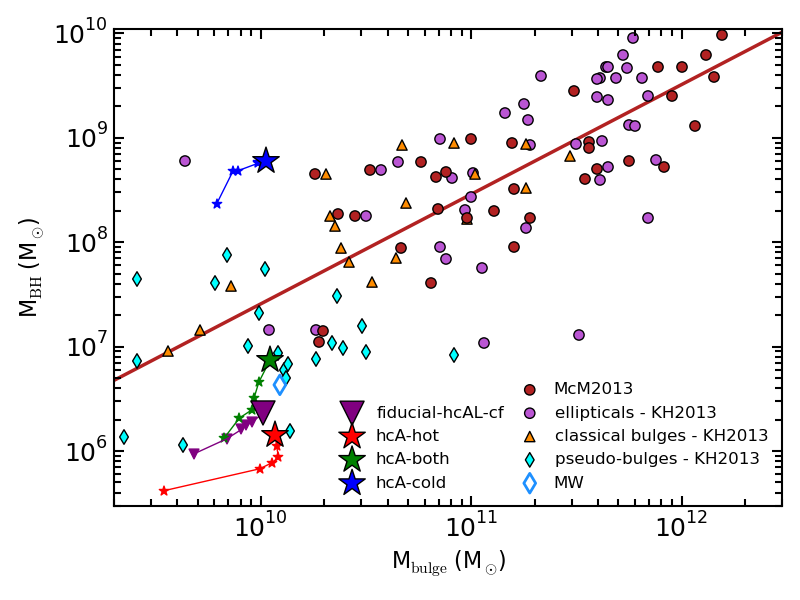} 
\caption[]{$M_{\rm bulge}$-$M_{\rm BH}$ relation 
	for BHs in the simulated galaxies fiducial--hcAL--cf (purple triangle), hcA--hot (red starlet), 
	hcA--both (green starlet), and hcA--cold (blue starlet). 
	We also show the evolution tracks of the four models in the $M_{\rm bulge}$-$M_{\rm BH}$ plane 
	from $z=3$ down to $z=0$. Symbols over each track pinpoint 
	$z = 3, \, 2, \, 1.5, \, 1, \, 0.5$, and $z = 0$. 
	Observations are from \citet[][KH2013]{KormendyHo2013} and 
	from \citet[][McM2013]{McConnell2013}. The red solid line depicts the best fit to the $35$ elliptical galaxies in the 
	sample of \citet[][]{McConnell2013}. 
	The light-blue empty diamond shows the position of the BH in our Galaxy \citep[][]{KormendyHo2013}.}
\label{ch10:mago_ref} 
\end{figure}

Figure~\ref{ch10:mago_ref} shows the position of the BHs of the four simulated galaxies on the plane of 
the $M_{\rm bulge}$-$M_{\rm BH}$ relation \citep{Magorrian1998}, that describes the correlation existing between 
the mass of the BH and that of the bulge of the host galaxy (see Section~\ref{sec:introduction}). 
For each simulated galaxy, we consider the mass of the BH, at $z=0$, and the mass of the bulge of the galaxy. 
The mass of the bulge is estimated by performing a kinematic decomposition and 
considering only dispersion supported stars. We thus consider the gas particles 
within $R_{\rm gal}$ and assume that half of the bulge mass is made up of all the 
counter-rotating ($J_{\rm z}/J_{\rm circ} < 0$) stars 
(see Section~\ref{ch10:Galaxyevo} and Figure~\ref{ch10:jcirc}, for details). 
We compare results from simulations with 
observations from the sample by \citet[][]{KormendyHo2013} and from the sample 
of \citet{McConnell2013}, which is made of $35$ early-type galaxies (and whose best fit is provided by the red solid line). 
In their sample, \citet[][]{KormendyHo2013} distinguish between elliptical galaxies, classical bulges in late-type galaxies 
and pseudo-bulges in late-type galaxies. Classical bulges are scaled-down versions of ellipticals, 
with which they share the formation scenario. On the other hand, pseudo-bulges are the outcome 
of the secular evolution they experienced within galaxy discs \citep{KormendyKennicutt2004} 
and do not obey the same relation as the elliptical galaxies \citep{Gadotti2009, KormendyBender2012}. 
This is evident from Figure~\ref{ch10:mago_ref}, where the majority of pseudo-bulges is located below the best-fit to 
ellipticals only (red solid line), and are responsible for the bending of the $M_{\rm bulge}$-$M_{\rm BH}$ relation 
at $M_{\rm bulge} \lesssim 5 \cdot 10^{10}$~M$_{\odot}$.

%%%%%%%%%%%%%%%%%%%%%%%%%%%%% mago ref
%%%%%%%%%%%%%%%%%%%%%%%%%%%%% mago ref

Predictions from simulations are indicated by the purple triangle (fiducial model) 
and stars in Figure~\ref{ch10:mago_ref}: we also show the tracks of the four models from $z=3$ down to $z=0$, 
to outline their evolution on the $M_{\rm bulge}$-$M_{\rm BH}$ plane. 
Symbols over each track highlight the position of the systems at 
$z = 3$, $z = 2$, $z = 1.5$, $z = 1$, $z = 0.5$, and $z = 0$.
The bulges of the galaxies that we have simulated have a formation history more similar to that of 
pseudo-bulges rather than to that of classical bulges:
they indeed have grown within the galaxy as the galaxy itself grew more 
massive\footnote{We postpone to a forthcoming work a proper classification and an extensive 
	investigation of the formation path of the bulges of the galaxies that we have simulated.}. 
Therefore, we consider as in agreement with observations those BHs that are located below the fit to the 
sample of elliptical galaxies only. 
The mass of the bulge for the considered galaxies at $z=0$ is as follows: 
$1.02 \cdot 10^{10}$~M$_{\odot}$ for fiducial--hcAL--cf,
$1.17 \cdot 10^{10}$~M$_{\odot}$ for hcA--hot, 
$1.10 \cdot 10^{10}$~M$_{\odot}$ for hcA--both, and 
$1.06 \cdot 10^{10}$~M$_{\odot}$ for hcA--cold. 
Figure~\ref{ch10:mago_ref} shows that the BHs of the fiducial--hcAL--cf and hcA--both galaxies 
are in good agreement with observations. The BH of hcA--hot is quite in keeping with observations, 
as it lies on the lower edge of the region occupied by pseudo-bulges. 
On the other hand, the hcA--cold galaxy hosts a BH that is too massive for the bulge (and thus, the 
galaxy) in which it resides. 

BHs are indeed expected to grow mainly at high-redshift ($z \gtrsim 2$), while at later times the AGN reaches a quasi 
self-regulated state, with AGN feedback roughly counterbalancing gas accretion and cooling. 
At approximately that point, the BH approaches the $M_{\rm bulge}$-$M_{\rm BH}$ relation, 
the BH accretion rate drops to lower values and gas accretion lies in the low-accretion mode regime. 
BHs of the simulated galaxies fiducial--hcAL--cf and hcA--both set on the $M_{\rm bulge}$-$M_{\rm BH}$ relation 
at $2 \gtrsim z \gtrsim1$. The way in which BHs climb the plane of the $M_{\rm bulge}$-$M_{\rm BH}$ relation 
proceeds along with the evolution of their mass shown in Figure~\ref{ch10:BHMA}. 
Indeed, the mass of the bulge of these galaxies approaches by $2 \gtrsim z \gtrsim1$ a position close 
to that where they are at $z=0$. Below this redshift range the bulge growth is not significant and mainly driven by 
processes which occur within the innermost region of the galaxy. 
For instance, considering the hcA--both galaxy, the mass of its bulge increases from 
$9.10 \cdot 10^9$~M$_{\odot}$ at $z=1.5$, to 
$9.27 \cdot 10^9$~M$_{\odot}$ at $z=1$, to 
$9.77 \cdot 10^9$~M$_{\odot}$ at $z=0.5$, reaching then 
$1.10 \cdot 10^{10}$~M$_{\odot}$ at $z=0$. 
Matching observations when the $M_{\rm bulge}$-$M_{\rm BH}$ relation is considered is a valuable benchmark for 
simulated galaxies, as this relation involves the BH mass and the stellar mass of the bulge of the host galaxy, 
that are quantities integrated throughout the galaxy evolution. In addition, the BH mass is also highly sensitive 
to details of the feedback process. 

The location of BHs on the $M_{\rm bulge}$-$M_{\rm BH}$ plane is quite sensitive to the adopted feedback 
efficiency $\epsilon_{\rm f}$. This value is expected to have a crucial impact on final properties of BHs and to affect 
the normalization of the $M_{\rm bulge}$-$M_{\rm BH}$ relation: indeed, the higher the feedback efficiency is, 
the smaller the final mass of BHs is expected to be, as a larger amount of energy is provided to the ISM 
to counterbalance AGN feeding.
We note that when the AGN feedback energy is entirely coupled to the cold phase of the ISM (hcA--cold) 
the BH has a final mass which grows beyond observations. We expect that a larger value of $\epsilon_{\rm f}$ 
can compensate for the low response of the ISM to AGN feedback in this model.

In Appendix~\ref{ch10:CalibMago}, we will show how final results are sensitive to BH seed mass, 
and how we took advantage of the $M_{\rm bulge}$-$M_{\rm BH}$ relation to choose 
the reference $M_{\rm BH, \, seed}$.

\subsection{BH-galaxy coevolution} 
\label{ch10:Galaxyevo}

\begin{figure}
\newcommand{\captionfonts}{\small}
%\vspace{-0.5ex}
\centering
\includegraphics[trim=0.4cm 0.4cm 0.35cm 0.2cm, clip, width=.475\textwidth]{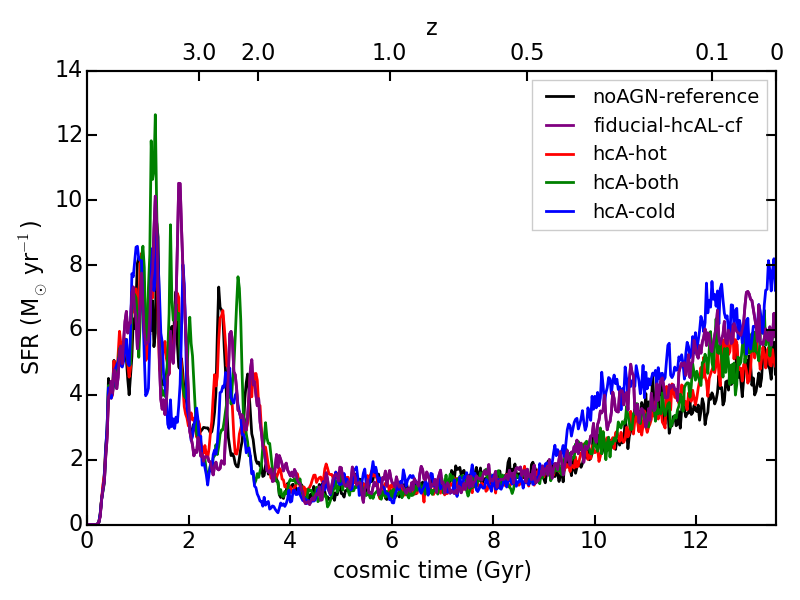} 
\caption[]{Star formation history for galaxies simulated with and without AGN 
	feedback, colour-coded as explained in the legend.}
\label{ch10:sfr} 
\end{figure}

In this section, we introduce the most important properties of the simulated galaxies presented in Section~\ref{ch10:GenRes}. 
Figure~\ref{ch10:sfr} shows the star formation history of the five galaxies. With respect to the simulation 
noAGN--reference, the four galaxies including AGN feedback 
experience a comparable star formation history at early epochs ($z \gtrsim 3$), when the 
galaxy bulge forms. The comparison between some of the most pronounced star formation peaks 
highlights how the AGN usually has a positive feedback: for instance, the peaks 
at $z \sim 4$ of fiducial--hcAL--cf and at $z \sim 2.3$ of hcA--both are a clear evidence of this. 
Also, bursts in the evolution of the star formation 
can be related, relatively easily for hcA--both and hcA--cold, to peaks in the evolution of 
the BH accretion rate (see Figure~\ref{ch10:BHMD}). Nonetheless, a number of episodes where the AGN is 
found to have a negative feedback, suppressing star formation, are also present. 
The inclusion of the AGN clearly produces a positive feedback at low redshift ($z \lesssim 0.5$), where 
the SFR is higher for simulations with AGN. The reason for the enhanced star formation stems from 
the fact that AGN feedback energy over-pressurises the gas (see equations~(\ref{eq:sfr})~and~(\ref{eq:f_mol})).

%%%%%%%%%%%%%%%%%%%%%%%%%%%%% sfr
%%%%%%%%%%%%%%%%%%%%%%%%%%%%% sfr

%%%%%%%%%%%%%%%%%%%%%%%%%%%%% Pressure
\begin{figure}
\newcommand{\captionfonts}{\small}
%\vspace{-0.5ex}
\centering
\includegraphics[trim=0.4cm 0.4cm 0.05cm 0.cm, clip, width=.48\textwidth]{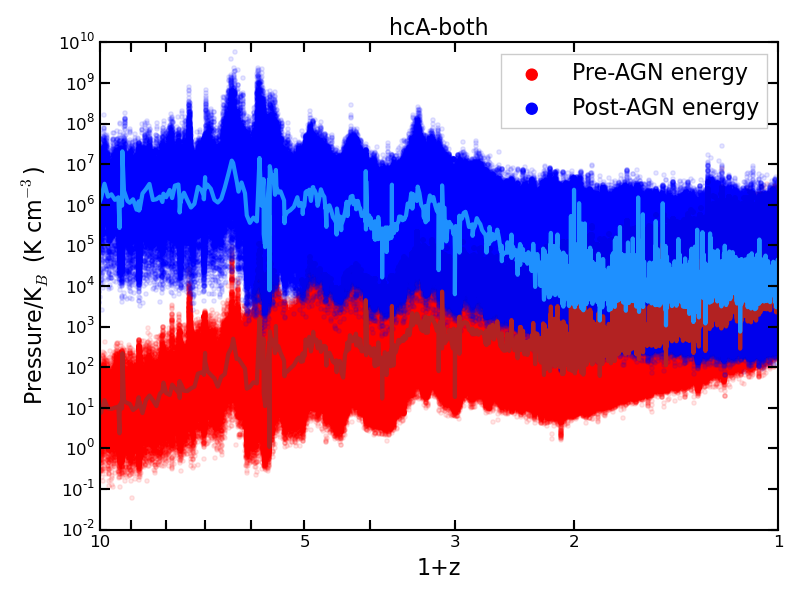} 
\includegraphics[trim=0.4cm 0.4cm 0.05cm 0.cm, clip, width=.48\textwidth]{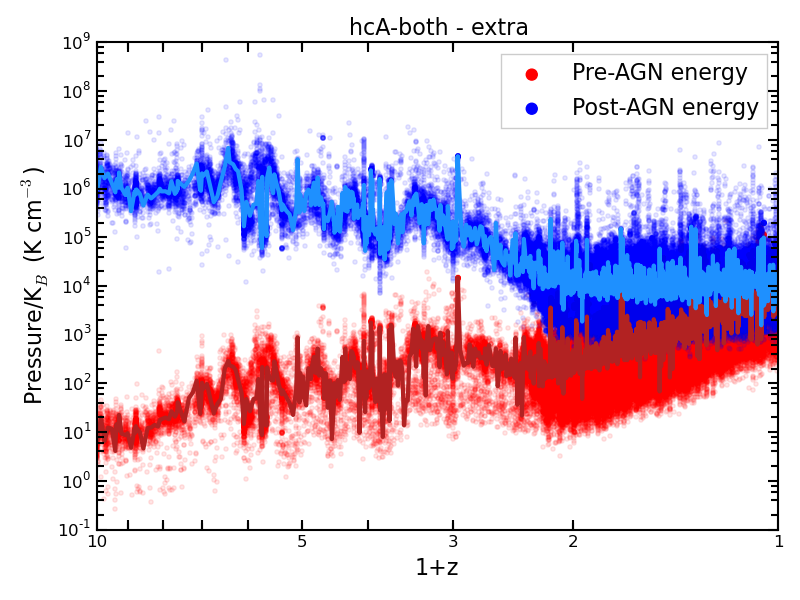} 
\caption[AGN positive feedback: over-pressurisation of gas]{Pressure of multiphase gas particles before and soon after 
	the feedback energy injection, as a function of redshift, for the simulation hcA--both. Solid curves show the 
	median evolution. {\sl {Top panel:}} 
	evolution of gas particles for which all the feedback energy supplied to the cold gas is used for the evaporation of the 
	cold gas mass. 
	{\sl {Bottom panel:}} evolution of gas particles for which a fraction of the energy initially allocated to the cold phase is 
	provided to the hot phase, as an additional contribution that ensures that no feedback energy is lost.}
\label{ch10:PrePostPressure} 
\end{figure}
%%%%%%%%%%%%%%%%%%%%%%%%%%%%% Pressure

AGN-induced over-pressurisation of gas can be quantified by comparing the pressure of multiphase gas 
particles that received feedback energy. Figure~\ref{ch10:PrePostPressure} considers the simulation hcA--both 
and shows the pressure of all the multiphase gas particles which have been provided with AGN feedback energy, 
as a function of redshift. Particles' pressure is evaluated at the beginning and at the end of each SPH time-step 
during which particles received feedback energy. 
The top panel of Figure~\ref{ch10:PrePostPressure} considers the multiphase particles 
for which all the feedback energy supplied to the cold gas is used for the evaporation of the 
cold gas mass. The bottom panel describes the evolution of gas particles for which a fraction of the energy initially 
allocated to the cold phase is provided to the hot phase (see equation~(\ref{ch9:extraEnergy}) and Section~\ref{AGNmuppi}), as an additional contribution that ensures that no feedback energy is lost. 
In the bottom panel of Figure~\ref{ch10:PrePostPressure} are thus considered those multiphase particles 
for which the cold phase is entirely evaporated by a single AGN feedback energy injection event. 
Solid curves depict the median evolution. 
The trend is the same for the two sub-samples of particles: nevertheless, in this way it is possible to appreciate the 
impact of the condition that guarantees that all the feedback energy is actually used (even if the multiphase 
gas particle that is provided with it has not enough mass of cold gas) and the number of particles that this condition 
involves ($\sim$$1/10$ of the total number of multiphase particles selected for receiving 
feedback energy over the whole simulation). 
For the sake of concision, we show how the pressure of gas particles increases due to AGN feedback for 
the simulation hcA--both only. Similar conclusions can be drawn when considering all the other simulations, 
with the AGN-induced over-pressurisation of gas particles always becoming less significant as the redshift decreases. 
Pressure increase is mainly driven by direct heating of the multiphase ISM 
(see equations~\ref{ch9:AGNmuppi4}~and~\ref{ch9:AGNmuppi6}).

%%%%%%%%%%%%%%%%%%%%%%%%%%%%% jcirc
\begin{figure}
\newcommand{\captionfonts}{\small}
%\vspace{-1.65ex}
\centering
\includegraphics[trim=0.4cm 0.4cm 0.35cm 0.cm, clip, width=.475\textwidth]{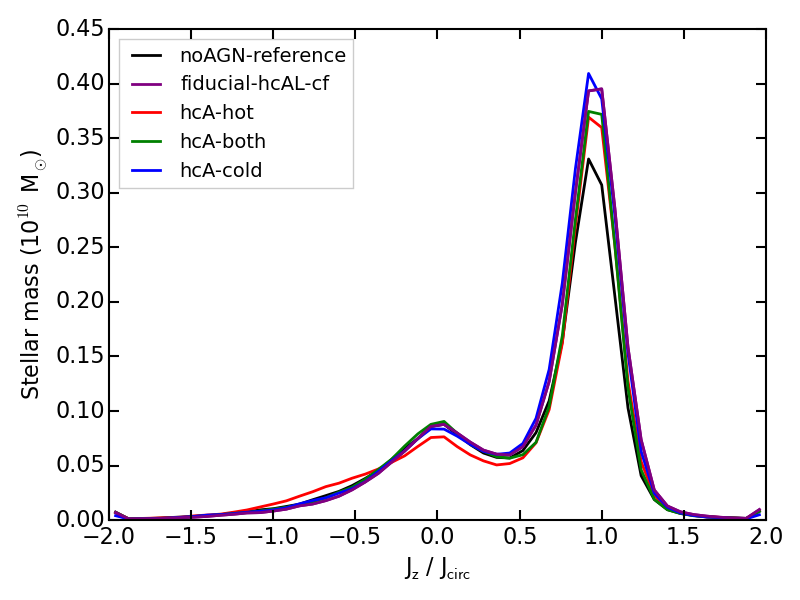} 
\caption[]{Circularity of star particles' orbits for galaxies simulated 
	with and without AGN feedback.}
\label{ch10:jcirc} 
\end{figure}
%%%%%%%%%%%%%%%%%%%%%%%%%%%%% jcirc

The over-pressurisation of gas shown in Figure~\ref{ch10:PrePostPressure} does not result in an enhanced SFR at all 
the redshifts (see Figure~\ref{ch10:sfr}: the star formation history of galaxies with and without the AGN feedback 
is comparable in the redshift range $2 \gtrsim z \gtrsim 0.5$). This is due to the complex effect of AGN feedback, that 
also produces a concurrent overall heating of the forming galaxy and promotes massive outflows, thus reducing the 
gas reservoir available for star formation. 
These findings show that in our BH feedback scheme the negative effect is also important, 
so BH accretion is (partly) self-regulated. 
Moreover, numerical simulations have shown that the 
surrounding large-scale environment can provide high 
angular momentum gas through cold flows \citep{Brooks2009, Pichon2011}, 
and that the star formation feeding through cold flows decreases with cosmic time \citep{Dubois2014, Codis2015}. 
Within this picture, star formation in galaxies is mainly fueled by cold gas accreted from outside 
the system itself at high redshift, this cold gas being able to reach the innermost regions of the forming galaxies 
where star formation occurs. 
This can explain why our galaxies with and without AGN feedback share a similar high-z star formation history, 
while the over-pressurisation of gas induced by the AGN since high-z turns out to be important at low z. At low redshift the star formation is indeed mainly sustained by gas within the galaxy and by gas which falls back after its previous 
expulsion driven by stellar feedback, while the channel of external fuel through cold flows 
is by far subdominant.

%%%%%%%%%%%%%%%%%%%%%%%%%%%%% Mstar
\begin{figure}
\newcommand{\captionfonts}{\small}
\vspace{-1.5ex}
\centering
\includegraphics[trim=0.1cm 0.1cm 1.7cm 0.2cm, clip, width=.475\textwidth]{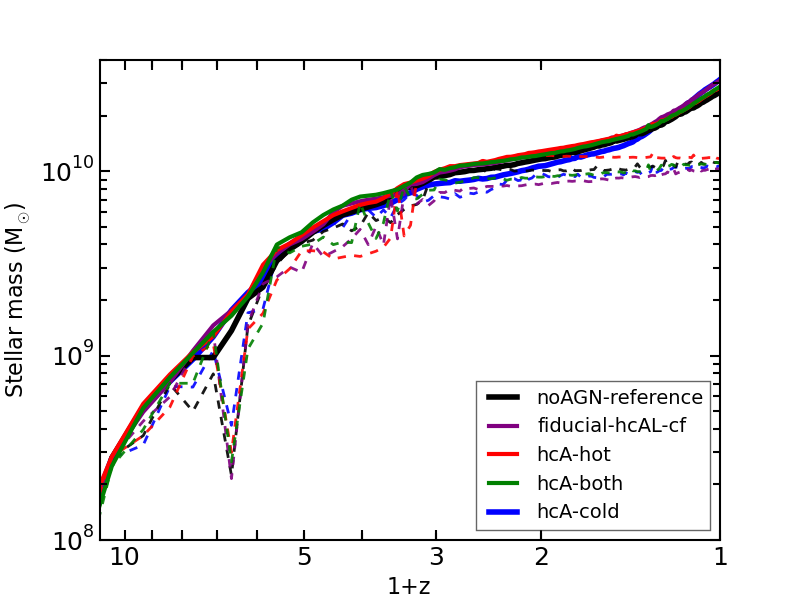} 
\caption[]{Evolution of the total stellar mass for galaxies simulated 
	with and without AGN feedback. Solid lines show the total stellar mass within the galactic radius of each 
	simulated galaxy, dashed curve highlight the contribution from the stellar mass in the bulge (as estimated from 
	kinematic decomposition, see text and Figure~\ref{ch10:jcirc}). }
\label{TotStellarMass} 
\end{figure}
%%%%%%%%%%%%%%%%%%%%%%%%%%%%% Mstar

\begin{table}
\centering
\begin{minipage}{85mm}
\caption[]{Relevant features of the galaxies analysed 
in Sections~\ref{ch10:GenRes},~\ref{ch10:BHevo},~and~\ref{ch10:Galaxyevo}. 
{\sl {Column~1:}} simulation label. 
{\sl {Column~2:}} bulge-over-total mass ratio. 
{\sl {Column~3:}} total stellar mass within $R_{\rm gal}$. 
{\sl {Column~4:}} stellar mass of the galaxy bulge. 
{\sl {Column~5:}} galactic radius $R_{\rm gal}$.} 
\renewcommand\tabcolsep{3.25mm}
\begin{tabular}{@{}lcccc@{}}
\hline
Simulation  & $B/T$  &   $M_{\rm \ast, \, tot}$   & $M_{\rm \ast, \, bulge}$    &    $R_{\rm gal}$         \\ 
               &              &  (M$_{\odot}$)    &   (M$_{\odot}$)            &       (kpc)                                                   \\ 
\hline
\hline
noAGN--reference &  $0.41$  &   $2.68 \cdot 10^{10}$  &  $1.11 \cdot 10^{10}$ &  $24.03$   \\  
\hline
fiducial--hcAL--cf &  $0.33$  &   $3.10 \cdot 10^{10}$  &  $1.02 \cdot 10^{10}$ &  $24.15$   \\  
\hline
hcA--hot &  $0.41$  &   $2.83 \cdot 10^{10}$  &  $1.17 \cdot 10^{10}$ &  $24.18$   \\  
\hline
hcA--both &  $0.38$  &   $2.87 \cdot 10^{10}$  &  $1.10 \cdot 10^{10}$ &  $24.15$   \\  
\hline
hcA--cold &  $0.34$  &   $3.14 \cdot 10^{10}$  &  $1.06 \cdot 10^{10}$ &  $24.16$   \\  
\hline
\hline
\end{tabular}
\label{ch10:AGNmasseStellari}
\end{minipage}
\end{table}

The low-redshift enhanced star formation in galaxies simulated with AGN feedback 
with respect to the noAGN--reference simulation is responsible for more 
extended galaxy stellar discs (see also Figure~\ref{StellarDensityMaps}). 
Figure~\ref{ch10:jcirc} shows the distribution of stellar mass as a function of the circularity of the orbits 
of star particles, at $z=0$, for the different simulations. The circularity of a stellar orbit is quantified by means of 
the ratio of specific angular momenta $J_{\rm z}$/$J_{\rm circ}$, 
where $J_{\rm z}$ is the specific angular momentum in the direction 
perpendicular to the disc, and $J_{\rm circ}$ is that of a reference circular orbit at a given 
distance from the galaxy centre \citep{Scannapieco2009}.
Stars in the disc and in the bulge mainly contribute to the peak 
where $J_{\rm z}/J_{\rm circ}=1$ and $J_{\rm z}/J_{\rm circ} \sim 0$, respectively. 
The stellar disc component is remarkably larger in the simulations including AGN feedback 
with respect to the noAGN--reference galaxy. 
The evolution of the total stellar mass of the five galaxies is shown in Figure~\ref{TotStellarMass}. 
We analyze the total stellar mass within the galactic radius $R_{\rm gal}$ of the simulated galaxies (solid curves), 
and the contribution to the total from the stars in the bulge (dashed lines). We rely on the kinematic decomposition 
of the galaxy stellar component to assess whether a star particle belongs to the galaxy bulge (assuming that 
all the counter-rotating $J_{\rm z}/J_{\rm circ} < 0$ stars make up half of the bulge mass, 
see also Figures~\ref{ch10:mago_ref}~and~\ref{ch10:jcirc}). 
The drop in the evolution of the stellar mass of the bulge at $z=5.6$ in Figure~\ref{TotStellarMass} is due to 
an interacting substructure, which perturbs the morphology of the main galaxy progenitor and later merges with it. 
Bulge-over-total (B/T) mass ratios are lower (or equal, at most) 
for galaxies simulated with the AGN feedback; their z=0 values\footnote{Note that 
	the B/T values that we quote should not be directly contrasted with observational photometric 
	ones, as our estimates for 
	the B/T ratio could also include satellites, stellar streams, and contribution from bars within $R_{\rm gal}$. 
	Halo stars are also included when estimasting the dispersion supported component. 
	Photometric determination for the value of B/T is lower than 
	the corresponding kinematic estimate \citep{Scannapieco2010}. }, along with 
total and bulge stellar masses within $R_{\rm gal}$ are detailed in Table~\ref{ch10:AGNmasseStellari}.

%%%%%%%%%%%%%%%%%%%%%%%%%%%%% MAH
\begin{figure}
\newcommand{\captionfonts}{\small}
%\vspace{-1.65ex}
\centering
\includegraphics[trim=0.4cm 0.4cm 0.35cm 0.2cm, clip, width=.475\textwidth]{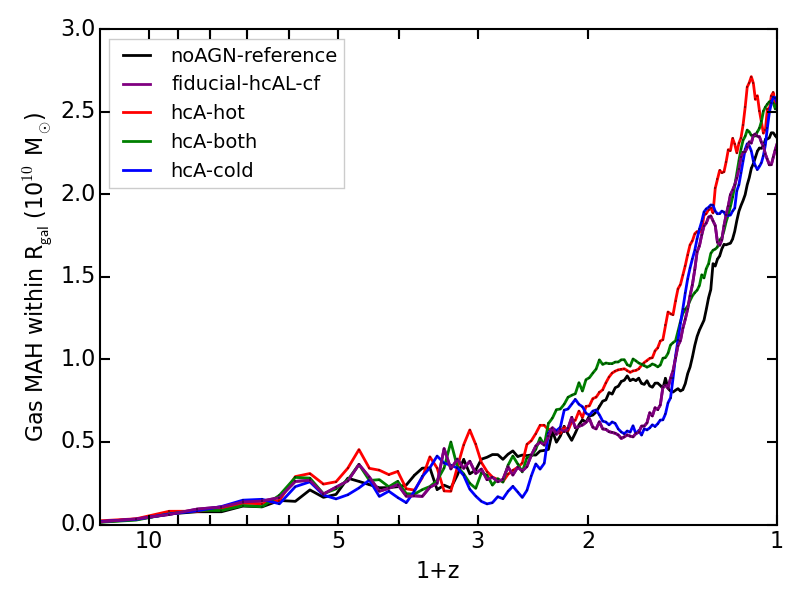} 
\caption[]{Gas mass accretion history for galaxies simulated 
	with and without AGN feedback. Evolution of the gas accreted within the galactic radius, 
	i.e. $R_{\rm gal} = 0.1 \, R_{\rm vir}$.}
\label{ch10:MAH} 
\end{figure}
%%%%%%%%%%%%%%%%%%%%%%%%%%%%% MAH

Figure~\ref{ch10:MAH} shows the mass accretion history of the five galaxies, i.e. the evolution of the mass 
of gas that is accreted within their galactic radius. Galaxies share comparable accretion histories, 
except for some episodes where differences are evident. These discrepancies are the result of the internal gas dynamics: 
powerful outflows fostered by the joint activity of stellar and AGN feedback can reduce or even suppress for a while 
the accretion of gas from the large scale environment towards the innermost regions of galaxies 
(see Section~\ref{ch10:galacticOutflows}). 

\begin{figure}
\newcommand{\captionfonts}{\small}
\vspace{-1.65ex}
\centering
\includegraphics[trim=0.4cm 0.4cm 0.35cm 0.2cm, clip, width=.475\textwidth]{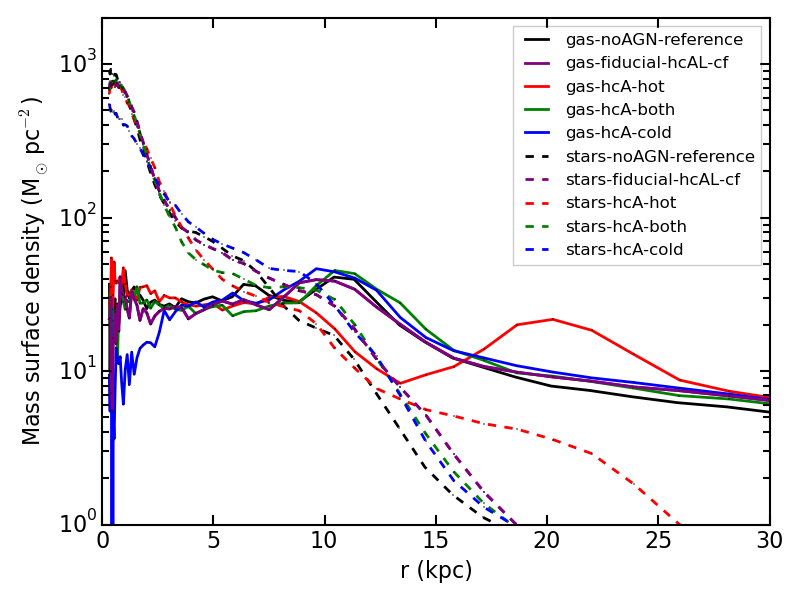} 
\caption[]{Gas (solid curves) and stellar (dashed lines) surface density 
	profiles for galaxies simulated with and without AGN feedback.}
\label{ch10:SurfaceDensity} 
\end{figure}

Finally, Figure~\ref{ch10:SurfaceDensity} shows the radial profiles of gas and stellar surface density for the set of simulated galaxies, 
and provides complementary evidence to the gas and stellar density maps discussed in Section~\ref{ch10:GenRes} 
(see Figure~\ref{StellarDensityMaps}). The most striking features are the drop in the gas 
density profiles of hcA--cold for $r \lesssim 2.5$~kpc, and the external bump ($ 25 \gtrsim  r \gtrsim 18$~kpc) 
in the gas density profile of hcA--hot. The latter property is the outcome of recent gas accretion 
\citep{Valentini2017}.

%%%%%%%%%%%%%%%%%%%%%%%%%%%%% stellar surface density
%%%%%%%%%%%%%%%%%%%%%%%%%%%%% stellar surface density

In summary, we find that the inclusion of the AGN mainly results in a positive feedback 
in our simulated spiral galaxies. Although we have shown that the SMBH negative effect is also important, 
AGN feedback is primarily positive: it pressurises the multiphase gas, enhances the low-redshift
star formation, and promotes the formation of more extended stellar discs. 
This results can be partly ascribed to our pressure-regulated star formation law. However, it goes 
beyond the numerical prescription adopted to estimate the molecular gas available for star formation, 
as there is accumulating (theoretical and observational) evidence that supports 
AGN-triggered star formation \citep[e.g.][]{Silk2013, Bieri2015, Wagner2016, Cresci2018Nat}. 
This conclusion is expected not to hold in simulations of elliptical galaxies, galaxy groups 
and clusters, as the AGN feedback is expected and observed to be mainly negative in these systems.

\subsection{Galactic outflows} 
\label{ch10:galacticOutflows}

%%%%%%%%%%%%%%%%%%%%%%%%%%%%% outflow geometry
\begin{figure}
\newcommand{\captionfonts}{\small}
\vspace{-1.65ex}
\raggedright
\includegraphics[trim=0.4cm 0.1cm 0.35cm 0.2cm, clip, width=.51\textwidth]{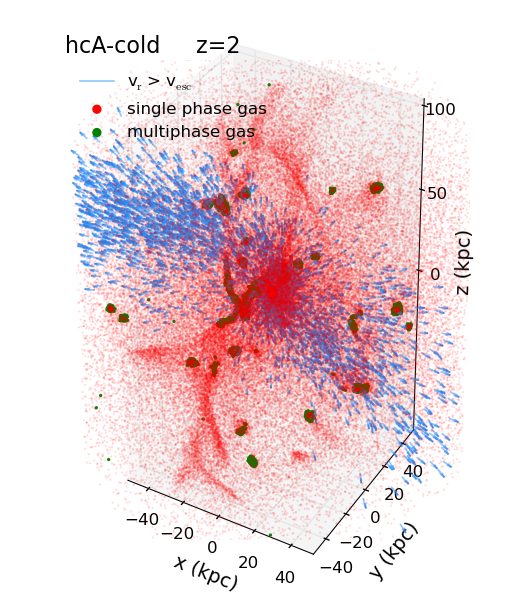} 
\includegraphics[trim=0.4cm 0.1cm 0.35cm 0.2cm, clip, width=.51\textwidth]{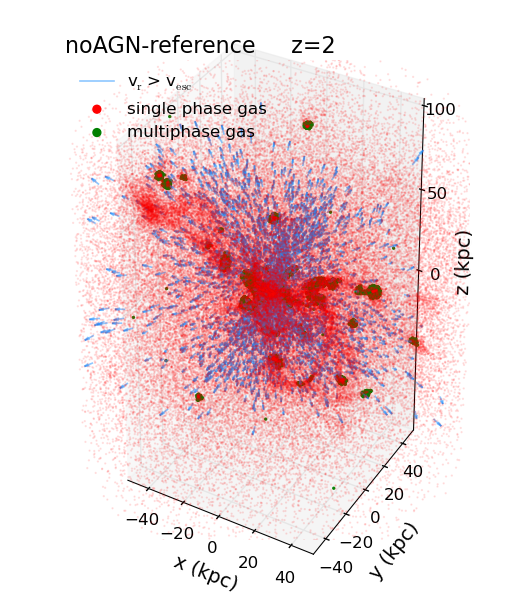}
\caption[]{Geometry of the outflowing gas in the hcA--cold model and in the reference simulation with no AGN, 
	at $z=2$. Single-phase gas is shown by red points, multiphase gas by green dots, and is mostly embedded 
	within the innermost regions of the forming galaxy. Light blue arrows pinpoint gas that is outflowing with 
	a radial velocity larger than the escape velocity of the halo (see Table~\ref{ch10:Outflow_details} for details).}
\label{ch10:OutflGeom} 
\end{figure}
%%%%%%%%%%%%%%%%%%%%%%%%%%%%% outflow geometry

Galactic outflows in our simulated galaxies are the result of the joint activity of stellar and AGN feedback. 
They have a key role in regulating the cosmological accretion of gas from the large scale and in 
shaping both the expulsion of gas from the galaxy and the circulation of gas within and around the galaxy 
\citep[see also][]{Valentini2017}.
As an example, strong outflows at $z=2$ in the hcA--cold simulation are responsible for the dip in the mass 
accretion history of this galaxy (Figure~\ref{ch10:MAH}), 
since they hinder gas accretion (note that we are now considering a different 
redshift range with respect to that analysed in Section~\ref{ch10:BHevo}). 
Outflow geometry in the aformentioned case is shown in 
Figure~\ref{ch10:OutflGeom}: from the forming galaxy, at $z=2$, a powerful bipolar outflow is launched, as 
shown by the light blue arrows pinpointing single-phase gas that is outflowing with a radial velocity 
that exceeds the escape velocity of the halo (i.e. $v_{\rm r} >268.5$~km~s$^{-1}$). 
The lower panel of the same figure illustrates the corresponding case for the reference simulation without AGN. 
The most striking feature that emerges from the comparison of the two panels in Figure~\ref{ch10:OutflGeom} 
is that the inclusion of AGN feedback promotes the formation of a (large-scale) bipolar outflow, 
while the geometry of the outflow in the noAGN--reference model is more isotropic. 
This result is independent of the details adopted in the AGN feeding and feedback modelling. 

Galactic outflows involve both single-phase and multiphase gas. We estimate the mass of gas which is 
outflowing (i.e. which has a positive radial velocity $v_{\rm r}$) to quantify the impact of stellar and AGN feedback 
in driving outflows. We distinguish between single-phase and multiphase outflowing gas. The total gas mass involved 
in the outflows is given by the sum of the former ones. Table~\ref{ch10:Outflow_details} provides outflowing gas 
masses for the simulations considered so far. 
The content of Table~\ref{ch10:Outflow_details} is displayed in Figure~\ref{ForRef}.
We focus at redshift $z=2$ and at $z=0$. Besides considering 
gas which simply has $v_{\rm r}>0$, we also estimate the mass of gas which is fostered to outflow with radial 
velocity exceeding $50$~km~s$^{-1}$ (a reference threshold to get rid of gas whose motion could not emerge from 
the bulk motion within the galaxy, in observations) and the escape velocity of the halo at the considered redshift 
($v_{\rm esc}$, detailed in Table~\ref{ch10:Outflow_details}). 
The contribution to the outflowing gas mass from the single-phase gas is larger by (at least) an order of magnitude 
than that coming from the multiphase gas. 
It is possible to quantify the impact of AGN-triggered outflows by contrasting the amount of gas which is outflowing 
in models with and without AGN feedback. 

By focussing on the mass accretion history of hcA--both and hcA--cold in Figure~\ref{ch10:MAH}, 
at $z \sim 1$, we see that outflows powered within galaxies that 
are experiencing a coevolution with SMBHs of different masses and accretion rates 
(see Figures~\ref{ch10:BHMA}~and~\ref{ch10:BHMD}, top panel) 
have a different interaction with the large scale environment. Also, from Figure~\ref{ch10:MAH} 
we note the relative role of stellar- and AGN-driven outflows in regulating the 
accretion of gas, thus controlling the reservoir for star formation.

\begin{table*}
\centering
\begin{minipage}{176mm}
\caption[]{Mass of single-phase and multiphase gas involved in galactic outflows, for simulations
fiducial--hcAL--cf, hcA--hot, hcA--both, hcA--cold, and noAGN--reference. 
{\sl {Column~1:}} simulation label. 
{\sl {Column~2:}} redshift. 
{\sl {Column~3:}} mass of multiphase gas outflowing with radial velocity $v_{\rm r} > 0$. 
{\sl {Column~4:}} mass of single-phase gas outflowing with radial velocity $v_{\rm r} > 0$. 
{\sl {Columns~5~and~6:}} same as Columns~3~and~4, but considering gas with radial velocity $v_{\rm r} > 50$~km~s$^{-1}$.
{\sl {Columns~7~and~8:}} same as Columns~3~and~4, but considering gas with radial velocity $v_{\rm r} > v_{\rm esc}$.
{\sl {Column~9:}} Escape velocity of the halo.
Total gas mass in outflow is given by the sum of single-phase and multiphase gas involved in galactic outflows. 
} 
\renewcommand\tabcolsep{3.8mm}
\begin{tabular}{@{}lcccccccc@{}}
\hline
Simulation  & $z$  &   $M_{\rm outf, \, mp}$    &    $M_{\rm outf, \, sp}$    &    $M_{\rm outf, \, mp}$    &    $M_{\rm outf, \, sp}$    &    $M_{\rm outf, \, mp}$    &    $M_{\rm outf, \, sp}$   &   $v_{\rm esc} $   \\ 
               &                 &   (M$_{\odot}$)                   &  (M$_{\odot}$)    &   (M$_{\odot}$)            &       (M$_{\odot}$)            &   (M$_{\odot}$)            &       (M$_{\odot}$)         & (km s$^{-1}$)                                           \\ 
\hline
\hline
 &  &    $v_{\rm r} > 0$ & &  $v_{\rm r} > 50$ km s$^{-1}$ & & $v_{\rm r} > v_{\rm esc}$   &  & \\
\hline
\hline
fiducial--hcAL--cf &  $z=2$  &   $1.94 \cdot 10^9$  &  $2.03 \cdot 10^{10}$ &  $7.41 \cdot 10^8$    & $1.52 \cdot 10^{10}$ &  $1.49 \cdot 10^7$  & $1.98 \cdot 10^9$  & $268.3$ \\ 
\hline
hcA--hot &  $z=2$  &  $3.27 \cdot 10^9$  &  $1.81 \cdot 10^{10}$ &  $1.28 \cdot 10^9$    & $1.26 \cdot 10^{10}$ &  $8.15 \cdot 10^6$  & $1.74 \cdot 10^9$ & $269.7$ \\  
\hline
hcA--both &  $z=2$  &   $1.92 \cdot 10^9$  &  $1.91 \cdot 10^{10}$ &  $7.72 \cdot 10^8$    & $1.36 \cdot 10^{10}$ &  $4.49 \cdot 10^5$  & $4.67 \cdot 10^8$  & $269.2$ \\ 
\hline
hcA--cold &  $z=2$  &   $1.50 \cdot 10^9$  &  $1.83 \cdot 10^{10}$ &  $7.65 \cdot 10^8$    & $1.29 \cdot 10^{10}$ &  $1.39 \cdot 10^7$  & $1.07 \cdot 10^9$  & $268.5$ \\  
\hline
noAGN--reference &  $z=2$  &   $2.20 \cdot 10^9$  &  $1.81 \cdot 10^{10}$ &  $7.33 \cdot 10^8$    & $1.26 \cdot 10^{10}$ &  $5.66 \cdot 10^5$  & $1.08 \cdot 10^9$  & $270.2$  \\  
\hline
\hline
fiducial--hcAL--cf &  $z=0$  &   $5.97 \cdot 10^9$  &  $5.46 \cdot 10^{10}$ &  $4.45 \cdot 10^8$    & $5.14 \cdot 10^{9}$ &  $1.73 \cdot 10^6$  & $1.92 \cdot 10^8$ & $250.6$  \\ 
\hline
hcA--hot &  $z=0$  &  $9.69 \cdot 10^9$  &  $6.03 \cdot 10^{10}$ &  $7.59 \cdot 10^8$    & $7.89 \cdot 10^{9}$ &  $1.61 \cdot 10^6$  & $9.17 \cdot 10^7$ & $250.9$  \\  
\hline
hcA--both &  $z=0$  &   $7.62 \cdot 10^9$  &  $5.27 \cdot 10^{10}$ &  $2.77 \cdot 10^8$    & $3.83 \cdot 10^{9}$ &  $2.46 \cdot 10^6$  & $1.32 \cdot 10^8$ & $250.5$ \\ 
\hline
hcA--cold &  $z=0$  &   $7.16 \cdot 10^9$  &  $4.90 \cdot 10^{10}$ &  $6.43 \cdot 10^8$    & $5.73 \cdot 10^{9}$ &  $1.07 \cdot 10^6$  & $3.42 \cdot 10^8$ & $250.7$ \\  
\hline
noAGN--reference &  $z=0$  &  $7.28 \cdot 10^9$  &  $5.19 \cdot 10^{10}$ &  $5.34 \cdot 10^8$    & $4.54 \cdot 10^{9}$ &  $2.18 \cdot 10^6$  & $1.22 \cdot 10^8$ & $249.3$ \\  
\hline
\hline
\end{tabular}
\label{ch10:Outflow_details}
\end{minipage}
\end{table*}

%%%%%%%%%%%%%%%%%%%%%%%%%%%%% outflow vel
\begin{figure*}
\vspace{1.05ex}
\newcommand{\captionfonts}{\small}
\centering
\includegraphics[trim=0.cm 8.cm 0.cm 0.cm, clip, width=1.\textwidth]{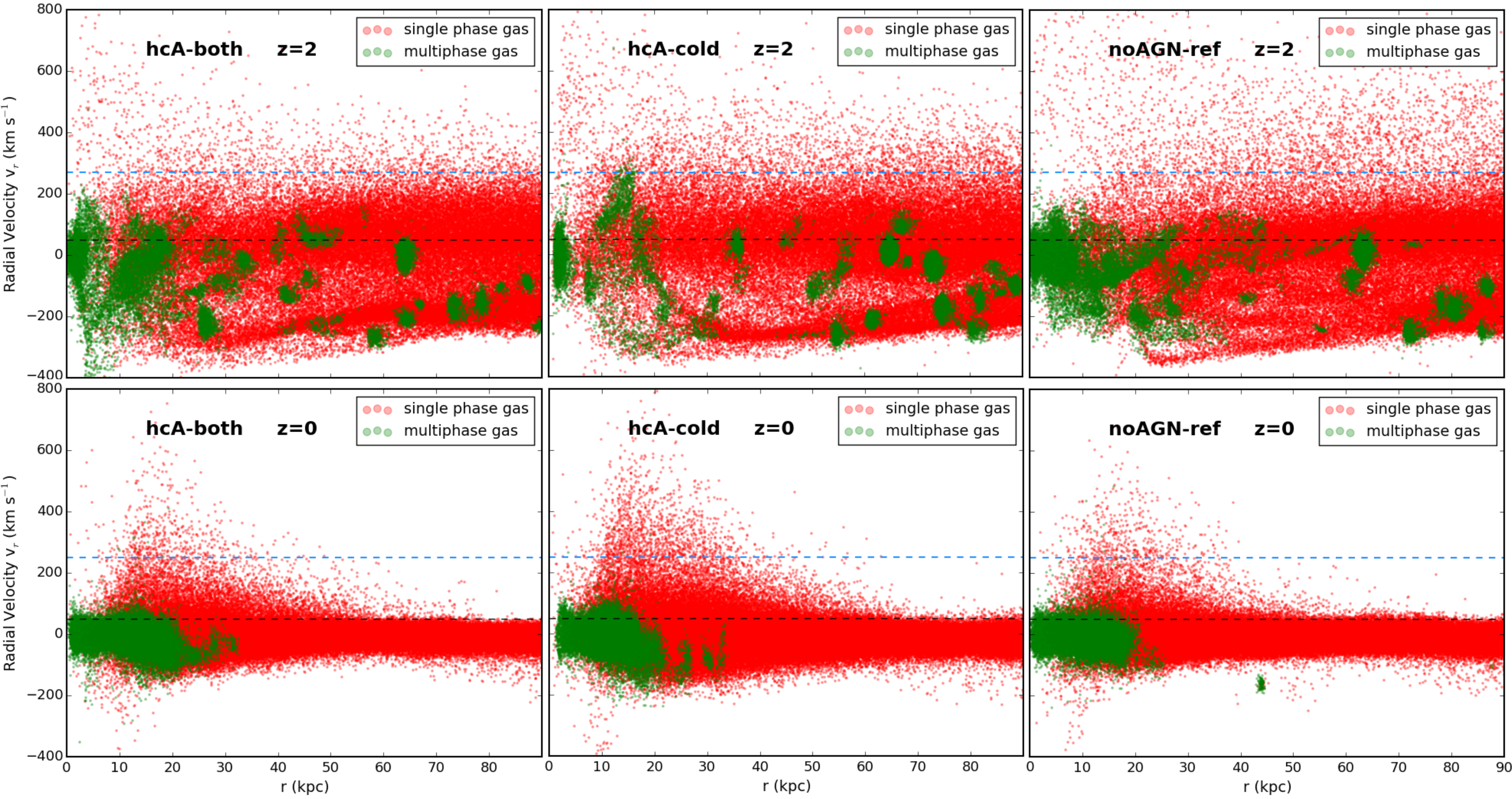}
\caption[]{Radial component of the velocity for single-phase (red) and multiphase (green) gas particles 
	as a function of their distance from the galaxy centre, at $z=2$ (top row) and at $z=0$ (bottom row). 
	{\sl{From left to right:}} simulations hcA--both, hcA--cold, and noAGN--reference. The horizontal, dashed 
	light blue line highlights the escape velocity of the halo for each model, while the dashed black line 
	pinpoints the reference velocity threshold of $v_{\rm r} = 50$~km~s$^{-1}$.}
\label{ch10:OutflVel} 
\end{figure*}
%%%%%%%%%%%%%%%%%%%%%%%%%%%%% outflow vel

On average, by analysing the mass of outflowing gas, the contribution of AGN to drive outflows in our simulations 
ranges between $20$~and~$50 \%$ at $z=2$ (see Table~\ref{ch10:Outflow_details}). 
The impact of AGN is even more sub-dominant with respect to star formation-driven outflows at $z=0$. 
Nevertheless, the role of AGN appears to be quite important in accelerating to high velocities the multiphase gas.  
Figure~\ref{ch10:OutflVel} shows 
the radial component of the velocity for both single-phase and multiphase gas particles, as a function 
of their distance from the galaxy centre, at $z=2$ (top row) and at $z=0$ (bottom row). 
For reference, in Figure~\ref{ch10:OutflVel}, we also show the radial velocity for single-phase and 
multiphase gas particles when the AGN is not included. 
A larger amount of gas is launched to higher velocities in the inner region of the galaxy 
($r \lesssim 30$~kpc), when the AGN feedback is considered. 
We do not find a significant difference in the temperature of galactic outflows according to whether the AGN 
feedback is included or not (the hot phase temperature of the bulk of outflowing gas spanning the range 
$\sim10^6$--$10^7$~K). We do not find a significant evolution of the outflow temperature between $z=2$ and $z=0$: 
the hot gas temperature of outflowing particles is higher by a factor of $\sim 2$ at most at $z=2$ 
in the innermost regions of the galaxy. 

To test the relative effect of SN- and AGN-triggered outflows in simulations, \citet{Costa2015}, for instance, 
investigated the impact of AGN-driven outflows in cosmological simulations of high-redshift quasars. Interestingly, they 
found that the combined action of SN and AGN feedback produces the largest mass of both cold and hot gas to the highest outflow speed, with respect to the case in which the AGN feedback is not included. 
\citet{Biernacki2018} found that AGN feedback is responsible for the formation of the hotter and lower density 
component of galactic outflows, and that it drives outflowing gas to larger distances from the galactic disc of 
simulated high-redshift galaxies.
At variance with our findings, \citet{Koudmani2019} find that the AGN activity promotes outflows to temperatures and 
velocities which are higher by up to two orders of magnitude in dwarf galaxies, by using isolated galaxy simulations. 

As highlighted by \citet{Veilleux2005}, it is hard to state whether
a galactic wind is powered either by starburst and star formation activity only
or by AGN activity alone. Recent observations have suggested 
correlations between outflow properties and ongoing AGN and star formation activity 
in systems with different mass: these relations can be exploited to distinguish between different 
mechanisms of outflow triggering \citep{ForsterSchreiber2019}. 
This topic represents a current challenge in cosmological simulations, too \citep[e.g.][]{Nelson2019}. 
We found that both stellar and AGN feedback contribute to trigger outflows. 
The disc galaxies that we simulate including AGN feedback do not have an AGN-driven outflow component 
that makes galactic outflows differ significantly from those of galaxies simulated without including AGN feedback, 
as we have quantified by analyzing outflow velocities and the mass of outflowing gas. 
The reason for this stems from the secondary role that AGN feedback is thought to play in systems 
of the same stellar mass as those that we are simulating. The picture emerging here is that AGN feedback 
in disc galaxies acts through a maintenance mode at low redshift, and provides a supporting role to stellar feedback.

%%%%%%%%%%%%%%%%%%%%%%%%%%%%% gas metallicity
\begin{figure}
\newcommand{\captionfonts}{\small}
%\vspace{-.5ex}
\centering
\includegraphics[trim=0.4cm 0.4cm 0.35cm 0.2cm, clip, width=.475\textwidth]{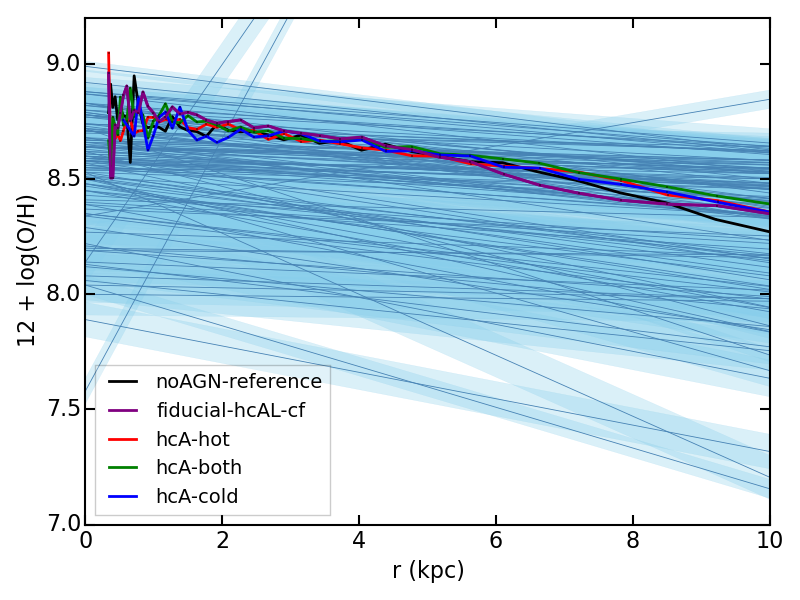} 
\caption[]{Oxygen abundance gradients of gas in the simulated galaxies. Light blue profiles are observations 
	from the sample of disc galaxies of \citet{Pilyugin2014}, with shaded envelopes depicting the scatter of 
	oxygen abundance around the trend.}
\label{ch10:IMF-1aB-met} 
\end{figure}
%%%%%%%%%%%%%%%%%%%%%%%%%%%%% gas metallicity

\subsection{Does AGN feedback affect metallicity profiles?} 
\label{ch10:Metals}

The goal of this section is to address the question of whether AGN feedback has an impact on the 
distribution of heavy elements within galaxies. 

%At variance with stellar feedback, the feedback from SMBHs is limited to energy injection, while it does not 
%supply metals to the surrounding medium. However, AGN feedback promotes powerful outflows: 
%they can foster the circulation of gas in and around galaxies, thus resulting in a modification of the heavy elements 
%distribution. 
AGN-triggered galactic outflows can indeed promote the circulation of gas in and around galaxies, 
thus resulting in a modification of the heavy elements distribution. In addition, while launching outflows, 
SMBHs can modify the chemo-galactic ecosystem because 
they could either expel pristine gas at high redshift, that later falls back and dilutes the local metal content of 
the galaxy; or they could eject outwards gas enriched from stellar feedback, depriving of metals the innermost regions 
of galaxies. In order to assess the possible importance of these processes, the slopes and the 
normalizations of metallicity gradients can reveal vital information.

%%%%%%%%%%%%%%%%%%%%%%%%%%%%% mappe met
\begin{figure*}
\newcommand{\captionfonts}{\small}
\centering
\includegraphics[trim=0.cm 2.cm 0.cm 5.5cm, clip, width=1.\textwidth]{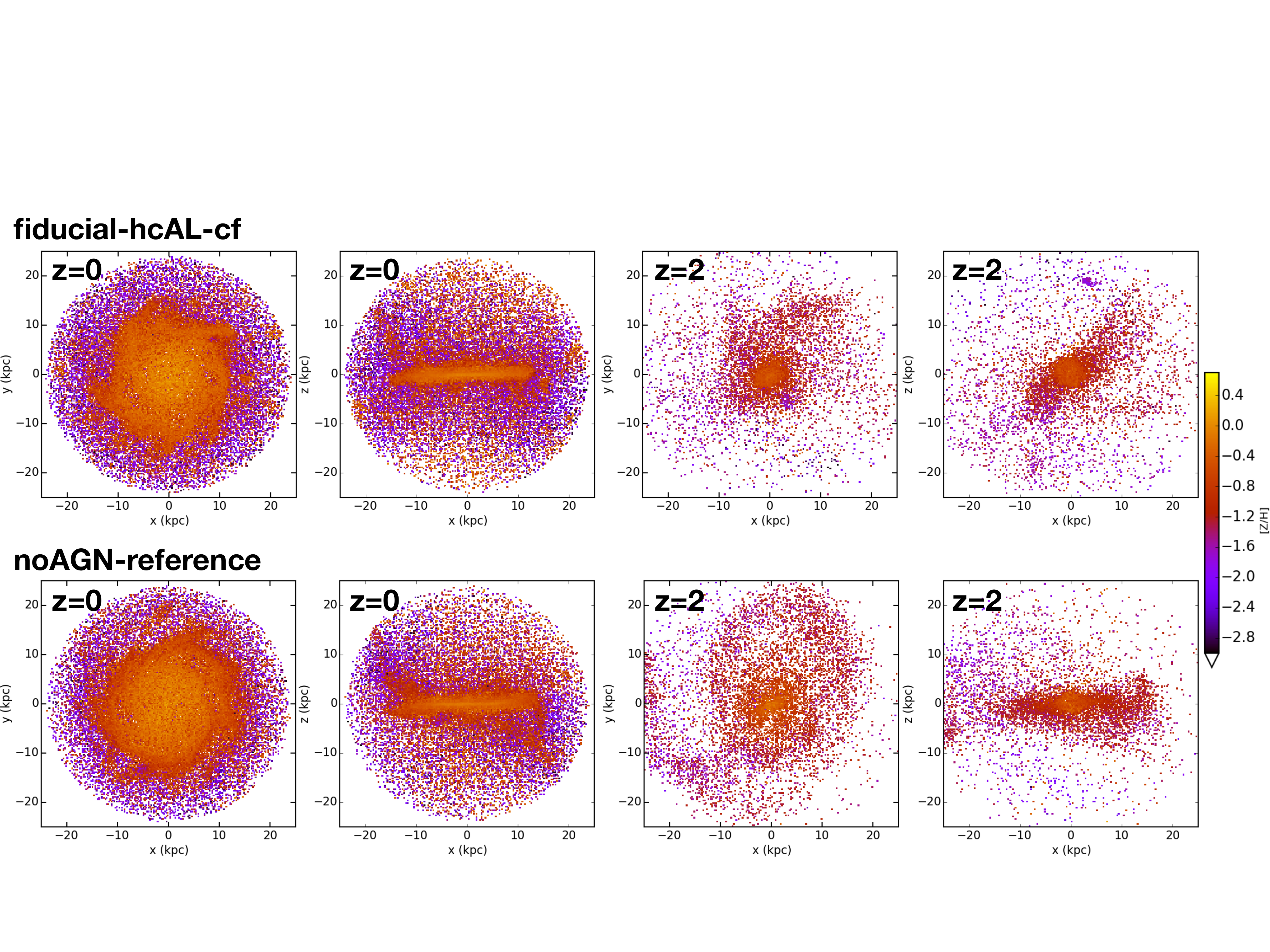} 
\caption[]{Face-on (first and third columns) and edge-on (second and fourth) binned distributions of all 
	the gas particles located within the galactic radius for the galaxy simulations fiducial--hcAL-cf and 
	noAGN--reference. The colour encodes the mean [Z/H] of the gas particles in the bin. 
	The distributions are shown at $z=0$ (first and second columns) and at $z=2$ (third and fourth ones).}
\label{ch10:MAppe_met} 
\end{figure*}
%%%%%%%%%%%%%%%%%%%%%%%%%%%%% mappe met

Figure~\ref{ch10:IMF-1aB-met} shows the oxygen abundance radial profiles of gas in the simulated 
galaxies, at $z=0$. 
Predictions from simulations are compared with observations 
from the sample of $130$ nearby late-type galaxies of \citet[][]{Pilyugin2014}. 
The present-day Sun's abundance in Oxygen is taken from \citet{Asplund2009}. 
Profiles of all the simulated galaxies are in agreement with observations. 
The simulations noAGN--reference and those including AGN feedback have almost 
indistinguishable metallicity profiles, that only show a slight difference beyond 
$r \gtrsim 6$~kpc from the galaxy centre (the discrepancy being as high as $0.1$~dex at $r=10$~kpc). 
The marginal difference at $r \gtrsim 8$~kpc between the noAGN--reference model and 
the other simulated galaxies is due to the AGN-stimulated star formation at $z\lesssim 0.1$ 
(see Figure~\ref{ch10:sfr}), that results in recent star formation occurring in the outer regions of the galaxy disc. 
The stellar mass surface density of both fiducial--hcAL--cf and hcA--both at $r=10$~kpc 
is twice as high as that of noAGN--reference (see Figure~\ref{ch10:SurfaceDensity}). 

The profiles of simulated galaxies share comparable slopes and normalization, this indicating that the 
AGN feedback does not affect significantly the distribution of heavy elements in the galaxy, at $z=0$. 
The negligible effect of AGN feedback in shaping the metallicity gradients stems from the inability of 
AGN feedback to significantly affect the circulation of metals at large distance from the galaxy centre 
(the outermost radius considered in Figure~\ref{ch10:IMF-1aB-met} is set by observational constraints, 
see \citet{Pilyugin2014}).

The picture emerging is that the normalization of the metallicity profiles is driven by the 
IMF \citep{Valentini2019}, while the AGN feedback has a negligible effect on them. 
The conclusions drawn in \citet{Valentini2019} as for the indication to prefer a \citet[][]{kroupa93} IMF 
(more top-light with respect to the Chabrier-like \citet{Kroupa2001}) for disc galaxies in the local Universe 
is confirmed and further corroborated when AGN feedback is included in our simulations. 

To further investigate the possible role of AGN feedback in affecting the distribution of metals also 
at larger distances from the galaxy centre with respect to those considered in Figure~\ref{ch10:IMF-1aB-met}, 
we analyse gas metallicity maps. Figure~\ref{ch10:MAppe_met} shows the face-on and edge-on 
distribution of all the gas particles located within 
the galactic radius $R_{\rm gal}$ (see Table~\ref{ch10:AGNmasseStellari}) of the fiducial--hcAL--cf and 
noAGN--reference galaxy simulations. We analyse the metallicity maps at redshift $z = 0$ and $z=2$. 
The colour encodes the mean metallicity [Z/H] of the gas particles in each spatial bin. 
As for the present-day Sun's metallicity, we adopt [Z/H]$_{\odot} = 0.0207\pm0.0015$ \citep{Bressan2012}. 
The two galaxies simulated with and without AGN feedback share a comparable metal content at $z=0$, 
the mean gas metallicity of the galaxy disc ranging from slightly super-solar to sub-solar 
($-0.4 \lesssim [Z/H] \lesssim 0.5$). At $z=2$, the galaxy fiducial--hcAL--cf which includes AGN feedback 
is characterised by lower-metallicity gas in the innermost regions of the forming galaxy with respect to the 
case noAGN--reference, as we can appreciate by comparing panels in the third and forth columns of 
Figure~\ref{ch10:MAppe_met}. Galactic outflows drive a larger amount of both single-phase 
and multi-phase gas in the galaxy fiducial--hcAL--cf with respect to the noAGN--reference model at $z=2$
(see Table~\ref{ch10:Outflow_details}): in this way, they expel gas previously enriched in heavy elements. 
As the metallicity profiles shown in Figure~\ref{ch10:IMF-1aB-met} highlight no significant differences at $z=0$ 
between galaxies simulated with and without AGN feedback, we can conclude that the role of AGN-driven outflows 
in driving enriched gas out of the sites of star formation is episodic and mainly confined 
at higher redshift.

\subsection{Angular momentum dependent accretion} 
\label{ch10:Mango}

In this section, we investigate how different models of gas accretion onto the SMBH impact on final results. 
In particular, we study the effect that limiting accretion of the cold gas that is supported by rotational velocity has 
on the evolution of the central BH and on its mass at $z=0$. 

\begin{figure}
\newcommand{\captionfonts}{\small}
%\vspace{-1.65ex}
\centering
\includegraphics[trim=0.1cm 0.1cm 0.45cm 1.cm, clip, width=.5\textwidth]{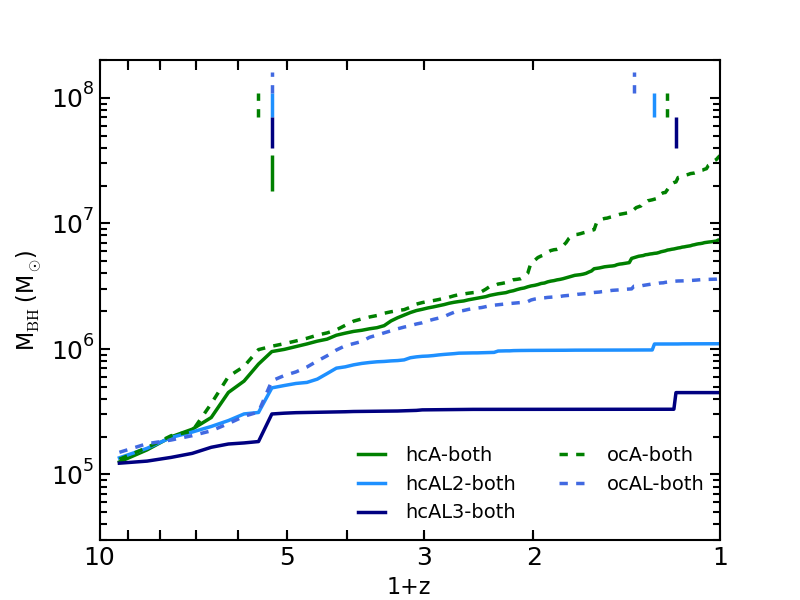} 
\caption[]{Evolution of the BH mass growth for the most massive BH in each of the 
	simulated galaxies. Segments at the top of the figure highlight the redshift at which the considered BH 
	experienced a BH merger, as in Figure~\ref{ch10:BHMA}. 
	Solid curves refer to simulations in which the SMBH accretes both hot and 
	cold gas ({\sl {hcA}}), while dashed curves identify the case of cold gas accretion only ({\sl {ocA}}). 
	All the simulations in this figure share the same AGN feedback model ({\sl {both}}). 
	See text for details on the labels.}
\label{ch10:BHMA_an_both} 
\end{figure}

Figure~\ref{ch10:BHMA_an_both} shows the evolution of the BH mass as a function of the redshift. 
As in Section~\ref{ch10:BHevo}, we consider the most massive BH within a distance of $100$~kpc from the 
galaxy centre. Vertical segments at the top of the figure record the redshift at which a BH merger 
occurred. 

All the simulations considered in Figure~\ref{ch10:BHMA_an_both} share the same AGN feedback model, 
with the AGN feedback energy which is provided to the multiphase medium being evenly distributed 
to the hot and cold phase. 
As for the AGN feeding, we consider four different models for BH gas accretion: 
$(i)$- Bondi accretion of both hot and cold gas (hcA--both), 
$(ii)$- Bondi accretion of cold gas only (ocA--both), 
$(iii)$- modified Bondi accretion of hot and cold gas, where the accretion of cold gas with high angular velocity 
is limited according to equation~(\ref{Mdot_limited_mango}) (hcAL2--both and hcAL3--both), 
$(iv)$- modified Bondi accretion of cold gas only, which is limited according to its rotational support (ocAL--both). 
In the latter simulation, all the gas that contributes to accretion is controlled by the limiter $ \mathcal{L}_{\rm AM} $. 
On the other hand, should the BH accrete both hot and cold gas, the limiter only controls cold gas accretion. 
The simulations hcAL2--both and hcAL3--both only differ from each other for the value of the 
parameter $C_{\rm visc}$ that describes the viscosity of the cold gas supported by rotational velocity, 
at the sub-resolution level. 
The values of $C_{\rm visc} $ that we explore are the following: 
$C_{\rm visc} / 2 \, \pi = 1$ (ocAL--both)
$C_{\rm visc} / 2 \, \pi = 10^2$ (hcAL2--both), and 
$C_{\rm visc} / 2 \, \pi = 10^3$ (hcAL3--both), 
and correspond to those considered by \citet{Schaye2015, Crain2015}. 

As discussed in \citet{RosasGuevara2015}, the parameter $C_{\rm visc} $ encodes 
the sub-resolution parametrisation of the viscosity. 
The factor $\, \mathcal{L}_{\rm AM} = C_{\rm visc} ^{-1} \, (c_{\rm s, \, c} / V_{\phi})^3 \, $  
(see equation~(\ref{AngMomLimiter})), which reduces the Bondi accretion rate (see equation~(\ref{Mdot_limited_mango})), is equivalent to the ratio between the Bondi timescale 
and the viscous timescale ($t_{\rm visc}$, see below). Indeed, the presence of gas with a 
given amount of angular momentum introduces a characteristic spatial scale when modelling BH accretion, 
that is the size of the disc on which gas orbits before infalling onto the BH. Depending on the angular momentum 
of the gas, a typical timescale is set. This viscous timescale enters the problem in addition to the Bondi timescale, 
that is valid for gas accretion under the assumption that accreting gas does not rotate while infalling. 
The Bondi timescale reads: 
$\, t_{\rm B} = r_{\rm B} / c_{\rm s} \,$, where $r_{\rm B}$ is the Bondi radius, i.e. the scale where the BH dominates 
over hydrodynamical processes (see Section~\ref{sec:introduction}), and $c_{\rm s}$ is the sound speed of gas. 
The viscous timescale is proportional to the dynamical time, and can be cast as: 
$\, t_{\rm visc} = [ \alpha_{\rm visc} \, (H/R)^2 ]^{-1} \, t_{\rm dyn} \,$ 
\citep[see][for further details]{RosasGuevara2015}. 
Here, $R$ and $H$ are the radius and the scale height of the accretion disc, respectively, while 
$ \alpha_{\rm visc} $ is a dimensionless parameter that is related to the kinematic 
viscosity. If the transport processes through the viscous accreting disc were fully understood, 
adequately accurate values for $R$, $H$, and $ \alpha_{\rm visc} $ 
would be inserted to model the effective accretion process. However, since we lack a full 
understanding of viscosity and accretion, and also considering that the accretion disc is far from being 
resolved in cosmological simulations, the ignorance is parametrised by 
means of $\, C_{\rm visc} = 2 \, \pi \, [ \alpha_{\rm visc} \, (H/R)^2 ]^{-1}  \,$. In this way, 
a parameter is introduced in order to numerically capture the viscosity of gas in rotational support 
on a notional accretion disc, at the sub-resolution level. 
The (highly uncertain) value of the effective viscosity parameter $C_{\rm visc}$ has been first explored by \citet{RosasGuevara2015}, 
who proposed for it the range $10^3 \div 10^8$. They adopted $C_{\rm visc} = 2 \cdot 10^6$ as fiducial value, 
and found that the larger is the value of $C_{\rm visc}$, the smaller is the mass of the hosted SMBH, 
for DM haloes as massive as $\sim 10^{12}$~M$_{\odot}$. 
Note that $C_{\rm visc} \rightarrow \infty$ would correspond to the case in which (cold) gas accretion onto 
the SMBH is not included. Also, note that large variations in $C_{\rm visc}$ are required in order to have 
remarkable differences in the final results, since the suppression of the gas accretion by angular momentum needs 
$\, C_{\rm visc} ^ {1/3} \, V_{\phi} > c_{\rm s} \, $ to result effective (see equation~(\ref{AngMomLimiter})). 
Decreasing $C_{\rm visc}$ corresponds to the situation in which the viscosity of the disc is higher and 
the gas accretion proceeds faster: indeed, viscosity transports angular momentum outward, enabling accreting gas
to spiral in towards the BH.

%%%%%%%%%%%%%%%%%%%%%%%%%%%%% BHMA   **both**   hcA, ocA, hcAL
%%%%%%%%%%%%%%%%%%%%%%%%%%%%% BHMA   **both**   hcA, ocA, hcAL

%%%%%%%%%%%%%%%%%%%%%%%%%%%%% BHMA   **cf** plus both   hcA, hcAL
\begin{figure}
\newcommand{\captionfonts}{\small}
%\vspace{-1.65ex}
\centering
\includegraphics[trim=0.1cm 0.1cm 0.45cm 1.cm, clip, width=.5\textwidth]{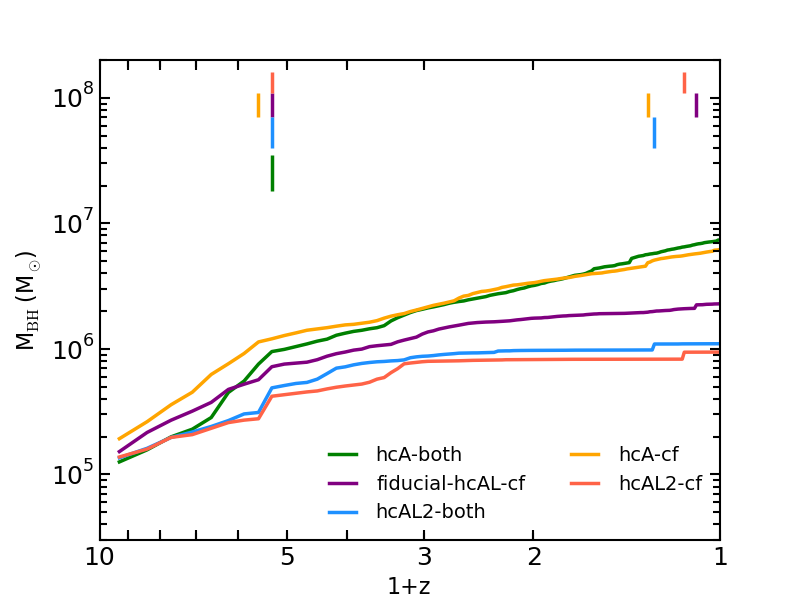} 
\caption[]{Same as Figure~\ref{ch10:BHMA_an_both}, but considering other simulations which 
	also include our fiducial model.}
\label{ch10:BHMA_an_cf} 
\end{figure}
%%%%%%%%%%%%%%%%%%%%%%%%%%%%% BHMA   **cf** plus both   hcA, hcAL

As a final remark, we note that state-of-the art cosmological simulations are still far from resolving 
the BH accretion disc ($\sim$~pc and sub-pc~scale) and fully capturing the physics of processes 
which occur in the proximity of the BH. 
Quantities that enter $\mathcal{L}_{\rm AM}$ (see equation~(\ref{AngMomLimiter})) are estimated by 
considering the gas particles within the smoothing length of the BH ($\sim$~kpc~scale, see Section~\ref{ch10:BHevo}). 
As a consequence, the value of $C_{\rm visc}$ does not only parameterize the loosely constrained properties of 
BH accretion discs: it also encodes our ignorance of the unresolved processes occurring below the resolution limit 
of the simulation.

The evolution of the BH mass of hcA--both has been already considered in Figure~\ref{ch10:BHMA}. 
Should only cold gas be accreted (ocA--both), the BH grows faster below $z \lesssim 1$. 
Despite the accretion of a reduced amount of gas (with respect to hcA--both; note that the BH of 
ocA--both does not accrete the hot gas that is located within its smoothing sphere), the BH of 
ocA--both grows more massive and faster because of the reduced feedback. The gas in the proximity of the 
BH is denser in hcA--both than in ocA--both (by a factor of $\sim 2$ at $z=1$, and 
by a factor of $\sim 1.5$ at $z=0$, at $r=1$~kpc from the galaxy centre). 

When the limiter to the accretion of cold gas that is supported by rotational velocity is accounted for, the 
evolution of the BH mass changes: reducing the accretion of cold gas delays or even suppresses the BH growth. 
The higher the values of $C_{\rm visc}$ that are adopted, the more significant is the BH growth reduction. 
Focussing on hcAL2--both, the BH mass growth is hindered below $z \sim 4$; finally, the BH is 
prevented from growing below $z \sim 2$, aside from a BH merger at $z\sim 0.2$. 
The BH growth by gas accretion is even stopped when $C_{\rm visc} / 2 \, \pi = 10^3$ (hcAL3--both) 
is adopted. With respect to the reference simulation hcA--both, the BH mass at $z=0$ is reduced by  
an order of magnitude or even more, the BH masses being $1.10 \cdot 10^6$~M$_{\odot}$ (hcAL2--both) 
and $4.48 \cdot 10^5$~M$_{\odot}$ (hcAL3--both).

The scenario that emerges from Figure~\ref{ch10:BHMA_an_both} has interesting implications for 
the formation and evolution of disc galaxies. Indeed, the fundamental connection between the $M_{\rm BH}$ 
and $M_{\rm bulge}$ implies that the growth of the BH is tightly linked to the formation of galaxy 
bulges \citep{KormendyHo2013}. 
Even if the coevolution with pseudo-bulges is not as close as for classical bulges, 
BHs do not correlate with the properties of galaxy discs \citep{Kormendy2001, Kormendy2011, KormendyHo2013}. 
As a consequence, the SMBH hosted in a late-type galaxy is expected to grow mainly at high-redshift, while 
the formation of the bulge of the galaxy proceeds. On the contrary, the BH growth is limited at low-redshift, 
when the formation of the galaxy disc occurs \citep[e.g.][]{Greene2008, Shankar2012}. 

Figure~\ref{ch10:BHMA_an_both} shows that this is the case for hcAL2--both: below redshift $z \sim 3 \div 2$ 
the growth of the BH through gas accretion is almost insignificant. Below redshift $z \sim 2$, the BH mass only grows 
as a consequence of mergers with other BHs. The effect of the limiter $ \mathcal{L}_{\rm AM} $ is clearly evident 
when considering the simulated galaxy hcAL3--both: below redshift $z \sim 4$, the growth of the BH by gas 
accretion is suppressed, and the BH mass evolution is driven by mergers alone. 
As the formation of the bulge and of the disc of our simulated galaxies typically occurs 
above and below $z \sim 3 \div 2$, respectively \citep[see Figure~\ref{ch10:sfr} and][]{Valentini2019}, 
what emerges from Figures~\ref{ch10:BHMA_an_both}~and~\ref{ch10:BHMA_an_cf} 
is that once the angular momentum of the cold gas that is accreting onto the central BH is taken into account, 
the aforementioned scenario for BH-disc galaxy coevolution is successfully reproduced. 

Figure~\ref{ch10:BHMA_an_cf} shows the evolution of the BH mass as a function of the redshift for 
another suite of simulations, including our fiducial model. 
The simulation hcA--cf has a SMBH which grows in a way that is similar to that of hcA--both. 
The effect of suppressing the accretion of high angular momentum cold gas reduces the BH mass 
of hcAL2--cf (wrt hcA--cf) at $z=0$ by an amount that is comparable to what obtained 
in hcAL2--both (wrt hcA--both). 
This figure illustrates that the conclusions drawn so far when discussing Figure~\ref{ch10:BHMA_an_both} 
are still valid when AGN feedback energy is distributed 
to the multiphase ISM according to the physical properties of the gas. 

The scenario of BH merging as the most viable channel for BH growth in galaxies within haloes 
of $\sim10^{12}$~M$_{\odot}$ has been pointed out by \citet{Bonoli2016}, who presented a
detailed study of the coevolution of a MW-sized simulated galaxy and its SMBH. 
Interestingly, they found that the SMBH growth is mainly due to the mergers with other BHs located within 
satellites approaching the forming galaxy, rather than to gas accretion. At $z=0$, their simulated galaxy hosts 
a BH as massive as $2 \cdot 10^6$~M$_{\odot}$.

The suppression of the low-redshift BH growth due to the angular momentum limiter is evident 
when the BH accretes both the hot and cold gas (solid curves in Figure~\ref{ch10:BHMA_an_both}), 
as well as when the BH only accretes cold gas (dashed curves). 
Should only cold gas be accreted, a lower value of $C_{\rm visc}$ is enough to have roughly the same 
BH mass reduction that is observed with a higher value of $C_{\rm visc}$ when the BH accretes 
both the hot and cold gas. Indeed, the difference between the final BH mass of ocAL--both 
($3.61 \cdot 10^6$~M$_{\odot}$) and of ocA--both ($3.46 \cdot 10^7$~M$_{\odot}$) 
is comparable to that which distinguishes hcAL2--both from hcA--both. 
Therefore, when only cold gas accretion is considered, the delay and suppression of the BH mass growth 
is obtained by assuming a higher viscosity for the disc on which the accreting gas is settled, i.e. by assuming 
an accretion not as impeded as if it involved both hot and cold gas. 

The values of $C_{\rm visc}$ for hcAL2--both and hcAL3--both are the ones adopted in the 
EAGLE simulations \citep{Schaye2015}. 
They adopt $C_{\rm visc} / 2 \, \pi = 10^2$ for the simulation of their suite at lower resolution (in which 
the initial mass of gas particles is larger than that of AqC5 by a factor $4.4$, see Section~\ref{simus}), 
and $C_{\rm visc} / 2 \, \pi = 10^3$ for their higher resolution run (where the initial mass of gas particles is smaller 
than that of AqC5 by a factor $1.8$). 
Indeed, \citet{Crain2015} explored the variation on galaxy scaling relations produced by different values 
of $C_{\rm visc}$ (they considered the following values: $C_{\rm visc}/ 2 \, \pi = 0.01 ,\, 1, \, 100 $), to 
quantify the impact of the sub-resolution viscosity in calibrating the EAGLE simulation. They found 
that their model for AGN feedback is primarily dependent on $C_{\rm visc}$. Also, they found that 
larger values of $C_{\rm visc}$ delay the BH growth via gas accretion and the quenching of star formation 
by AGN feedback. Also, they observed that the higher is the value of $C_{\rm visc}$, the lower is 
the BH mass corresponding to a determined stellar mass (of the bulge) of the host galaxy. Our results 
are thus in keeping with \citet{Crain2015}. 
As a future direction of investigation, it would be interesting to explore how the aforementioned 
scenario of low-redshift suppression of gas acccretion fits within the framework of Seyfert galaxy evolution, 
being gas accretion the most viable channel for BH mass growth at low redshift in these systems.

\subsection{Overview of previous works and comparison}
\label{uffachepizza}

In this Section, we focus on the modelling of AGN feedback in a multiphase ISM and on the positive 
AGN feedback which enhances star formation. The physical idea behind our modelling is in line with 
results from several higher-resolution ($\sim$~pc) simulations which explicitly model the clumpy, 
multiphase ISM \citep[e.g.][]{Wagner2012, Wagner2013, Gaibler2012}.
For instance, \citet{Wagner2013} resolve a two-phase ISM, where the cold clouds are in pressure equilibrium 
with the hot ambient medium. Once the cold gas receives AGN feedback energy, it is progressively eroded 
and disperses; the hot gas receives the bulk of the feedback energy, which is deposited mostly in thermal form, 
and is pushed to further distances. Interestingly, they find that when the cold clouds are distributed in a disc, 
the AGN feedback uplifts them from the galactic plane and compresses the gas, while the warm gas in the 
disc inflows towards the galaxy centre shortly after. The interaction between AGN-driven outflows and the 
multiphase ISM often results in a positive feedback, with jet-induced and pressure-triggered star formation \citep{Wagner2012, Gaibler2012}. Final results are not sensitive to the opening angle of the AGN jet 
\citep{Wagner2013}, while the cross-section of cold clouds (to be related to the covering factor in our modelling) 
determines how effective is the response of the cold ISM to the AGN feedback energy injection  \citep{Wagner2012}. 
High-resolution simulations by \citet{Bourne2014, Bourne2015} investigating the impact of outflows 
from a SMBH on the ISM of the innermost region of a galaxy show that the bulk of feedback energy is coupled to 
the hot phase and carried away from the centre of the host through paths of least resistance within the clumpy 
multiphase ISM; on the other hand, the bulk of the momentum is coupled to the cold component. Interestingly, 
\citet{Bourne2015} discuss how an adequately high resolution and a detailed modelling of the ISM is crucial to 
recover the AGN-induced compression, this result being missed in lower-resolution simulations and the 
negative effect of the AGN feedback being thus over-predicted. 

High-resolution simulations with idealised and simplified initial conditions which study the impact of AGN outflows 
on a multiphase ISM are also important to understand where the AGN-triggered star formation is expected to occur. 
For instance, \citet{Nayakshin2012} find that star formation bursts are produced within thin and dense layers of 
cold gas surrounding the shocked and compressed hot gas out of which they form \citep[see also][]{Zubovas2013}. 
\citet{Zubovas2017} discuss the twofold role of AGN-triggered outflows, which can either suppress or enhance 
star formation, in line with our results. They find that spots of AGN-induced star formation are located in cold and 
dense knots of gas which get compressed within hot feedback bubbles \citep[see also][]{Zubovas2014}; also, 
star formation occurs at the outer edge of the hot outflow bubble, highlighting that AGN outflows compress 
swept up gas and locally induce star formation. Clumps with ongoing star formation can also form along 
the backflow of AGN inflated bubbles, the backflow ultimately reaching the galaxy disc, 
where further star formation is triggered \citep{Silk2013, Bieri2016}.

\section{Conclusions}
\label{sec:conclusions}

We introduced a novel model for AGN feedback and implemented it within 
the sub-resolution model MUPPI, already featuring cooling, star formation, stellar feedback and chemical enrichment. 
We carried out a suite of cosmological hydrodynamical simulations of disc galaxies, with zoomed-in 
initial conditions leading to the formation of a halo of mass $M_{\rm halo, \, DM} \simeq 2 \cdot 10^{12}$~M$_{\odot}$ 
at redshift $z=0$. These simulations have been designed to investigate: 
\begin{itemize}
\item [-] the effect of different ways of coupling AGN feedback energy to the hot and cold phases of the 
multiphase ISM; 
\item [-] the impact of different models of gas accretion onto SMBHs, namely only cold gas, both cold and hot gas, 
with the additional possibility of limiting gas accretion from cold gas with high angular momentum; 
\item [-] how different models of gas accretion and AGN feedback energy coupling affect the overall BH-galaxy coevolution. 
\end{itemize}

\noindent
The most relevant results of this work can be summarised as follows.
\begin{itemize}
\item We investigated the effect that coupling AGN feedback energy to the different phases of the ISM 
has on the evolution of SMBHs. Providing to the hot phase the entire budget of feedback energy, or a considerable 
fraction of it, produces a SMBH that grows in mass by up to one or two order of magnitudes from $z \sim 8.5$ to 
$z =0$. Its final mass ranges between $\sim10^6$~M$_{\odot}$ and $\sim10^7$~M$_{\odot}$. 
On the other hand, when the AGN feedback energy is entirely supplied to the cold phase, the multiphase 
ISM is not promoted to outflow and remains close to the BH: as a consequence, 
the BH experiences a rapid phase of mass growth (in the redshift range $3 \lesssim z \lesssim 5$) 
during which it deprives of gas the centre of the host galaxy, and then its growth is suppressed. 
The BH mass at $z =0$ can be as high as $\sim10^9$~M$_{\odot}$ in this case. 
Therefore, a prediction of our model is that at least a share of the AGN feedback energy has to couple with 
the diffuse hot gas. 

\item We examined the effect of coupling the AGN feedback energy injected in a multiphase ISM 
to its phases according to their physical properties. We considered a model where 
feedback energy coupling is driven by the covering factors of the hot and cold phases, assuming 
that the larger is the volume occupied by the cold gas clumps, the larger is the amount of energy 
that the cold gas absorbs. 

\item Gas accretion is the process that contributes the most to the BH growth, rather than mergers with 
other BHs (Section~\ref{ch10:BHevo}). Throughout BH evolution, the total accretion rate is dominated 
by cold gas accretion, with respect to hot. Remarkably, this is in line with \citet{Gaspari2019}, 
even if they derived such a conclusion by means of a phenomenological analysis focussed on more massive systems. 
Our conclusion that gas accretion is a more viable channel for BH growth than BH mergers is true 
unless cold gas which is supported by rotational velocity is prevented from accreting (see below). 
The quiet merging history that characterises our simulated disc galaxies also contributes to the minor role 
of BH mergers to the BH mass growth. The contribution to the BH growth by BH-BH mergers is expected 
to become more significant for increasing BH mass \citep{Fanidakis2011, Dubois2014b, Weinberger2018}.

\item We find that when the BH only accretes cold gas, it experiences a growth by gas accretion that is 
faster than (or at most comparable to) the case in which both cold and hot gas are accreted. 
As for the galaxy $M_{\rm bulge}$-$M_{\rm BH}$ scaling relation, predictions from 
simulations are in keeping with observations, considering a BH seed mass as large as 
$\sim10^5$~M$_{\odot}$.

\item When the accretion of cold gas that is supported by rotational velocity is reduced, the 
BH mass growth is delayed and the BH mass at $z=0$ is reduced by up to an order of magnitude with 
respect to the case in which both hot and cold gas accretion proceed unimpeded. Within the scenario 
emerging from this model, the SMBH in MW-sized galaxies is prevented from growing by gas accretion 
below $z \lesssim 2$, aside from possible BH mergers. 
Also, the lower is the viscosity assumed for the cold gas which is accreting (i.e. the larger $C_{\rm visc}$), 
the more the BH mass growth is reduced. 

\item Simulations that include AGN feedback produce spiral galaxies with a more 
extended stellar disc component. The SMBH mainly 
produces a positive feedback in our simulated late-type galaxies, pressurising the multiphase gas 
and ultimately enhancing the low-redshift star formation. 
We expect that this conclusion changes when we will simulate elliptical galaxies, galaxy groups and clusters, 
where the AGN feedback is supposed to be mainly negative. 

\item 
Including AGN feedback does not affect the slope nor the normalization of metallicity gradients at low redshift. 
As a consequence, we can conclude that the energy injection and especially the outflows 
driven by the SMBH (Section~\ref{ch10:galacticOutflows}) do not alter significantly the circulation of metals 
within galaxies (Section~\ref{ch10:Metals}). 
\end{itemize}

AGN feedback operates in a wide class of systems, from galactic bulges to massive galaxy clusters. 
In this work, we studied the way in which AGN feedback energy is coupled to the different 
components of the ISM focussing on late-type galaxies: however, AGN feedback is expected to play a supporting role 
in MW-sized galaxies, where the central BH contributes to determine the properties of the bulge 
and shape the galaxy innermost regions. 
A forthcoming extension of this study is to include elliptical galaxies in the analysis, so as to investigate how the 
AGN affects the formation and evolution of systems characterised by more massive BHs accreting at higher rates.
Also, we plan to investigate the effect of stellar and AGN feedback on 
the properties of simulated galaxy populations, instead of those of single galaxies.

At the same time, it would be interesting to further investigate hot gas haloes in massive spiral galaxies. 
\citet{Valentini2015} studied the hot gas cooling process triggered by AGN feedback in models of low mass 
ellipticals: provided the similarity between the hot ISM of these systems and detected hot coronae 
around disc galaxies, it appears likely that any outburst from the central BH could promote cooling of the hot corona, 
generating cold clouds that would accrete on the galactic disc.

As for the modelling of AGN feedback, a further direction for improvement 
is provided by the additional work that is required to numerically account for the mechanical component of 
AGN-triggered outflows. In the model that we have introduced, all the AGN feedback energy is coupled 
thermally and isotropically to the surrounding medium, with a constant efficiency. 
Besides the radiative feedback, our implementation 
is thus still missing the modelling of the mechanical AGN feedback, that is considered to be the dominant 
channel in which the AGN operates for low-activity stages (i.e. low accretion rates) of the BH (radio-mode). 
Moreover, we have here followed a rather simplified approach also for the modelling of gas accretion 
onto SMBHs. Albeit the Bondi-like accretion is adopted in the majority of cosmological simulations 
nowadays, and despite having corrected it by taking into account the angular momentum of cold gas which is 
accreted, we aim at achieving a more accurate description of the BH accretion in the near future. 
A desirable possibility to attain is the so-called chaotic cold accretion 
\citep{gaspari2013, Gaspari2017}, 
where cold blobs condensed out of the hot ambient medium fall onto the SMBH and are 
accreted after they undergo chaotic inelastic collisions. In this way, the BH accretion rate is boosted and 
the AGN activity consists of frequent bursts. 
The modelling of this regime for AGN feeding will be a key one especially once processes 
such as cooling, heating, turbulence, and rotation are consistently accounted for and the resolution is increased. 
This will be extremely important when considering more massive systems. 

Another quite interesting challenge would be to compare predictions from these simulations to observations 
at high redshift, in order to better constrain the BH growth across cosmic time and interpret possible scenarios 
of galaxy evolution.

\section*{Acknowledgments}
We thank the anonymous referee for the careful and constructive report that helped 
improving the presentation of results. 
We thank Volker Springel for making the GADGET3 code available to us. 
We are grateful to Lucio Mayer, Gabriella De Lucia, Klaus Dolag, Annalisa Pillepich, Massimo Gaspari, 
and Filippo Mannucci for constructive discussions and feedback. 
SB acknowledges financial support from PRIN-MIUR 2015W7KAWC, 
the agreement ASI-INAF n.2017-14-H.0, 
the INFN INDARK grant. 
SB, GM, GG, and LT are also supported by the EU H2020 Research and Innovation Programme under the 
ExaNeSt project (Grant Agreement No. 671553). 
AB and AL acknowledge support by PRIN MIUR 2017 prot.20173ML3WW 002 ``Opening the ALMA window on the cosmic evolution of gas, stars and supermassive black holes''. 
Simulations were carried out using ULISSE at SISSA, Marconi at CINECA, 
and the Trieste ``baastet'' cluster at INAF-Osservatorio Astronomico di Trieste (Italy). CPU time has been 
assigned through the project Sis18\_bressan under Convenzione SISSA,
through the project INA17\_C1A00, 
and through Italian Super-Computing Resource Allocation (ISCRA) proposals and an agreement 
with the University of Trieste. 
The post-processing has been performed using the PICO HPC cluster at CINECA through our expression of interest.

%%%%%%%%%%%%%%%%%%%%%%%%%%%%%%%%%%%%%%%%%%%%%%%%%%
%%%%%%%%%%%%%%%%%%%%%%%%%%%%%%%%%%%%%%%%%%%%%%%%%%

%%%%%%%%%%%%%%%%%%%% REFERENCES %%%%%%%%%%%%%%%%%%

% The best way to enter references is to use BibTeX:

\bibliographystyle{mnras} 
\bibliography{cool_ref}

%%%%%%%%%%%%%%%%%%%%%%%%%%%%%%%%%%%%%%%%%%%%%%%%%%
%%%%%%%%%%%%%%%%%%%%%%%%%%%%%%%%%%%%%%%%%%%%%%%%%%

%%%%%%%%%%%%%%%%% APPENDICES %%%%%%%%%%%%%%%%%%%%%
%%%%%%%%%%%%%%%%%%%%%%%%%%%%%%%%%%%%%%%%%%%%%%%%%%

\appendix

%%%%%%%%%%%%%%%%%%%%%%%%%%%%%%%%%%%%%%%%%%%%%%%%%%%%%%%%%%

\section{Constant coupling parameters}
\label{ch10:ConstantcouplingFactors}

We consider three different possibilities as test cases to investigate how AGN feedback energy 
couples to the surrounding multiphase ISM: 
\begin{enumerate}
\item [-] All the energy is provided to the hot gas phase ($\mathcal{A}_{\rm h}=1$ and $\mathcal{A}_{\rm c} = 0$). 
\item [-] AGN feedback energy is entirely supplied to the cold component ($\mathcal{A}_{\rm h}=0$ and $\mathcal{A}_{\rm c} = 1$). 
\item [-] The energy assigned to each multiphase particle is evenly shared among the hot 
and cold gas ($\mathcal{A}_{\rm h}=0.5$ and $\mathcal{A}_{\rm c} = 0.5$).
\end{enumerate}

%%%%%%%%  Hot Only
When $\mathcal{A}_{\rm h}=1$ and $\mathcal{A}_{\rm c}=0$, the system of 
equations~(\ref{ch9:AGNmuppi1}),~(\ref{ch9:AGNmuppi2}),~(\ref{ch9:AGNmuppi3}),~and~(\ref{ch9:AGNmuppi4}) 
reduces to: 
\begin{align}
\dot{M}_{\rm h}  &=  - \dot{M}_{\rm cool} + \dot{M}_{\rm ev} 
				  \,\,\,, \label{ch9:AGNmuppi1h} \\
\dot{M}_{\rm c}  &=   \dot{M}_{\rm cool}  - \dot{M}_{\rm sf} - \dot{M}_{\rm ev} 
				  \,\,\,,  \label{ch9:AGNmuppi2h}  \\
\dot{M}_{\ast}  &=  \dot{M}_{\rm sf}  \,\,\,,  \label{ch9:AGNmuppi3h} \\
\dot{E}_{\rm h}  &=  \dot{E}_{\rm fb, local} - \dot{E}_{\rm cool} + \dot{E}_{\rm hydro} 
			    + \dot{E}^{\rm AGN}_{\rm h}  \,\,\,, \label{ch9:AGNmuppi4h} 
\end{align}
and there is no need anymore for integrating equation~(\ref{ch9:AGNmuppi5}). 

%%%%%%%%  Cold Only
On the other hand, if $\mathcal{A}_{\rm h}=0$ and $\mathcal{A}_{\rm c}=1$, the 
system of equations to be integrated is: 
\begin{align}
\dot{M}_{\rm h}  &=  - \dot{M}_{\rm cool} + \dot{M}_{\rm ev} 
				+  \dot{M}^{\rm AGN}_{\rm c \rightarrow h}  \,\,\,, \label{ch9:AGNmuppi1c} \\
\dot{M}_{\rm c}  &=   \dot{M}_{\rm cool}  - \dot{M}_{\rm sf} - \dot{M}_{\rm ev} 
				-  \dot{M}^{\rm AGN}_{\rm c \rightarrow h} \,\,\,,  \label{ch9:AGNmuppi2c}  \\
\dot{M}_{\ast}  &=  \dot{M}_{\rm sf}  \,\,\,,  \label{ch9:AGNmuppi3c} \\
\dot{E}_{\rm h}  &=  \dot{E}_{\rm fb, local} - \dot{E}_{\rm cool} + \dot{E}_{\rm hydro} 
			     + \dot{E}^{\rm AGN}_{\rm c \rightarrow h} \,\,\,, \label{ch9:AGNmuppi4c} \\
\dot{E}^{\rm AGN}_{\rm c, \, used}  &=  \dot{E}^{\rm AGN}_{\rm c \rightarrow h} \,\,\,, \label{ch9:AGNmuppi5c} 
\end{align}
where the only source term $ \dot{E}^{\rm AGN}_{\rm h}$ is missing in equation~(\ref{ch9:AGNmuppi4c}). 

%%%%%%%%  0.5_0.5 
When $\mathcal{A}_{\rm h}=0.5$ and $\mathcal{A}_{\rm c} = 0.5$, the general description of the model 
outlined in Section~\ref{AGNmuppi} is valid. 
%%%%    conti con Granato nel caso 0.5_0.5
Interestingly, in this case when 
\begin{equation} 
E^{\rm AGN}_{\rm c }  = E^{\rm AGN}_{\rm h }  =  \frac{1}{2} \, E^{\rm AGN}_{\rm fb } \,\,\,,  % E_AGN tot 
\label{ch9:GL0}
\end{equation}
it is worth to analytically quantify the mass of initially cold gas that can be 
evaporated and brought to the hot phase, i.e. $M^{\rm AGN}_{\rm c \rightarrow h}$, 
and cast it as a function of the initial mass of the hot gas in the 
multiphase particle, $M_{\rm h, \, init}$, i.e. before receiving AGN feedback energy. 

Under the simplified assumptions that there is enough cold gas to receive all the feedback 
energy $E^{\rm AGN}_{\rm c }$, so that $M^{\rm AGN}_{\rm c \rightarrow h} = M^{\rm AGN}_{\rm c, \, th}$ 
(see equation~(\ref{ch10:sph4k})) and $E^{\rm AGN}_{\rm c, \, extra} =0$ (see equation~(\ref{ch9:extraEnergy})), 
and that contributions from cooling and evaporation are neglected, in order to focus on the mass flow induced 
by the AGN feedback, it is possible to proceed as follows. 
From equation~(\ref{ch9:AGNmuppi6_th}): 
\begin{equation}     
M^{\rm AGN}_{\rm c \rightarrow h} = E^{\rm AGN}_{\rm c } 
\, \frac{(\gamma -1) \, \mu \, m_{\rm p}}{k_{\rm B} \, (T_{\rm h, \, fin} - T_{\rm c})}  \,\,\,;
\label{ch9:GL1}
\end{equation}
%the final temperature of the hot phase after the energy contribution by the AGN-induced 
%evaporation of the cold gas alone (without considering the further contribution $E^{\rm AGN}_{\rm h }$) reads: 
the final temperature of the hot phase after the energy contribution by the AGN, $E^{\rm AGN}_{\rm h }$, reads:
\begin{equation}     
T_{\rm h, \, fin} = T_{\rm h, \, init} + E^{\rm AGN}_{\rm h } 
\, \frac{(\gamma -1) \, \mu \, m_{\rm p}}{k_{\rm B} \, M_{\rm h, \, init}}  \,\,\,. 
\label{ch9:GL2}
\end{equation}

By approximating $(T_{\rm h, \, fin} - T_{\rm c}) \simeq T_{\rm h, \, fin}$ and 
plugging equation~(\ref{ch9:GL2}) into equation~(\ref{ch9:GL1}): 
\begin{align}     
M^{\rm AGN}_{\rm c \rightarrow h} & = E^{\rm AGN}_{\rm c } 
\, \frac{(\gamma -1) \, \mu \, m_{\rm p}}{k_{\rm B} \, \bigl( T_{\rm h, \, init} + E^{\rm AGN}_{\rm h } 
\, \frac{(\gamma -1) \, \mu \, m_{\rm p}}{k_{\rm B} \, M_{\rm h, \, init}} \bigr) }  \nonumber \\
& = E^{\rm AGN}_{\rm c } 
\, \frac{(\gamma -1) \, \mu \, m_{\rm p} \, M_{\rm h, \, init}}{k_{\rm B} \, T_{\rm h, \, init} \, M_{\rm h, \, init} + 
E^{\rm AGN}_{\rm h } \, (\gamma -1) \, \mu \, m_{\rm p}} \nonumber \\
& = E^{\rm AGN}_{\rm c } 
\, \frac{M_{\rm h, \, init}}{k_{\rm B} \, T_{\rm h, \, init} \, \frac{M_{\rm h, \, init}}{(\gamma -1) \, \mu \, m_{\rm p}} + 
E^{\rm AGN}_{\rm h }} \nonumber \\
& = M_{\rm h, \, init} \, \frac{E^{\rm AGN}_{\rm c }}{ E^{\rm AGN}_{\rm h } + M_{\rm h, \, init} \, 
\frac{k_{\rm B} \, T_{\rm h, \, init}}{(\gamma -1) \, \mu \, m_{\rm p}}}  \,\,\,.
\label{ch9:GL3}
\end{align}
Then, using equation~(\ref{ch9:GL0}): 
\begin{equation}     
M^{\rm AGN}_{\rm c \rightarrow h}  = M_{\rm h, \, init} \, \frac{E^{\rm AGN}_{\rm fb }}{ E^{\rm AGN}_{\rm fb } 
+ 2 \, M_{\rm h, \, init} \, \frac{k_{\rm B} \, T_{\rm h, \, init}}{(\gamma -1) \, \mu \, m_{\rm p}}}   < M_{\rm h, \, init} \,\,\,. 
\label{ch9:GL4}
\end{equation}

As a consequence, assuming $\mathcal{A}_{\rm h}=\mathcal{A}_{\rm c} = 0.5$, the hot gas mass and thus 
the hot gas density can increase by up to a factor of $\lesssim 2$, at most. Therefore, the hot gas phase will not 
experience a runaway cooling, and the SPH temperature of the multiphase particle is not expected to change 
significantly due to the AGN-induced transfer of cold gas to the hot phase.  

%%%%%%%%%%%%%%%%%%%%%%%%%%%%%%%%%%%%%%%%%%%%%%%%%%%%%%%%%%

%%%%%%%%%%%%%%%%%%%%%%%%%%%%%%%%%%%%%%%%%%%%%%%%%%%%%%%%%%

\section{Effect of BH seed mass} 
\label{ch10:CalibMago}

The initial mass assumed for BHs in cosmological simulations (see Section~\ref{BHseeding}) is rather important, 
and has fundamental implications for theoretical models, as it is linked to the mass of SMBH progenitors and to viable 
scenarios of SMBH formation. 
The value adopted for the BH seed mass is crucial when simulating MW-sized galaxies 
(i.e. $M_{\rm halo, \, DM} \simeq 10^{12}$~M$_{\odot}$ at redshift $z=0$) in a cosmological 
context: indeed, since BH growth due to gas accretion is relatively moderate in these galaxies, 
final results are quite sensitive to the value adopted for $M_{\rm BH, \, seed}$. 

The adopted value of $M_{\rm BH, \, seed}$ is closely connected to the choice of $M_{\rm DM, thresh}$ 
(see Section~\ref{BHseeding}). Indeed, a lower mass threshold 
$M_{\rm DM, thresh}$ for the DM halo within which BHs can be seeded translates directly to the introduction of 
the BH at higher redshift. In this section, we explore the impact that the value assumed for $M_{\rm BH, \, seed}$ has 
on final results. 
We consider the following values for BH seed masses: 
$M_{\rm BH, \, seed} = 1.1 \cdot 10^5$~M$_{\odot}$ (reference value), 
$M_{\rm BH, \, seed} = 5.5 \cdot 10^4$~M$_{\odot}$ ({\sl {S0.5x}}), and 
$M_{\rm BH, \, seed} = 2.7 \cdot 10^5$~M$_{\odot}$ ({\sl {S2x}}) 
(see Table~\ref{simList}). 
BH seeds as massive as $\sim 10^5$~M$_{\odot}$ would correspond to a formation scenario for SMBHs 
by direct collapse \citep[e.g.][]{Begelman2006}.

We consider a first set of three simulations: hcA--both, hcA--both--S0.5x, 
and hcA--both--S2x (see Table~\ref{simList}). They share the same setup and physics, and they 
only differ for the assumed $M_{\rm BH, \, seed}$. 

\begin{figure}
\newcommand{\captionfonts}{\small}
%\vspace{-.2ex}
\centering
\includegraphics[trim=0.4cm 0.4cm 0.35cm 0.2cm, clip, width=0.47\textwidth]{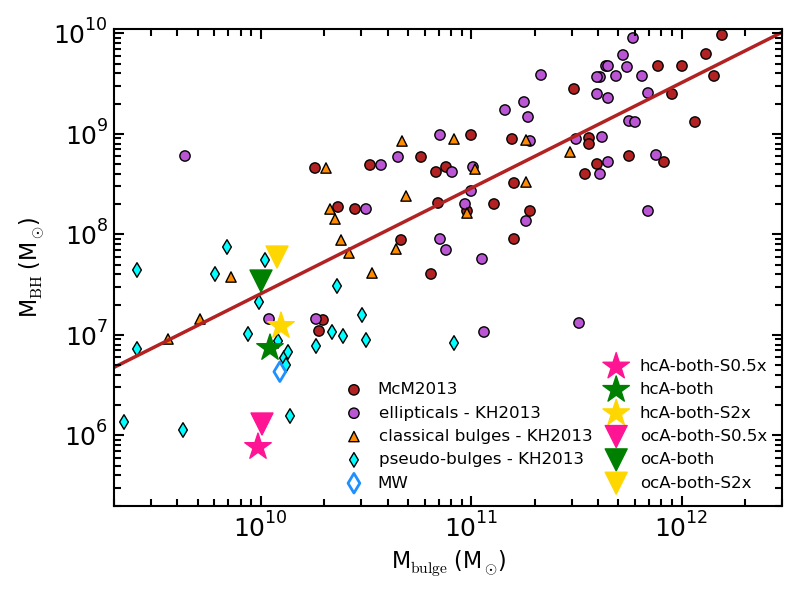} 
\caption[]{$M_{\rm bulge}$-$M_{\rm BH}$ relation 
	for BHs in the simulated galaxies where different BH seed masses are considered (see Table~\ref{simList}). 
	Simulations pinpointed by starlets assume both cold and hot gas accretion ({\sl {hcA}}), while simulations 
	identified by triangles assume only cold gas accretion ({\sl {ocA}}). Observations are from \citet[][KH2013]{KormendyHo2013} and 
	from \citet[][McM2013]{McConnell2013}, as in Figure~\ref{ch10:mago_ref}.}
\label{ch10:mago_seed}            
\end{figure}

Masses of their BHs at $z=0$ are as follows: 
$M_{\rm BH} = 7.4 \cdot 10^6$~M$_{\odot}$ (hcA--both), 
$M_{\rm BH} = 7.7 \cdot 10^5 $~M$_{\odot}$ (hcA--both--S0.5x), and 
$M_{\rm BH} = 1.2 \cdot 10^7$~M$_{\odot}$ (hcA--both--S2x). 
Figure~\ref{ch10:mago_seed} shows the $M_{\rm bulge}$-$M_{\rm BH}$ relation for the simulated galaxies. 
We compare the outcome of the three simulations (identified by stars) to observations 
(see Section~\ref{ch10:BHevo}). 
The seed mass of the BHs is indeed commonly calibrated in order to reproduce observed scaling relations 
at redshift $z=0$. 
The simulation adopting $M_{\rm BH, \, seed} = 1.1 \cdot 10^5$~M$_{\odot}$ is the one that best 
agrees with observations. A BH seed mass as large as twice the reference value also leads to a good agreement 
with observations. On the other hand, decreasing $M_{\rm BH, \, seed}$ by a factor of $\sim 2$ with respect to 
the fiducial value, would decrease the $M_{\rm BH}$ at $z=0$ by an order of magnitude. This worsens significantly 
the matching with observations in Figure~\ref{ch10:mago_seed}. 
It is not straightforward to relate the BH seed mass to other properties of the simulated galaxies, and 
to highlight definite trends. For instance, hcA--both, hcA--both--S0.5x, and hcA--both--S2x have the 
following stellar mass: $2.87 \cdot 10^{10}$~M$_{\odot}$, $2.58 \cdot 10^{10}$~M$_{\odot}$, 
and $2.07 \cdot 10^{10}$~M$_{\odot}$, respectively. As for their bulge-over-total mass ratios, the 
$B/T$ of hcA--both, hcA--both--S0.5x, and hcA--both--S2x is as follows: $0.38$, $0.37$, and $0.60$, respectively.

%%%%%%%%%%%%%%%%%%%%%%%%%%%%% mago seed
%%%%%%%%%%%%%%%%%%%%%%%%%%%%% mago seed

We also consider three additional simulations: ocA--both, ocA--both--S0.5x, and ocA--both--S2x. 
They are analogous to the first set as for the adopted BH mass seeds and model of the coupling of AGN feedback 
energy, but the BH in these simulations only accretes cold gas (see Table~\ref{simList}). 
In this way, we investigate whether the prediction for the most suitable value of $M_{\rm BH, \, seed}$ is 
unchanged when the details of the gas accretion modelling are varied. 
At $z=0$, the most massive BH within each of the simulated galaxy has the following mass: 
$3.5  \cdot 10^7$~M$_{\odot}$ (ocA--both), 
$1.3 \cdot 10^6$~M$_{\odot}$ (ocA--both--S0.5x), and 
$6.0 \cdot 10^7$~M$_{\odot}$ (ocA--both--S2x).
The stellar mass of the bulge of simulated galaxies does not depend on whether only cold or 
both hot and cold gas is accreted. 

For this second set of simulations (triangles), the lowest value for $M_{\rm BH, \, seed}$ leads to a 
simulated galaxy on the edge of the region of the $M_{\rm bulge}$-$M_{\rm BH}$ relation where observations 
are found. The reference and the 
highest values for $M_{\rm BH, \, seed}$ predict a SMBH that is located in the upper edge of the region 
of the plane occupied by pseudo-bulges.
When only cold gas accretion is assumed, BHs grow more massive than the case in which 
both hot and cold gas accretion is considered.  

The location at $z=0$ of a SMBH on the plane of the $M_{\rm bulge}$-$M_{\rm BH}$ relation loosely constrains 
the way in which it coevolved with its host galaxy. 
However, when $M_{\rm BH, \, seed}=1.1 \cdot 10^5$~M$_{\odot}$ is adopted, 
the BH is required to roughly increase its mass by an order of magnitude or slightly more 
between the redshift $z$ at which it has been seeded and $z=0$. 
Such a requirement seems to favour the scenario according to which SMBHs accrete both hot and cold gas, 
at least when the reference seed mass is adopted and when the AGN feedback energy provided to the multiphase 
ISM is evenly shared by the hot and the cold phase. 
All the BHs in the simulations considered are seeded at $z \sim 8.5$: this redshift is 
closely related to the (fixed) value of $M_{\rm DM, thresh}$. 

As a consequence, we adopt $M_{\rm BH, \, seed}=1.1 \cdot 10^5$~M$_{\odot}$ as the fiducial value 
for the BH mass seed. Albeit the exploration of the parameter space for $M_{\rm BH, \, seed}$ has been 
carried out for a single galaxy rather than for galaxies in a cosmological box, and even if resolution effects 
can enter the calibration, the reference value for $M_{\rm BH, \, seed}$ can be considered as representative 
of typical progenitors of MW-sized BHs at $z=0$. 
According to predictions from the simulations considered here, SMBH progenitors as massive 
as~$\sim10^5$~M$_{\odot}$ should already be in place at redshift $z \gtrsim 8$. This poses a challenging 
question from a theoretical perspective \citep[][]{Begelman2006}, 
given the age of the Universe at that time ($\sim 0.6 \div 0.7$~Gyr).

%%%%%%%%%%%%%%%%%%%%%%%%%%%%%%%%%%%%%%%%%%%%%%%%%%%%%%%%%%

\section{Effect of $\ell_{\rm MC}$} 
\label{CalibLmc}

%%%%%%%%%%%%%%%%%%%%%%%%%%%%% mago ell_MC
\begin{figure}
\newcommand{\captionfonts}{\small}
%\vspace{-.2ex}
\centering
\includegraphics[trim=0.4cm 0.4cm 0.35cm 0.2cm, clip, width=0.47\textwidth]{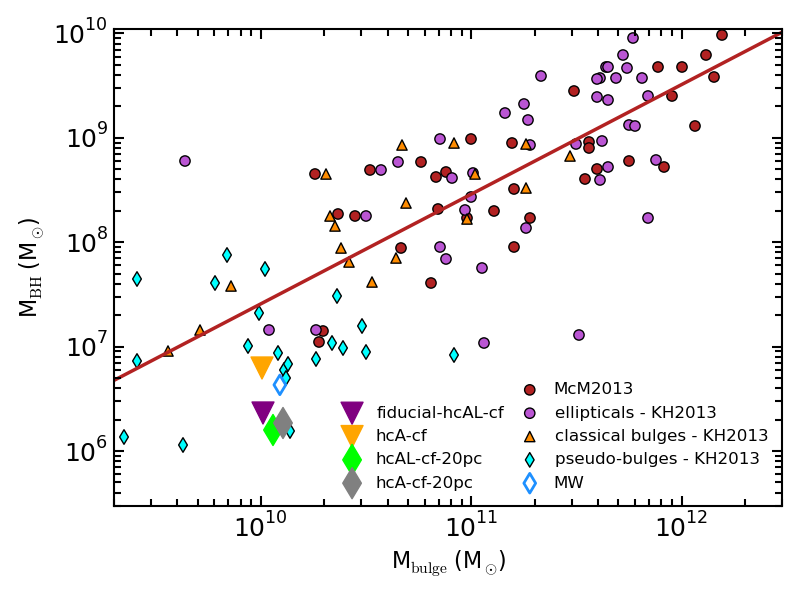} 
\caption[]{$M_{\rm bulge}$-$M_{\rm BH}$ relation 
	for BHs in the simulated galaxies where different $\ell_{\rm MC}$ are considered (see Table~\ref{simList}). 
	Simulations pinpointed by triangles assume the fiducial value $\ell_{\rm MC}=1$~pc, while simulations 
	identified by diamonds assume $\ell_{\rm MC}=20$~pc. Observations as in Figure~\ref{ch10:mago_ref}.}
\label{ch10:mago_lmc}            
\end{figure}
%%%%%%%%%%%%%%%%%%%%%%%%%%%%% mago ell_MC

In this section we investigate the impact of the parameter $\ell_{\rm MC}$, describing the typical size 
assumed for clumps within molecular clouds (see Section~\ref{ch10:CoveringFactors}). 
It enters in the sharing of AGN feedback energy among the hot and the cold 
phase of the multiphase ISM: the lower $\ell_{\rm MC}$, the larger $\mathcal{C}_{\rm c}$, 
when a multiphase particle with given physical properties is considered (see equation~(\ref{ch9:GL10})). 

We consider four simulations: hcA--cf, fiducial--hcAL--cf, hcA--cf--20pc, and hcAL--cf--20pc. 
They adopt either $\ell_{\rm MC}=1$~pc or $\ell_{\rm MC}=20$~pc (see Table~\ref{simList}). 
Further test runs carried out adopting $\ell_{\rm MC}=5$~pc and $\ell_{\rm MC}=30$~pc confirm the trends 
outlined here. 
For instance, the mean values for $\mathcal{C}_{\rm h}$ and $\mathcal{C}_{\rm c}$ in the 
simulation hcAL--cf--20pc are $\sim 0.76$ and $\sim 0.24$, respectively. 
The fiducial model fiducial--hcAL--cf has the following mean values for 
$\mathcal{C}_{\rm h}$ and $\mathcal{C}_{\rm c}$: $0.41$ and $0.59$ (see Section~\ref{ch10:GenRes}).

Figure~\ref{ch10:mago_lmc} shows the position of the BHs of the simulated galaxies on the plane of 
the $M_{\rm bulge}$-$M_{\rm BH}$ relation. The comparison with observations highlights that hcA--cf--20pc and 
hcAL--cf--20pc lie on the lower edge of the region occupied by pseudo-bulges. 
This implies that a smaller value of $\ell_{\rm MC}$ has to be preferred, that is also in better agreement 
with what observations suggest 
\citep[e.g.][and references therein; see Section~\ref{ch10:CoveringFactors}]{Williams1994, Bergin2007, Munoz2007, Gomez2014}.

%%%%%%%%%%%%%%%%%%%%%%%%%%%%%%%%%%%%%%%%%%%%%%%%%%%%%%%%%%

\section{Evolution of outflowing gas} 
\label{ForReferee}

\begin{figure}
\newcommand{\captionfonts}{\small}
%\vspace{-2.ex}
\raggedright
\includegraphics[trim=0.4cm 0.1cm 0.35cm 0.2cm, clip, width=.45\textwidth]{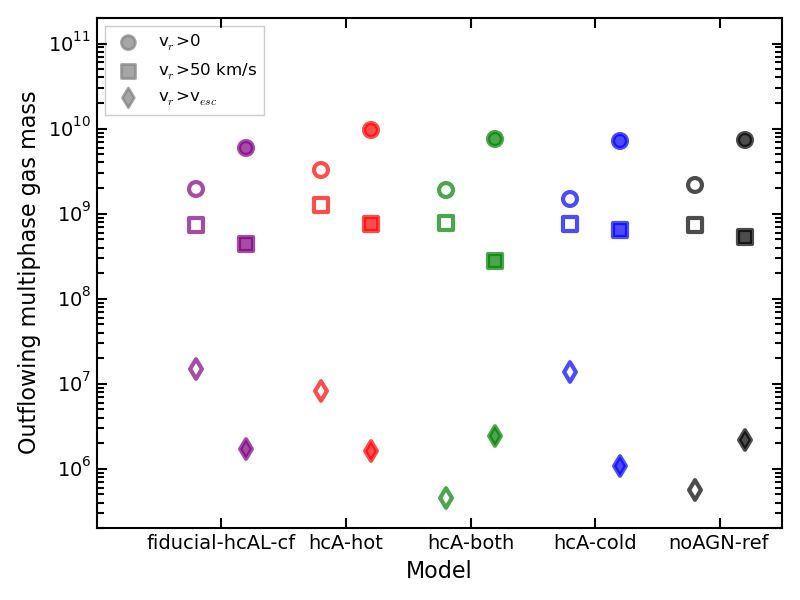} 
\includegraphics[trim=0.4cm 0.1cm 0.35cm 0.2cm, clip, width=.45\textwidth]{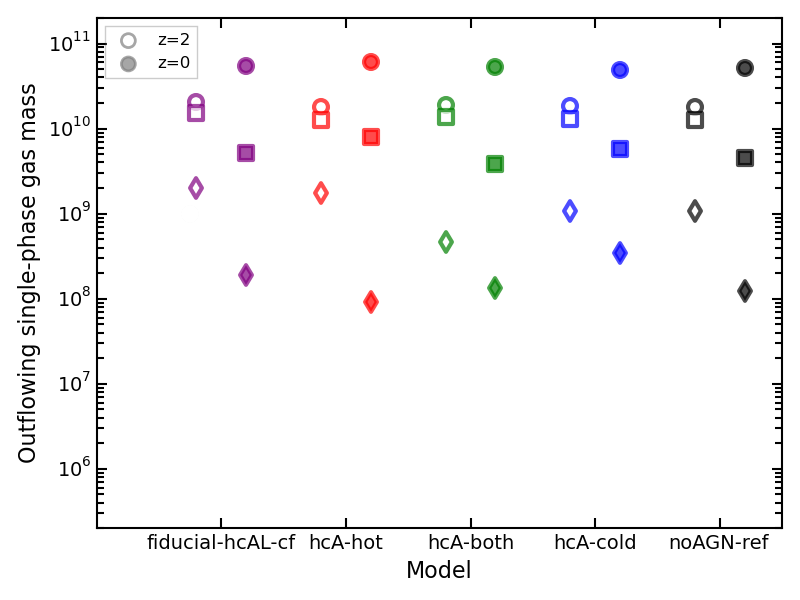}
\caption[]{Mass of multiphase (top panel) and single-phase (bottom panel) gas which is outflowing 
with positive radial velocity (circles) or radial velocity exceeding $50$~km~s$^{-1}$ (squares) 
and the escape velocity of the halo (diamonds; see Table~\ref{ch10:Outflow_details} for details). 
Quantities are analyzed at redshift $z=2$ (empty symbols) and $z=0$ (filled symbols). }
\label{ForRef} 
\end{figure}

In this section we show the evolution of the mass of multiphase and single-phase gas involved in outflows. 
Figure~\ref{ForRef} displays the content of Table~\ref{ch10:Outflow_details} (see Section~\ref{ch10:galacticOutflows} 
for details).

%%%%%%%%%%%%%%%%%%%%%%%%%%%%% outflow geometry
%%%%%%%%%%%%%%%%%%%%%%%%%%%%% outflow geometry

%%%%%%%%%%%%%%%%%%%%%%%%%%%%%%%%%%%%%%%%%%%%%%%%%%%%%%%%%%
%%%%%%%%%%%%%%%%%%%%%%%%%%%%%%%%%%%%%%%%%%%%%%%%%%%%%%%%%%

%%%%%%%%%%%%%%%%%%%%%%%%%%%%%%%%%%%%%%%%%%%%%%%%%%
%%%%%%%%%%%%%%%%%%%%%%%%%%%%%%%%%%%%%%%%%%%%%%%%%%

% Don't change these lines
\bsp	% typesetting comment
\label{lastpage}
\end{document}